\documentclass[aps,prd,reprint,floatfix,superscriptaddress,showkeys,nofootinbib]{revtex4-1}

\setcitestyle{authoryear,round}
\usepackage[utf8]{inputenc}
\usepackage[T1]{fontenc}

\usepackage{graphicx}

\usepackage{savesym}
\savesymbol{tablenum}
\usepackage{siunitx}
\restoresymbol{SIX}{tablenum}

\usepackage{mathtools}
\savesymbol{corresponds}
\usepackage{mathabx}  
\restoresymbol{ABX}{corresponds}
\usepackage{bm}

\usepackage{verbatim}
\usepackage{rotate}
\usepackage{color}
\usepackage{aas_macros}
\usepackage{tikz}
\usetikzlibrary{calc}
\usepackage{ulem}
\DeclareFontFamily{OT1}{pzc}{}
\DeclareFontShape{OT1}{pzc}{m}{it}%
            {<-> s * [1.10] pzcmi7t}{}
\DeclareMathAlphabet{\mathscr}{OT1}{pzc}%
                                {m}{it}

\definecolor{RedWine}{rgb}{0.743,0,0}
\definecolor{green(pigment)}{rgb}{0.0,0.65,0.31}
\definecolor{RoyalBlue}{rgb}{0.25,0.41,0.88}

\newcommand{\be}{\begin{equation}}
\newcommand{\ee}{\end{equation}}
\newcommand{\bea}{\begin{eqnarray}}
\newcommand{\eea}{\end{eqnarray}}
\def\ba#1\ea{\begin{align}#1\end{align}}

\def\({\left(}
\def\){\right)}
\def\<{\left\langle}
\def\>{\right\rangle}

\newcommand{\vs}{\nonumber\\}

\def\vr{{\bm{r}}}

\def\vk{{\bm{k}}}

\def\vq{{\bm{q}}}

\def\khat{{\hat{\bm{k}}}}

\def\qhat{{\hat{\bm{q}}}}
\def\rhat{{\hat{\bm{r}}}}

\def\nbar{{\bar{n}}}

\def\orderof{\mathcal{O}}

\DeclareSIUnit \parsec {pc}
\DeclareSIUnit \h {\text{$h$}}
\DeclareSIUnit \year {yr}
\DeclareSIUnit \solarmass {M_\odot}
\DeclareSIUnit \Mpc {\mega\parsec}

\def\obs{\mathrm{obs}}

\def\true{\mathrm{true}}

\def\min{\mathrm{min}}
\def\max{\mathrm{max}}

\def\dd{\mathrm{d}}

\def\myapp#1#2{%
  \mathrel{%
    \setbox0=\hbox{$#1\sim$}%
    \setbox2=\hbox{%
      \rlap{\hbox{$#1\propto$}}%
      \lower1.1\ht0\box0%
    }%
    \raise0.25\ht2\box2%
  }%
}

\newcommand{\incgraph}[2][0.49]{\includegraphics[width=#1\textwidth]{#2}}

\usepackage[colorlinks]{hyperref}
\usepackage[capitalise]{cleveref}

\hypersetup{
colorlinks=true,
citecolor=cyan,
pagebackref=true,}

\begin{document}

\title{SuperFaB: a fabulous code for Spherical Fourier-Bessel decomposition}

\author{Henry S. \surname{Grasshorn Gebhardt}}
\email{henry.s.gebhardt@jpl.nasa.gov}
\thanks{NASA Postdoctoral Program Fellow}
\affiliation{Jet Propulsion Laboratory, California Institute of Technology, Pasadena, CA 91109, USA}
\affiliation{California Institute of Technology, Pasadena, CA 91125, USA}

\author{Olivier \surname{Dor\'e}}
\affiliation{Jet Propulsion Laboratory, California Institute of Technology, Pasadena, CA 91109, USA}
\affiliation{California Institute of Technology, Pasadena, CA 91125, USA}

\begin{abstract}
  The spherical Fourier-Bessel (SFB) decomposition is a natural choice for the
  radial/angular separation that allows extraction of cosmological
  information from large volume galaxy surveys, taking into account all wide-angle effects.
  In this paper we develop a SFB
  power spectrum estimator that allows the measurement of the largest angular
  and radial modes with the next generation of galaxy surveys. The code
  measures the pseudo-SFB power spectrum, and takes into account mask,
  selection function, pixel window, and shot noise. We show that the local
  average effect (or integral constraint) is significant only in the largest-scale mode, and we provide
  an analytical covariance matrix. By imposing boundary conditions at the
  minimum and maximum radius encompassing the survey volume, the estimator does
  not suffer from the numerical instabilities that have proven challenging for SFB analyses in
  the past. The estimator is demonstrated on simplified but realistic \emph{Roman}-like,
  \emph{SPHEREx}-like, and \emph{Euclid}-like mask and selection functions. For
  intuition and validation, we also explore the SFB power spectrum in the
  Limber approximation. We release the associated public code written in
  \texttt{Julia}.
\end{abstract}

\keywords{cosmology; large-scale structure}

\maketitle

\newcommand{\codename}{\emph{SuperFaB}}

\section{Introduction}
One of the aims of future galaxy surveys such as the \textit{Nancy Grace Roman
Space Telescope}, \textit{SPHEREx}, \textit{Euclid}, DESI, PFS
\citep{Spergel+:2015arXiv150303757S, Dore+:2014arXiv1412.4872D,
Amendola+:2018LRR....21....2A, DESICollaboration:2016arXiv161100036D,
Takada+:2014PASJ...66R...1T} is to answer questions that require measurement of
the galaxy overdensity power spectrum on very large cosmological scales.
Chiefly among those is the study of modified theories of gravity
\citep[e.g.][]{Munshi+:2016MNRAS.456.1627M} and the measurement of primordial
non-Gaussianity that manifests itself in the power spectrum as a
scale-dependent galaxy bias $\propto \, k^{-2}$ in the simplest models
\citep{Salopek+:1990PhRvD..42.3936S, Komatsu+:2001PhRvD..63f3002K,
Dalal+:2008PhRvD..77l3514D, Desjacques+:2018PhR...733....1D}. However, the
measurement of these large scales is not without challenge.

Large \emph{angular} scales are difficult to exploit fully with a standard 3D
power spectrum analysis due to line-of-sight effects such as redshift-space
distortions (RSD). When the angular separation between galaxies is large, the
assumption that a single line of sight can be used for both galaxies breaks
down, which results in a loss of information from the measurement. For example,
for a full-sky survey, a fixed LOS estimator is expected to measure a vanishing
quadrupole. On medium large scales the problem can be mitigated by choosing a
common line of sight for each pair of galaxies
\citep{Yamamoto+:2006PASJ...58...93Y, Bianchi+:2015MNRAS.453L..11B,
Scoccimarro:2015PhRvD..92h3532S}. However, on very large angular scales, we
expect that the Yamamoto estimator suffers from the same problem as a fixed-LOS
estimator, because at least one of the lines of sight for each galaxy pair is
being projected, and that projection likely leads to a loss of information. An
optimal power spectrum measurement, therefore, needs to allow for a different
line of sight for every galaxy in the survey.

Large \emph{radial} scales pose a different kind of challenge, and in the past have
mostly been treated by splitting surveys into redshift bins
\citep[e.g.,][]{Beutler+:2017MNRAS.464.3409B}. The advantage is that it makes the
analysis simple. However, modes larger than the redshift bin are not measured
in the radial direction, and that information is lost by such an analysis.
\citep[See][however.]{Zaroubi+:1996ApJ...462...25Z}

In this paper, we study and implement a method that enables accurate
measurements of the largest radial and angular scales mapped by coming surveys.
It relies on the spherical Fourier-Bessel (SFB) transform. Most past
measurements of the galaxy overdensity power spectrum rely on Fourier
decomposition as it decreases the computation cost of near optimal statistical
estimators. While a standard Fourier transform decomposes a field into a linear
composition of eigenfunctions to the Laplacian in Cartesian coordinates, the
SFB transform we consider does the same but in spherical coordinates. Not only
does it maintain the statistical and computational (except for speed) advantages of
Fourier methods, but it is also the natural coordinate system for the
angular/radial separation over the sphere. The radial line of sight for every
single galaxy is built into the method, and the modeling of redshift
evolution of galaxy bias and growth factor is straightforward. An overview of
the SFB power spectrum is given in \citet{Pratten+:2013MNRAS.436.3792P} and a
mathematical treatment clarifying the relation between the configuration-space
correlation function, Fourier space correlation function, SFB correlation
function, and mixed-space correlation functions can be found in
\citet{Reimberg+:2016JCAP...01..048R}.

The spherical Fourier-Bessel transform for the analysis of galaxy surveys has
been considered multiple times in the past. \citet{Binney+:1991MNRAS.249..678B}
used a SFB decomposition to characterize overdensities deep in the nonlinear
regime. \citet{Lahav:1993cvf..conf..205L} applied the SFB analysis to local
galaxies on larger scales. \citet{Heavens+:1995MNRAS.275..483H} applied the SFB
analysis to the IRAS 1.2-Jy galaxy catalogue,
\citet{Tadros+:1999MNRAS.305..527T} applied it to the PSCz galaxy catalogue,
and \citet{Percival+:2004MNRAS.353.1201P} use it in the context of the 2dF
Galaxy Redshift Survey.
\citet{Leistedt+:2012A&A...540A..60L} have provided a public SFB code,
3DEX, that performs the SFB decomposition first in the radial direction for
each galaxy individually, then performs the angular transform using HEALPix.
More recently, \citet{Wang+:2020JCAP...10..022W} have built a
combined SFB/$P(k)$ estimation code that uses SFB on very
large scales, and a Cartesian multipole power spectrum estimator on smaller
scales.

SFB power spectrum measurements tended to be plagued by numerical instabilities
and computational complexity. The source of the numerical instability is the
incomplete coverage of the analysis volume by the survey. For example,
typically a boundary condition is applied at some distance $r_\max$, and the
analysis is performed for the entire volume inside a sphere of radius $r_\max$.
However, most surveys will leave most of that volume unexplored, and the
de-convolution of the window function or the inversion of the covariance matrix
become numerically unstable \citep[e.g., see][for a
solution]{Wang+:2020JCAP...10..022W}. The numerical complexity stems from the
large number of modes that need to be calculated even when a large fraction of
the analysis volume remains empty. Another source of computational complexity
is the combined estimation of the real-space power spectrum and redshift-space
distortion parameters that requires repeated estimations of the power spectrum
\citep{Wang+:2020JCAP...10..022W}.

The spherical Fourier Bessel decomposition code presented in this paper,
\codename, combines several approaches to address these problems. For the
first time, we limit the redshift range by introducing a boundary condition at
$r_\min$ as suggested by \citet{Samushia:2019arXiv190605866S}. We also use the
3DEX approach by \citet{Leistedt+:2012A&A...540A..60L} that does not suffer
from pixel window effects in the radial direction. The 3DEX approach also
allows separation of the angular and radial transforms, and for the angular
transform we use HEALPix
\citep{Gorski+:2005ApJ...622..759G,Zonca+:2019JOSS....4.1298Z}. Our bandpower
binning is done similar to that for CMB measurements
\citep{Hivon+:2002ApJ...567....2H,Alonso+:2019MNRAS.484.4127A}, should
a window-decoupled SFB power spectrum be desired, e.g., for comparison with
other surveys. We test our code on \emph{Roman}-like, \emph{SPHEREx}-like, and
\emph{Euclid}-like survey simulations with \numrange{\sim6}{\sim70}~million
galaxies per simulation.

In our approach the SFB power spectrum is measured directly, and parameter
estimation is left as a second step in the analysis pipeline (to be
developed in a future paper).
Evolution with redshift of parameters or the power spectrum is encoded in the
SFB power spectrum itself.
Thus, our vision is to construct the likelihood with parameters modeling the
deviations from a reference cosmology. For example, the deviation of the
distance-redshift relation would be modeled as a low-order polynomial relative
to the reference $r_\mathrm{ref}(z)$ so that the true distance is
$r_\true(z)=\(a_0 + a_1\,r_\mathrm{ref}(z)\)r_\mathrm{ref}(z)$, and parameters
$a_0$ and $a_1$ are to be measured.

An alternative to the SFB analysis that also naturally performs the
angular/radial separation is spherical harmonic tomography \citep[SHT, see
e.g.][]{Camera+:2018MNRAS.481.1251C,Nicola+:2014PhRvD..90f3515N} where an
angular spherical harmonic analysis is performed on shells of redshift bins.
\citet{Lanusse+:2015A&A...578A..10L} conclude that SFB yields similar
constraints as SHT, but when it comes to marginalizing over systematic biases
such as evolving scale-dependent galaxy bias, SFB performs better. Additionally, 
\citet{Castorina+:2018MNRAS.476.4403C} developed various approaches to incorporating 
wide-angle effects in Fourier based estimators. \citet{Beutler+:2019JCAP...03..040B} 
implement a small-angle expansion for the standard multipole power spectrum.

Another point to be made about the choice of the SFB basis is that RSD are
readily modeled \citep[see Eq.(16) in][]{Heavens+:1995MNRAS.275..483H}, because
they are ultimately sourced by the gravitational potential described by
Poisson's equation.

In \cref{sec:sfb_power_spectrum} we review the SFB power spectrum and
develop intuition in the Limber approximation. \cref{sec:sfb_discrete} details
the approach taken for the SFB decomposition, window de-convolution, shot
noise, bandpower binning, local average effect, and covariance matrix. We show
comparisons with log-normal simulations in \cref{sec:sfb_applications} for
\emph{Roman}, \emph{SPHEREx}, and \emph{Euclid}, and we conclude in
\cref{sec:sfb_conclusion}. We leave to the appendices a collection of useful
formulae in \cref{app:sfb_useful_formulae}, review the Laplacian in an
expanding universe in \cref{app:metric}, derive the radial potential boundary
conditions in \cref{app:gnl}, and simplify the covariance matrix in
\cref{sec:sfb_covariance_simplification,sec:w_chains}. Our \codename{} code is
publically available at
\url{https://github.com/hsgg/SphericalFourierBesselDecompositions.jl}.

\section{SFB Power Spectrum}
\label{sec:sfb_power_spectrum}
In this section we briefly review the SFB formalism. We start with the basic
transformation between configuration space and SFB space as well as between
Fourier space and SFB space. We then briefly show the power spectrum in a
completely homogeneous and isotropic universe before adding in selection
function, linear growth factor, galaxy bias, and RSD. We develop intuition by
applying Limber's approximation.

The spherical Fourier-Bessel decomposition expresses a field $\delta(\vr)$ in
terms of eigenfunctions of the Laplacian in spherical coordinates. For more
details, we refer the reader to \cref{sec:sfb_basis_functions}.
We define the spherical Fourier-Bessel modes $\delta_{\ell m}(k)$ by
\ba
\delta(\vr) &= \int\dd k\,\sum_{\ell m}
\left[\sqrt{\frac{2}{\pi}}\,k\,j_\ell(kr)\,Y_{\ell m}(\theta,\phi)\right]
\delta_{\ell m}(k)\,,
\label{eq:sfb_fourier_pair_a}
\\
\delta_{\ell m}(k) &=
\int\dd^3r
\left[\sqrt{\frac{2}{\pi}}\,k\,j_\ell(kr)\,Y^*_{\ell m}(\rhat)\right]
\delta(\vr)\,,
\label{eq:sfb_fourier_pair_b}
\ea
where $\vr=r\rhat$ is the position vector, $r$ is the comoving angular diameter
distance from the origin, and $\rhat$ is the direction on the sky.
Here, we assume that the universe is approximately flat. If the
curvature is significant, then the spherical Bessels need to be replaced by
ultra-spherical Bessels \citep{Abbott+:1986ApJ...308..546A,
Zaldarriaga+:2000ApJS..129..431Z, Kitching+:2007MNRAS.376..771K}.
The
orthogonality relations \cref{eq:jljlDelta,eq:YlmYlmDelta} for the spherical
Bessel functions and spherical harmonics are used to prove that
\cref{eq:sfb_fourier_pair_a,eq:sfb_fourier_pair_b} are inverses of each other.
The factor $2k^2/\pi$ can be split between
\cref{eq:sfb_fourier_pair_a,eq:sfb_fourier_pair_b} as pleased. Here we use the
convention in \citet{Nicola+:2014PhRvD..90f3515N}, because for a non-evolving,
homogeneous, and isotropic universe the SFB power spectrum then equals $P(k)$,
see \cref{eq:sfb_power_spectrum_homiso} below.

The relation between the SFB coefficients $\delta_{\ell m}(k)$ and the Fourier
modes $\delta(\vk)$ is obtained by expressing $\delta(\vr)$ in terms of its
Fourier transform in \cref{eq:sfb_fourier_pair_b},
\ba
\delta_{\ell m}(k)
&= \sqrt{\frac2\pi}\,k\int\dd^3r\,j_\ell(kr)\,Y^*_{\ell m}(\rhat)
\int\frac{\dd^3q}{(2\pi)^3}\,e^{i\vq\cdot\vr}\,\delta(\vq)\,.
\ea
With Rayleigh's formula \cref{eq:rayleigh} this turns into
\ba
\delta_{\ell m}(k)
&=
\int\frac{\dd^3q}{(2\pi)^3}\,
\sqrt{\frac{\pi}{2q^2}}
\,4\pi\sum_{\ell'm'} i^{\ell'}
\,Y^*_{\ell'm'}(\qhat)
\,\delta(\vq)
\vs&\quad\times
\frac{2kq}{\pi}
\int\dd{r}\,r^2\,j_\ell(kr)\, j_{\ell'}(qr)
\vs&\quad\times
\int\dd^2\rhat\,Y^*_{\ell m}(\rhat) \,Y_{\ell'm'}(\rhat)
\vs
&=
\frac{k}{(2\pi)^\frac32}
\,i^{\ell}
\int\dd^2\khat
\,Y^*_{\ell m}(\khat)
\,\delta(\vk)
\,,
\label{eq:deltak2deltaklm}
\ea
where the orthogonality relations \cref{eq:jljlDelta,eq:YlmYlmDelta} were used.
\cref{eq:deltak2deltaklm} shows that SFB is a spherical harmonic transform of
Cartesian Fourier modes with an additional phase factor $i^\ell$. And
\ba
\label{eq:deltaklm2deltak}
\delta(\vk)
&=
\frac{(2\pi)^\frac32}{k}
\sum_{\ell m}
i^{-\ell}
\,Y_{\ell m}(\khat)
\,\delta_{\ell m}(k)
\ea
is the inverse of \cref{eq:deltak2deltaklm}.

\subsection{The Homogeneous and Isotropic Universe}
In a homogeneous and isotropic universe in real-space (with no line-of-sight
effects), we have
\ba
\<\delta(\vk)\,\delta^*(\vk')\>
&= (2\pi)^3 \delta^D(\vk-\vk')\,P(k)\,.
\ea
Therefore, applying \cref{eq:deltak2deltaklm} gives the SFB power spectrum as
\ba
\label{eq:sfb_power_spectrum_homiso}
\< \delta_{\ell m}(k)\, \delta^*_{\ell'm'}(k') \>
&=
\delta^D(k-k')\,
\delta^K_{\ell\ell'}\,\delta^K_{mm'}\,
P(k)\,,
\ea
where we used \cref{eq:dirac3D} for the three-dimensional Dirac-delta function
in spherical coordinates.
That is, in a homogeneous and isotropic universe with no observational effects
the SFB power spectrum equals the 3D power spectrum $P(k)$.

\subsection{The Linear Universe}
We now generalize to include line-of-sight effects, a linearly evolving power
spectrum, and a radial window function. The galaxy density contrast we consider
is
\ba
\label{eq:delta_gobs}
\delta^\obs_g(\vr)
&=
W(\vr)\,D(r)
\int\frac{\dd^3q}{(2\pi)^3}\,e^{i\vq\cdot\vr}
\,\widetilde A_\mathrm{RSD}(\mu,q\mu,r)
\vs&\quad\times
b(r,q)
\,\delta(\vq)
\,,
\ea
where $\delta(\vq)$ is the matter density contrast in Fourier space, $W(\vr)$
is the survey window function, $D(r)$ is the linear growth factor, $b(r,q)$ is
the possibly scale-dependent linear galaxy bias, $\mu=\qhat\cdot\rhat$, and the
redshift-space distortions are encoded in
\citep[e.g.,][]{Kaiser:1987MNRAS.227....1K, GrasshornGebhardt+:2020PhRvD.102h3521G}
\ba
\label{eq:Arsd}
\widetilde A_\mathrm{RSD}(\mu,q\mu,r)= \(1+\beta\mu^2\) \widetilde A_\mathrm{FoG}(q\mu)\,,
\ea
with $\beta=f/b$, where $f=\dd\ln D/\dd\ln a$ is the linear growth rate, and we
assume a Gaussian fingers-of-God term \citep{Peacock+:1994MNRAS.267.1020P}
\ba
\label{eq:Afog}
\widetilde A_\mathrm{FoG}(q\mu)
&= e^{-\frac12 \sigma_u^2 q^2\mu^2}\,,
\ea
with $\sigma_u=\sigma_v/aH$ the pair-wise velocity dispersion in units of
length. The tilde on $A_\mathrm{RSD}$ signifies that it is a Fourier-space function.

The RSD term $\widetilde A_\mathrm{RSD}$ in \cref{eq:delta_gobs} can be
expressed as a function of derivatives on the complex exponential. That is, by
performing a Taylor series expansion we can replace $\mu\to-i\partial_{qr}$, or
\ba
\widetilde A_\mathrm{RSD}(\mu,q\mu,r)\,e^{i\vq\cdot\vr}
&=
\sum_n \frac{a_n(q,r)}{n!}\, \mu^n \,e^{iqr\mu}
\vs
&=
\sum_n \frac{a_n(q,r)}{n!} \(-i\,\frac{\partial}{\partial(qr)}\)^n \,e^{iqr\mu}
\vs
&=
\widetilde A_\mathrm{RSD}(-i\partial_{qr},-iq\partial_{qr},r) \,e^{i\vq\cdot\vr}\,.
\ea
Furthermore, the complex exponential is expanded using Rayleigh's formula
\cref{eq:rayleigh} so that the derivatives in $\widetilde A_\mathrm{RSD}$ only
act on the spherical Bessel function from Rayleigh's formula. Further
expressing the Fourier-space density contrast in terms of its SFB modes
\cref{eq:deltaklm2deltak}, the observed density contrast \cref{eq:delta_gobs}
now becomes
\begin{widetext}
\ba
\delta^\obs_g(\vr)
&=
W(\vr)\,D(r)
\int\frac{\dd^3q}{(2\pi)^3}
\,b(r,q)
\left[\widetilde A_\mathrm{RSD}(-i\partial_{qr},-iq\partial_{qr},r)
\,4\pi\sum_{L_1M_1}i^{L_1}j_{L_1}(qr)\,Y^*_{L_1M_1}(\qhat)\,Y_{L_1M_1}(\rhat)
\right]
\vs&\quad\times
\frac{(2\pi)^\frac32}{q}
\sum_{LM}
i^{-L}
\,Y_{LM}(\qhat)
\,\delta_{LM}(q)
\\
&=
W(\vr)\,D(r)
\,\sqrt{\frac{2}{\pi}}
\int\dd q\,q^2
\,b(r,q)
\left[\widetilde A_\mathrm{RSD}(-i\partial_{qr},-iq\partial_{qr},r)
\,\sum_{LM}j_{L}(qr)\,Y_{LM}(\rhat)
\right]
\,\frac{1}{q}
\,\delta_{LM}(q)\,.
\ea
\end{widetext}
Using \cref{eq:sfb_fourier_pair_b} to transform into SFB space,
\ba
\label{eq:sfb_delta_gobs}
\delta^{g,\obs}_{\ell m}(k)
&=
\int\dd q
\sum_{LM}
\mathcal{W}_{\ell m}^{LM}(k,q)
\,\delta_{LM}(q)\,,
\ea
where
\ba
\mathcal{W}_{\ell m}^{LM}(k,q)
&=
\int\dd^2\rhat
\,Y_{LM}(\rhat)
\,Y^*_{\ell m}(\rhat)
\,\mathcal{W}_\ell^L(k,q,\rhat)\,,
\label{eq:sfb_wlmLMkq}
\ea
and
\ba
\label{eq:sfb_wlLkqrhat}
\mathcal{W}_\ell^L(k,q,\rhat)
&=
\frac{2qk}{\pi}
\int\dd r\,r^2
\,W(\vr) \,D(r)\,b(r,q)
\,j_\ell(kr)
\vs&\quad\times
\widetilde A_\mathrm{RSD}(-i\partial_{qr},-iq\partial_{qr},r) j_{L}(qr)\,.
\ea
The SFB correlation function is, therefore,
\ba
&\<\delta^{g,\obs}_{\ell m}(k)\,\delta^{g,\obs,*}_{\ell'm'}(k')\>
\vs
&=
\int\dd q
\sum_{LM}
\mathcal{W}_{\ell m}^{LM}(k,q)
\,\mathcal{W}_{\ell'm'}^{LM,*}(k',q)
\,P(q)\,,
\label{eq:sfb_power_spectrum_window}
\ea
where we used \cref{eq:sfb_power_spectrum_homiso,eq:sfb_delta_gobs}.

Here we will only consider a radial selection function, as the angular mask
will be handled in the estimator. Then,
\ba
\label{eq:phi}
W(\vr) &= \phi(r)\,,
\ea
and we define the simplification of \cref{eq:sfb_wlLkqrhat}
\ba
\label{eq:sfb_wlkq}
\mathcal{W}_\ell(k,q)
&=
\mathcal{W}_\ell^\ell(k,q,\rhat)\,,
\ea
which is then independent of the direction $\rhat$. \cref{eq:sfb_wlmLMkq} and
\cref{eq:sfb_power_spectrum_window} then simplify to
\ba
\label{eq:sfb_2point_isotropic}
\<\delta^{g,\obs}_{\ell m}(k)\,\delta^{g,\obs,*}_{\ell'm'}(k')\>
&=
\delta^K_{\ell\ell'}
\delta^K_{mm'}
\,C_\ell(k,k')\,,
\ea
with the SFB power spectrum defined as
\ba
\label{eq:sfb_power_spectrum}
C_\ell(k,k')
&=
\int\dd q
\,\mathcal{W}_{\ell}(k,q)
\,\mathcal{W}^*_{\ell}(k',q)
\,P(q)\,.
\ea
\cref{eq:sfb_wlLkqrhat,eq:sfb_wlkq,eq:sfb_power_spectrum} show that RSD and
linear growth can be taken into account by a change in the radial window
function.

\cref{eq:sfb_2point_isotropic} shows that the SFB power spectrum is
non-zero only when $\ell=\ell'$, $m=m'$, and it is independent of $m$. This is
a consequence of the isotropy on the sky, or the rotational invariance around
the observer, as can be easily shown in general for spherical harmonic
transforms.

For a homogeneous and isotropic universe without selection function,
$W(\vr)=D(r)=\widetilde A_\mathrm{RSD}=1$ and $b(r,q)=\mathrm{const}$, the
window becomes $\mathcal{W}_\ell(k,q)\,\propto\,\delta^D(k-q)$, and
\cref{eq:sfb_power_spectrum_homiso} is reproduced.
Also, $\mathcal{W}_\ell(k,q)$ is real, because the imaginary
arguments to $\widetilde A_\mathrm{RSD}$ are only ever
raised to even powers.

To develop some intuition for \cref{eq:sfb_power_spectrum} we evaluate the SFB
power spectrum in a Limber-like approximation. However, we defer to
\cref{sec:sfb_limber} in order not to distract from the main content of this
paper. Other treatments are in
\citet{Munshi+:2016MNRAS.456.1627M,Yoo+:2013PhRvD..88b3502Y}.

\section{SFB Decomposition}
\label{sec:sfb_discrete}
This section describes our SFB decomposition for a galaxy survey with mask and
selection function. We largely follow \citet{Samushia:2019arXiv190605866S} for
the radial basis functions and \citet{Leistedt+:2012A&A...540A..60L} for the
angular/radial split in the estimator.

We start by giving a review of the basis functions, then we add window and
selection functions, we model the discrete galaxy distribution, and estimate
the covariance matrix.

\subsection{Spherical Fourier-Bessel basis with potential boundary conditions}
\label{sec:sfb_basis_functions}
\begin{figure*}
  \centering
  \incgraph{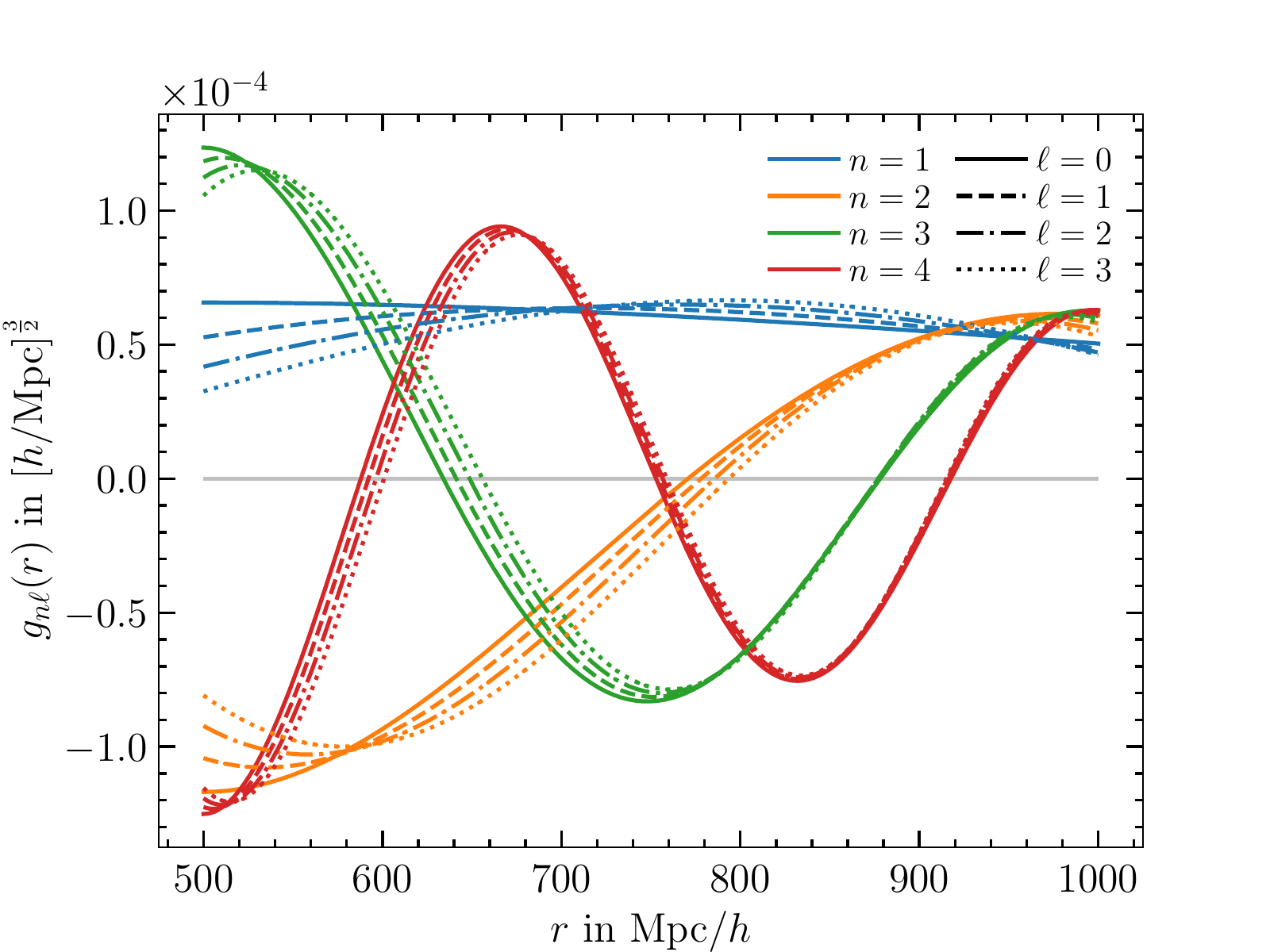}
  \incgraph{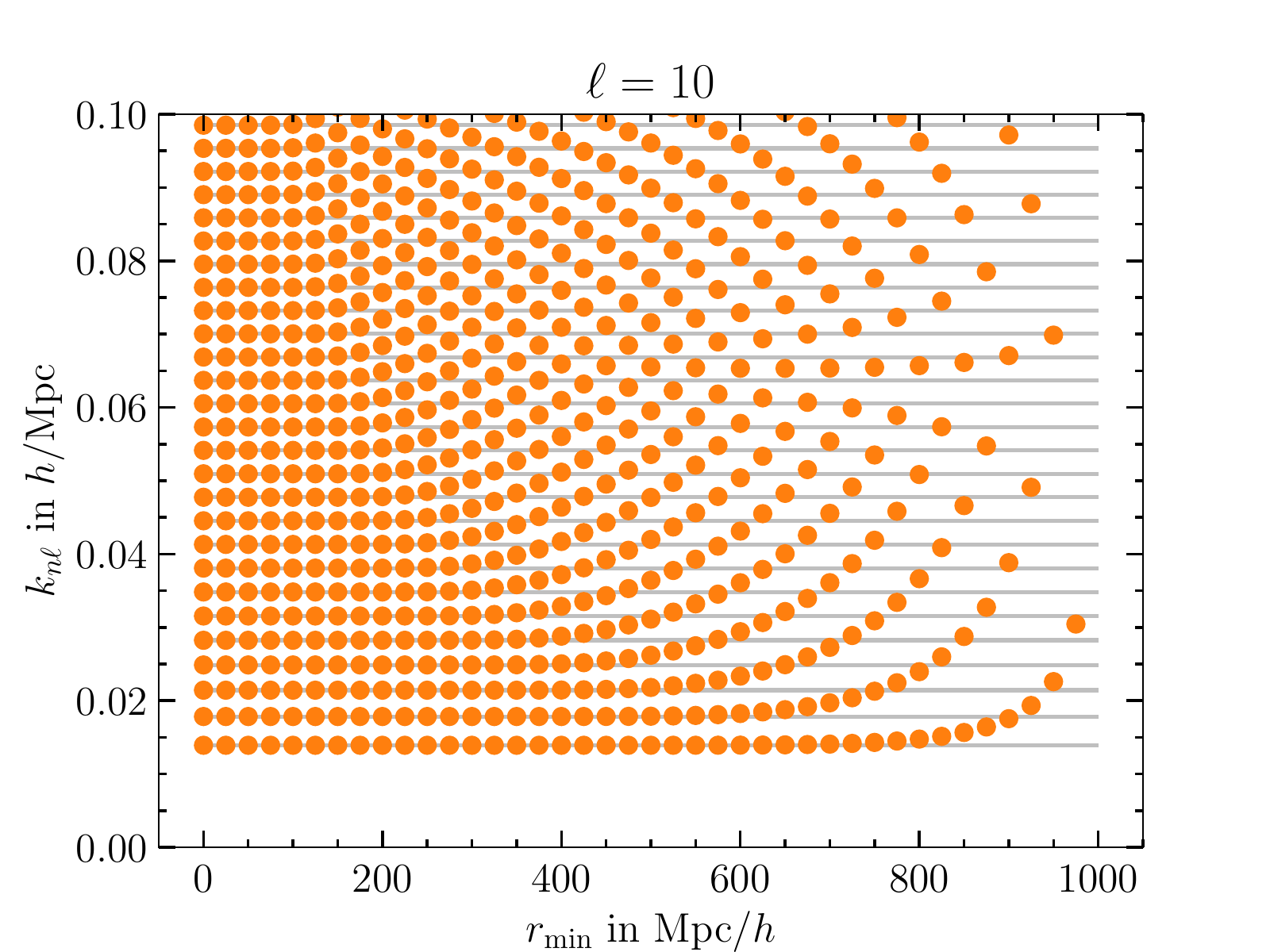}
  \caption{
    Left: The radial basis functions for potential boundary conditions as a
    function of $r$. Color indicates the modes $n$, line style indicates $\ell$
    as shown in the legend. Here, $r_\min=\SI{500}{\per\h\mega\parsec}$ and
    $r_\max=\SI{1000}{\per\h\mega\parsec}$.
    Right: $k_{n\ell}$ for potential boundary conditions as a function of
    $r_\min$ when $\ell=10$. The grey lines are for $r_\min=0$, and we fix
    $r_\max=\SI{1000}{\per\h\mega\parsec}$.
  }
  \label{fig:sfb_gnl_basis_potential}
\end{figure*}
We choose the eigenbasis of the Laplacian as it captures the rotational
invariance of the observed large-scale structure, and that leads to a
compressed summary statistic which is also rotationally invariant
while including all wide-angle effects.
Here, we lay out the boundary
conditions we consider similar to \citet{Samushia:2019arXiv190605866S}.
However, as observers fixed in one location, using light that travels at a
finite speed, it is more natural to use spherical polar coordinates that
separate the radial and angular observations. Then, the Laplacian on a scalar
function $f$ becomes
\begin{align}
    \nabla^2f &= \frac{1}{r^2}\,\frac{\partial}{\partial r}\left(r^2\,\frac{\partial f}{\partial r}\right)
    + \frac{1}{r^2\sin\theta}\,\frac{\partial}{\partial\theta}\left(\sin\theta\,\frac{\partial f}{\partial\theta}\right)
    \nonumber \\
    &\quad
    + \frac{1}{r^2\sin^2\theta}\,\frac{\partial^2f}{\partial\phi^2}\,,
    \label{eq:laplacian_spherical}
\end{align}
where $r$, $\theta$, and $\phi$ are the comoving angular diameter distance,
zenith angle, and azimuthal angle, respectively (see \cref{app:metric}
for a derivation). The eigenbasis to \cref{eq:laplacian_spherical} that
satisfies
\begin{align}
    \nabla^2f &= -k^2 f
\end{align}
for some mode $k$ is of the form \citep[see e.g.][]{Samushia:2019arXiv190605866S}
\begin{align}
    f_{\ell\mu}(k; r,\theta,\phi)
    &= \big[c_j\,j_\ell(kr) + c_y\,y_\ell(kr)\big]
    \vs&\quad\times
    \big[c_p\,P_\ell^\mu(\cos\theta) + c_q\,Q_\ell^\mu(\cos\theta)\big]
    \vs&\quad\times
    \big[c_+\,e^{i\mu\phi} + c_-\,e^{-i\mu\phi}\big]\,,
\end{align}
where the $c_i$ are constants, and $j_\ell$ and $y_\ell$ are spherical
Bessel functions of the first and second kind, and $P_\ell^\mu$ and
$Q_\ell^\mu$ are Legendre functions of the first and second kind.

The constants $c_i$ are set by boundary conditions. First, the spherical
Bessel of the second kind, $y_\ell$ diverges with vanishing argument; hence,
typically $c_y=0$. Typically, the functions also need to be periodic about the
azimuthal angle $\phi$; therefore, $\mu=0,1,2,\ldots$ is an integer. Then, the
functions also need to be finite for $\cos\theta=\pm1$, typically; therefore,
$c_q=0$, and $\ell=0,1,2,\ldots$ is an integer, and $-\ell\leq\mu\leq\ell$.

Effectively, the preceding paragraph imposed boundary conditions at $r_\min=0$
and assumed coverage of the whole sky. Typically
\citep[e.g.,][]{Heavens+:1995MNRAS.275..483H,Fisher+:1995MNRAS.272..885F,Tadros+:1999MNRAS.305..527T,Percival+:2004MNRAS.353.1201P},
one would then go ahead and also impose boundary conditions at some $r_\max$
such that the survey volume is contained within a sphere of radius $r_\max$.
This restricts the SFB volume, i.e., the volume on which the SFB transform is
performed, as the region from $0\leq r\leq r_\max$. Demanding the basis
functions to be orthogonal then leads to a discrete spectrum of modes
$k=k_{n\ell}$.

Realistic galaxy surveys do not occupy the entire SFB volume, but are
restricted in both redshift and angular area, and, therefore, they leave large
fractions of the SFB volume unobserved. This leads to the deconvolution of the
window function to be numerically unstable. It also results in wasted
computational resources if the survey covers only a (potentially thick) shell
at high redshift. The analogous picture for a standard Fourier transform would
be to have a transform box that is much larger than the survey volume.
Therefore, the selection function will vanish for part of the SFB volume, and,
because in that case some modes are not well constrained, the inversion of the
window function becomes numerically unstable.

In this paper, we employ two strategies to deal with this problem. First, we
follow \citet{Hivon+:2002ApJ...567....2H,Alonso+:2019MNRAS.484.4127A} and bin
the pseudo SFB power spectrum into bandpowers. This combines several
poorly-constrained modes into one well-constrained mode. We rely on this
strategy especially for the angular mask so that we can leverage the full-sky
spherical harmonic algorithms from the \texttt{HealPy} software
\citep{Gorski+:2005ApJ...622..759G, Zonca+:2019JOSS....4.1298Z}.

For the second strategy, we follow \citet{Samushia:2019arXiv190605866S} and
move the boundary at the origin to some $r_\min$ so that the SFB volume extends
from $r_\min\leq r\leq r_\max$. For galaxy surveys that start at some
minimum redshift this eliminates from the SFB tranform volume a hole around
the origin. As a result, the inversion of the window function is numerically
well behaved even without resorting to bandpower binning. Furthermore, the
number of SFB modes is reduced not just by the
boundary condition at $r_\max$, but the boundary condition at $r_\min$ also
reduces the number of modes further by the fraction $r_\min^3/r_\max^3$. In all
cases considered in this paper, this eliminates the need for bandpower binning
in the radial direction.

We differ from \citet{Samushia:2019arXiv190605866S} in that we use potential
boundary conditions \citep{Fisher+:1995MNRAS.272..885F} that ensure the field
represented by the SFB decomposition is continuous and smooth at the boundary.
These boundary conditions lead to a spectrum of modes $k_{n\ell}$, as shown in
\cref{app:gnl}. In the appendix we also derive that the radial basis functions
with such boundary conditions become a linear combination of spherical Bessels
of the first and second kind,
\ba
g_{n\ell}(r) &= c_{n\ell}\,j_\ell(k_{n\ell}\,r) + d_{n\ell}\,y_\ell(k_{n\ell}\,r)\,,
\label{eq:gnl_definition}
\ea
which satisfy an orthonormality relation
\ba
\label{eq:gnl_orthonormality}
\int_{r_\min}^{r_\max}\dd r\,r^2\,g_{n\ell}(r)\,g_{n'\ell}(r)
&=
\delta^K_{nn'}
\,,
\ea
where $\delta^K_{nn'}$ is a Kronecker delta, and the coefficients $c_{n\ell}$
and $d_{n\ell}$ are derived in \cref{app:gnl}. With
\cref{eq:gnl_orthonormality}, the Fourier pair
\cref{eq:sfb_fourier_pair_a,eq:sfb_fourier_pair_b} remains a Fourier pair with
the discrete $k_{n\ell}$-spectrum, and the pair becomes
\ba
\delta(\vr) &= \sum_{n \ell m}
\Big[g_{n\ell}(r)\,Y_{\ell m}(\rhat)\Big]
\delta_{n \ell m}\,,
\label{eq:sfb_discrete_fourier_pair_a}
\\
\delta_{n \ell m} &=
\int\dd^3\vr
\Big[g_{n\ell}(r)\,Y^*_{\ell m}(\rhat)\Big]
\delta(\vr)\,,
\label{eq:sfb_discrete_fourier_pair_b}
\ea
where the integral goes over the volume within $r_\min\leq r\leq r_\max$. Note
that our choice to normalize $g_{n\ell}(r)$ as in \cref{eq:gnl_orthonormality}
changes the units of $\delta_{n\ell m}$ compared to
\cref{eq:sfb_fourier_pair_a,eq:sfb_fourier_pair_b}. In effect, this choice of
units takes into account the survey volume at this stage rather than at the
stage of forming the correlation function.

Examples of the resulting basis functions and modes $k_{n\ell}$ are show in
\cref{fig:sfb_gnl_basis_potential}. We point out that the $\ell=0$ modes are
closely related to taking the average of the transformed field $\delta(\vr)$.
Also, a larger $r_\min$ results in a smaller volume, and, therefore, fewer
modes that can be constrained. 

\subsection{Window and selection function}
The observed number density $n(\vr)$ of galaxies is subject to the window and
selection function $W(\vr)$ of the survey, which we define as the fraction of
galaxies observed at position $\vr$. For a random catalogue subject to the same
window function, with density $n_r(\vr)$, and with $1/\alpha$ as many galaxies
as the survey, we then have
\ba
\label{eq:window_definition}
\alpha\<n_r(\vr)\> = W(\vr) \, \bar n\,,
\ea
where $\<n_r(\vr)\>=\alpha^{-1}\bar n(\vr)$ is the average number density of
the ensemble of random catalogs, and $\bar n$ is the average number density in
the survey. Note that \cref{eq:window_definition} can equivalently be expressed
in terms of the limit $\lim_{\alpha\to0}\alpha\,n_r(\vr)=\bar
n(\vr)=W(\vr)\,\bar n$. With this definition of the window function, we define
the effective volume as
\ba
\label{eq:Veff}
  V_\mathrm{eff} &= \int\dd^3\vr\,W(\vr)\,,
\ea
so that the average number density $\bar n$ becomes
\ba
\label{eq:nbar}
  \bar n &= \frac{N_\mathrm{gal}^\obs}{V_\mathrm{eff}}\,,
\ea
and $N_\mathrm{gal}^\obs$ is the observed number of galaxies in the survey. Any
variation across the survey in the actual average number density, e.g., due to
an evolving luminosity function, is absorbed into $W(\vr)$.
Our treatment is in line with \citet{Taruya+:2021PhRvD.103b3501T}, and our
$W(\vr)$ takes the role of the function $G(\vr)$ in
\citet{Feldman+:1994ApJ...426...23F}, except that we do not at present include
a weighting scheme.

In a sense, there are two window functions here: first, the one defined by the
SFB procedure and limited by $r_\min \leq r \leq r_\max$, and second, $W(\vr)$,
which defines the geometry and selection of the survey. However, the first one
should be irrelevant as long as the survey volume is entirely inside $r_\min
\leq r \leq r_\max$ and as along as sufficient number of modes are included in
the SFB analysis.

The observed density fluctuation field is, then,
\ba
  \delta^\obs(\vr)
    &= \frac{n(\vr) - \alpha\,n_r(\vr)}{\nbar}
    = \frac{n(\vr)}{\nbar} - W(\vr)\,,
  \label{eq:density_constrast_discrete_points}
\ea
where \cref{eq:window_definition} was used in the limit that the random
catalogue has an infinite number of galaxies, or $\alpha\to0$. Because the
observed density $n(\vr)$ is also subject to the window function $W(\vr)$, the
observed and true density contrasts are related by
\ba
\delta^\obs(\vr) &= W(\vr)\, \delta^A(\vr)\,,
\ea
where we attach the superscript `A' to refer to the local average effect (see
\cref{sec:sfb_local_average_effect} below).
Transforming to SFB-space and expressing $\delta^A(\vr)$ in terms of its SFB
decomposition
\cref{eq:sfb_discrete_fourier_pair_a,eq:sfb_discrete_fourier_pair_b}, we get
\ba
\delta^\obs_{n \ell m}
&=
\sum_{n'\ell'm'}
W_{n \ell m}^{n'\ell'm'}
\,\delta^A_{n'\ell'm'}\,,
\label{eq:delta_mixing}
\ea
where
\ba
\label{eq:delta_mixing_matrix_discrete}
W_{n \ell m}^{n'\ell'm'}
&=
\int\dd r\,r^2
\,g_{n\ell}(r)
\,g_{n'\ell'}(r)
\vs&\quad\times
\int\dd^2\rhat
\,Y^*_{\ell m}(\rhat)
\,Y_{\ell'm'}(\rhat)
\,W(r,\rhat)\,.
\ea

\subsubsection{Properties and implementation}
From \cref{eq:delta_mixing_matrix_discrete} follows the symmetry
\ba
\label{eq:sfb_window_parity}
W_{n,\ell,m}^{n',\ell',-m'}
&=
(-1)^{m+m'}
W_{n,\ell,-m}^{n',\ell',m',*}
\,,
\ea
and the Hermitian property
\ba
\label{eq:sfb_window_symmetry}
W_{n\ell m}^{n' \ell' m'}
&=
W^{n\ell m,*}_{n' \ell' m'}\,.
\ea
In the special case that $W(\vr)=1$ everywhere,
\ba
W_{n \ell m}^{n'\ell'm'}
&=
\delta^K_{nn'}
\delta^K_{\ell\ell'}
\delta^K_{mm'}\,,
\ea
which follows from \cref{eq:gnl_orthonormality} and \cref{eq:YlmYlmDelta}.

In all generality, \cref{eq:delta_mixing_matrix_discrete} can be simplified for computational
convenience by expressing the window function in terms of an angular
transform. That is, introduce
\ba
\label{eq:win_r_lm}
W_{LM}(r)
&=
\int\dd^2\rhat\,Y^*_{LM}(\rhat)\,W(r,\rhat)\,.
\ea
Then,
\ba
W_{n \ell m}^{n'\ell'm'}
&=
(-1)^m
\sum_{L}
\mathcal{G}^{\ell\ell'L}_{-m,m',m-m'}
\vs&\quad\times
\int\dd r\,r^2
\,g_{n\ell}(r)
\,g_{n'\ell'}(r)
\,W_{L,m-m'}(r)
\label{eq:delta_mixing_matrix_discrete_sht}
\,, 
\ea
where we used \cref{eq:Ylm_conjugate} and introduced the Gaunt
factor \cref{eq:gaunt_factor}.
In writing \cref{eq:delta_mixing_matrix_discrete_sht} we performed the angular
transform of the window function only as that leads to a computationally
suitable form. Had we performed a full SFB transform, we would have been left
with an infinite sum over $n$ that converges slowly, in addition to the need of
computing integrals over three spherical Bessel functions.

\subsection{Discrete points}
Now we specialize the SFB decomposition to the case that we have galaxies
represented by discrete points. That is, we assume the number density is given
by
\ba
\label{eq:number_density}
  n(\vr) &= \sum_p \delta^D(\vr - \vr_p)\,,
\ea
where the sum is over all points (galaxies) in the survey.

In the 3DEX approach \citep{Leistedt+:2012A&A...540A..60L}, which we adopt
here, \cref{eq:sfb_discrete_fourier_pair_b} is decomposed into its radial and
angular integrals, and the radial integration is performed first. That is,
\ba
\delta^\obs_{n \ell m} &=
\int\dd^2\Omega_\rhat
\,Y^*_{\ell m}(\rhat)
\,\delta^\obs_{n\ell}(\rhat)\,,
\ea
where
\ba
\delta^\obs_{n\ell}(\rhat)
&=
\int_{r_\min}^{r_\max}\dd r\,r^2
\,g_{n\ell}(r)
\,\delta^\obs(r, \rhat)
\ea
represents an angular field for each $n$ and $\ell$, and $g_{n\ell}$ is defined
in \cref{eq:gnl_definition}. For the density contrast
\cref{eq:density_constrast_discrete_points} with number density
\cref{eq:number_density},
\ba
\delta^\obs_{n\ell}(\rhat)
&=
\frac{1}{\nbar}\sum_p
\delta^D(\rhat-\rhat_p)
\,g_{n\ell}(r_p)
-
\, W_{n\ell}(\rhat)
\,,
\label{eq:delta_nl_theta_phi_exact}
\ea
where
\ba
W_{n\ell}(\rhat)
&= \int_{r_\min}^{r_\max}\dd r\,r^2 \,g_{n\ell}(r)\,W(\vr)\,.
\ea
\cref{eq:delta_nl_theta_phi_exact} is an exact expression for the observed
density contrast $\delta^\obs_{n\ell}(\rhat)$. However, for the angular
transform we wish to make use of the fast HEALPix
scheme\footnote{\url{https://healpix.jpl.nasa.gov/index.shtml}}$^,$\footnote{\url{https://healpy.readthedocs.io/en/latest/}},
and we need the density contrast in pixel $i$ averaged over the pixel area
$\Delta\Omega_i$,
\ba
\bar\delta^\obs_{n\ell}(\rhat_i)
&=
\frac{1}{\Delta\Omega_i}\int_{\Delta\Omega_i}\dd\Omega_\rhat\,\delta^\obs_{n\ell}(\rhat)
\\
&=
\frac{1}{\nbar\Delta\Omega_i}\sum_{p\in\Delta\Omega_i} g_{n\ell}(r_p)
- W_{n\ell}(\rhat_i)
\,,
\ea
where we assumed that $W_{n\ell}(\rhat)$ varies slowly over the size of an
angular pixel. Then, the angular transform is performed:
\ba
\label{eq:sfb_spherical_harmonic_transform}
\bar \delta^\obs_{n\ell m}
&=
\sum_i\Delta\Omega_i\,Y_{\ell m}^*(\rhat_i)\,\bar\delta^\obs_{n\ell}(\rhat_i)\,.
\ea
In what follows we will generally drop the bar indicating the
angular-pixel-averaging.

\subsection{Power spectrum estimation}
\citet{Wandelt+:2001PhRvD..64h3003W,Hivon+:2002ApJ...567....2H} use a
\emph{pseudo-$C_\ell$} method to estimate the power spectrum. Translating to
the SFB decomposition, the pseudo-$C_\ell$ method assumes that much of the
information about the power spectrum is contained in the pseudo-power spectrum
\ba
\label{eq:sfb_clnnobs}
\hat C^\obs_{\ell nn'}
&= \frac{1}{2\ell+1}\sum_m \delta^\obs_{n\ell m} \delta^{\obs,*}_{n'\ell m}\,.
\ea
That is, we ignore off-diagonal terms $L\neq\ell$ and $M\neq m$, and average
over $m$. The effect of the window is then described by a mixing matrix between
the $\hat C^\obs_{\ell nn'}$ and $\hat C^A_{\ell nn'}$,
\ba
\label{eq:cell_mixing_matrix}
\hat C^\obs_{\ell nn'}
&=
\sum_{LNN'} \mathcal{M}_{\ell nn'}^{LNN'}\,\hat C^A_{LNN'}
\,,
\ea
where we used \cref{eq:delta_mixing} and defined
\ba
\label{eq:sfb_cmix}
\mathcal{M}_{\ell nn'}^{LNN'}
&=
\frac{1}{2\ell+1}
\sum_{mM}
W_{n\ell m}^{NLM} W_{n'\ell m}^{N'LM,*}
\,,
\ea
and the index `A' on $C^A_{\ell nn'}$ indicates the local average effect, see
\cref{sec:sfb_local_average_effect}. Next, with
\cref{eq:win_r_lm,eq:delta_mixing_matrix_discrete_sht} we get
\ba
\mathcal{M}_{\ell nn'}^{LNN'}
&=
\frac{2L+1}{4\pi}
\sum_{L_1}
\begin{pmatrix}
  \ell & L & L_1 \\
  0 & 0 & 0
\end{pmatrix}^2
\sum_{M_1}
\vs&\quad\times
\int\dd r\,r^2
\,g_{n\ell}(r)
\,g_{NL}(r)
\,W_{L_1M_1}(r)
\vs&\quad\times
\int\dd r'\,r'^2
\,g_{n'\ell}(r')
\,g_{N'L}(r')
\,W^*_{L_1M_1}(r')
\label{eq:cell_mixing_matrix_explicit}
\,,
\ea
and we used the orthogonality of the Gaunt factor
\cref{eq:gaunt_3j,eq:gaunt_orthogonality}. (The sum over $M_1$ could be
performed first. However, that approach is much more memory intensive, so that
computing the integrals first ends up being faster. We have also avoided
expressing the result in terms of a full SFB transform, as that would require a
slowly-converging sum over $n$.)
Note that the matrix $(2L+1)^{-1}\mathcal{M}_{\ell nn'}^{LNN'}$ is symmetric under exchange of the set of indices $(LNN')$ and $(\ell nn')$,
but $\mathcal{M}$ by itself is not.

\subsubsection{Separable mask and radial selection}
\begin{figure*}
  \centering
  \incgraph{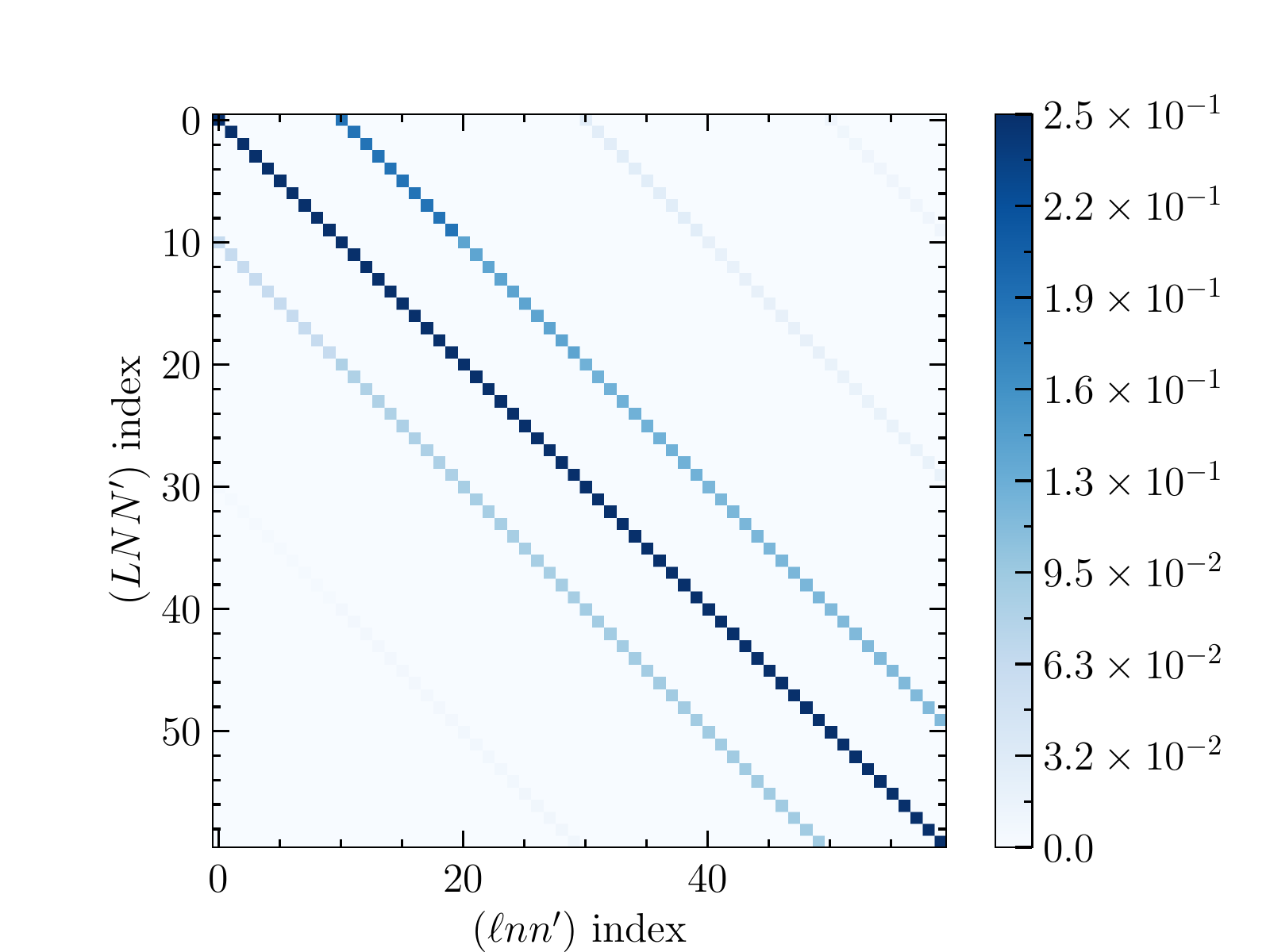}
  \incgraph{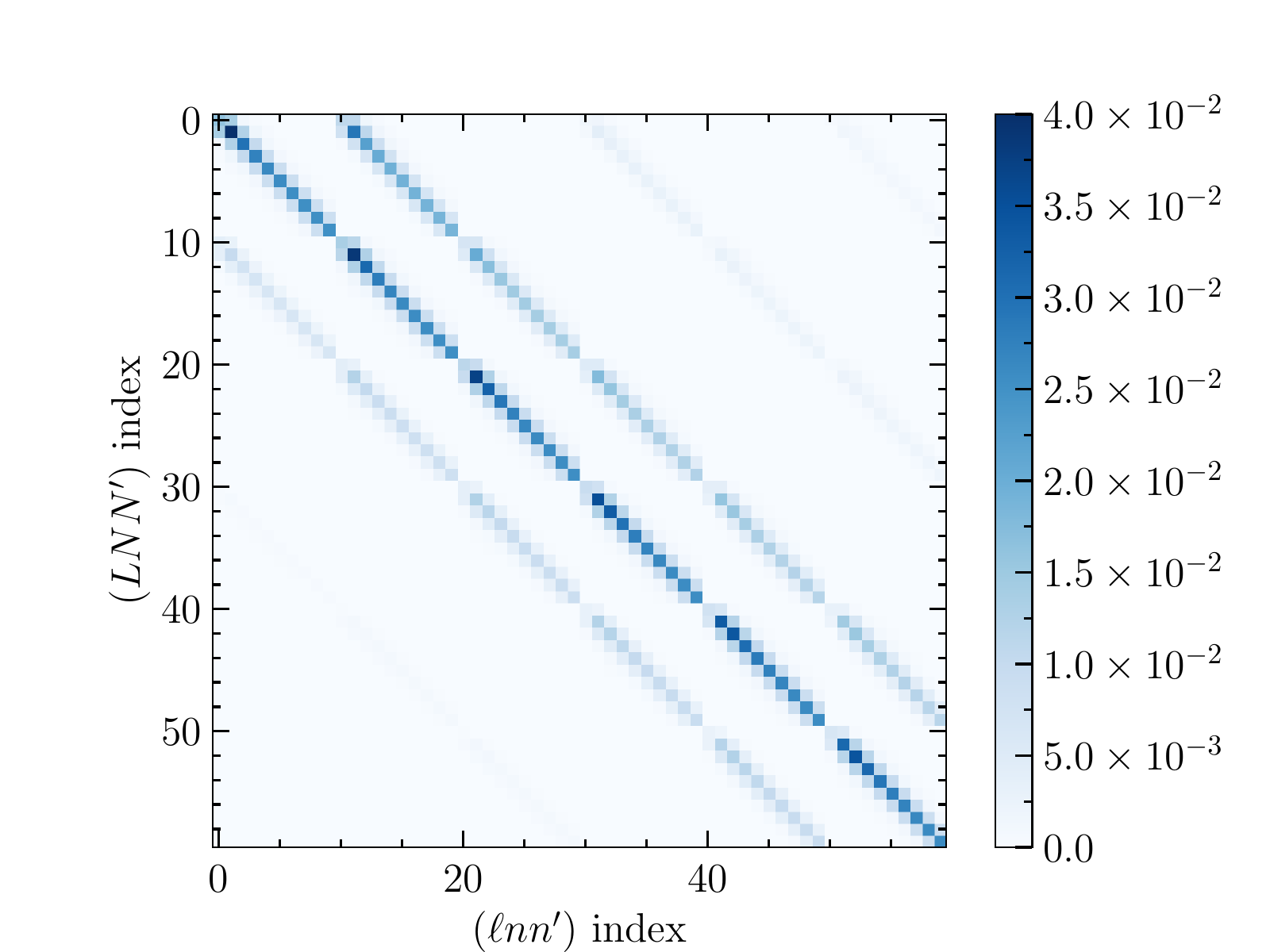}
  \caption{
    In the left panel we show the mixing matrix $\mathcal{M}_{\ell nn'}^{LNN'}$
    for a half-sky mask, and in the right panel we add a radial selection
    function. The ordering of the $(\ell nn')$ modes is such that $n=n'$
    increases first from 1 to 10, and $\ell$ increases by one for every ten
    $n$-modes. The half-sky mask in the left panel exhibits couplings between
    neighboring $\ell$-modes. On the right, the radial selection function
    decreases with distance, which leads to neighboring $n$-modes being coupled
    as well.
  }
  \label{fig:bcmix}
\end{figure*}
It is quite common that the window function is separable into a radial and an
angular term,
\ba
W(\vr) &= \phi(r) \, M(\rhat)\,.
\ea
If the flux limit in a blind
survey is near $L^*$, then the selection could change dramatically as a
function of angular depth variations that are due to, e.g., atmospheric
variations, and the separation of angular and radial selection would be a poor
approximation.
However, eBOSS, for example, had more targets selected in regions where two or
more plates overlapped \citep[e.g.][]{deMattia+:2021MNRAS.501.5616D}.
Similarly, PFS will have higher target numbers where pointings overlap
\citep{Sunayama+:2020JCAP...06..057S}.

When the window function is separable, then \cref{eq:win_r_lm} is separable as
well,
\ba
W_{LM}(r)
&=
\phi(r) \, W_{LM}\,,
\ea
where
\ba
\label{eq:sfb_Wlm}
W_{LM} &= \int\dd^2\rhat\,Y^*_{LM}(\rhat)\,M(\rhat)\,,
\ea
and \cref{eq:cell_mixing_matrix_explicit} becomes
\ba
\mathcal{M}_{\ell nn'}^{LNN'}
&=
\frac{2L+1}{4\pi}
\sum_{L_1}
\begin{pmatrix}
  \ell & L & L_1 \\
  0 & 0 & 0
\end{pmatrix}^2
\sum_{M_1}
\left|W_{L_1M_1}\right|^2
\vs&\quad\times
\int\dd r\,r^2
\,g_{n\ell}(r)
\,g_{NL}(r)
\,\phi(r)
\vs&\quad\times
\int\dd r'\,r'^2
\,g_{n'\ell}(r')
\,g_{N'L}(r')
\,\phi(r')
\,,
\ea
which is also separable, and therefore significantly reduces computation cost.
\cref{eq:delta_mixing_matrix_discrete_sht} simplifies in a similar manner.

In the special case that $W(\vr)=1$ everywhere, we recover the unit matrix
\ba
\mathcal{M}_{\ell nn'}^{LNN'}
&=
\delta^K_{\ell L}
\delta^K_{nN}
\delta^K_{n'N'}
\,,
\ea
as expected.

We give two further examples in \cref{fig:bcmix}. In the left panel, we show
the mixing matrix for a mask covering half the sky, and this leads to coupling
of neighboring $\ell$-modes. On the right, we add a radial selection decreasing
with redshift, and this additionally leads to the coupling of neighboring
$n$-modes.

\subsection{Shot noise}
The sampling of the density field by a limited number of points leads to a shot
noise component in the power spectrum. To estimate the shot noise, we start
with \citep{Peebles:1973ApJ...185..413P,Feldman+:1994ApJ...426...23F}
\ba
\<n(\vr)\,n(\vr')\>
&=
\bar n(\vr) \, \bar n(\vr') \left[1 + \xi(\vr,\vr')\right]
\vs&\quad
+ \bar n(\vr)\,\delta^D(\vr - \vr')\,,
\\
\<n(\vr)\,n_r(\vr')\>
&=
\alpha^{-1}\,\bar n(\vr) \, \bar n(\vr')\,,
\\
\<n_r(\vr)\,n_r(\vr')\>
&=
\alpha^{-2}\,\bar n(\vr) \, \bar n(\vr')
+ \alpha^{-1}\,\bar n(\vr)\,\delta^D(\vr - \vr')\,.
\ea
The density contrast is given by \cref{eq:density_constrast_discrete_points},
and the ensemble average becomes
\ba
\<\delta^\obs(\vr)\,\delta^\obs(\vr')\>
&=
W(\vr)\,W(\vr')\,\xi(\vr,\vr')
\vs&\quad
+ (1+\alpha)\,\frac{W(\vr)\,\delta^D(\vr'-\vr)}{\bar n}\,,
\ea
where we used \cref{eq:window_definition}.
Therefore, the SFB transform of the shot noise term becomes (see
\cref{eq:sfb_discrete_fourier_pair_b})
\ba
N^\obs
&= 
\frac{1}{\bar n}\,\mathrm{SFB}^2[W(\vr)\delta^D(\vr'-\vr)]
\\
&= \frac{1}{\bar n}\,W_{n\ell m}^{n'\ell'm'}\,,
\label{eq:shotnoise}
\ea
in the limit $\alpha\to0$, and the $W$ matrix is defined in
\cref{eq:delta_mixing_matrix_discrete}. The window-corrected shot noise,
therefore, is, in matrix form, $W^{-1}/\bar n$.

For the pseudo-SFB-power-spectrum estimator the shot noise simplifies significantly.
Averaging over the modes $m=m'$ and assuming $\ell=\ell'$, \cref{eq:shotnoise}
becomes
\ba
\label{eq:Nshot_lnn}
N_{\ell nn'}^{\obs}
&=
\frac{1}{\nbar}
\,\frac{1}{\sqrt{4\pi}}\,
\int\dd r\,r^2
\,g_{n\ell}(r)
\,g_{n'\ell}(r)
\,W_{00}(r)
\,,
\ea
where we used \cref{eq:delta_mixing_matrix_discrete_sht}. \cref{eq:Nshot_lnn}
can be implemented very efficiently.

\subsection{Pixel window}
The pixel window refers to a distortion of the power spectrum due to binning
galaxies into pixels. In the radial direction, we do not bin the galaxies, see
\cref{eq:delta_nl_theta_phi_exact}, and, therefore, we do not have a radial
pixel window \citep{Leistedt+:2012A&A...540A..60L}.

However, the signal in \cref{eq:sfb_clnnobs} is still affected by the pixel
window from the spherical harmonic transform. We correct this by subtracting
the shot noise from the observed power spectrum, then using the \texttt{pixwin}
function of \texttt{HealPy} to correct for the pixel window. We confirm the
accuracy of this procedure with simulations in \cref{sec:sfb_applications}.

\subsection{Bandpowers}
\label{sec:sfb_bandpowers}
We use a similar approach as
\citet{Hivon+:2002ApJ...567....2H,Alonso+:2019MNRAS.484.4127A} to bin the SFB
power spectrum into bandpowers. This is necessary if one wants to estimate the
SFB power spectrum itself, as the mixing matrices in
\cref{eq:delta_mixing_matrix_discrete,eq:cell_mixing_matrix} are, in general,
not invertible with finite-precision arithmetic. Compared to those authors our
situation is complicated, but not significantly changed, by the fact that we
may need to bin not only in $\ell$, but also in the $k$-modes $n$ and $n'$.

We define the bandpower-binned pseudo-$C_\ell$ SFB power spectrum as a weighted
sum over modes,
\ba
\label{eq:sfb_bandpowers}
\hat B^\obs_{LNN'}
&=
\sum_{\ell nn'} \widetilde w_{LNN'}^{\ell nn'}\,\hat C^\obs_{\ell nn'}
\,,
\ea
where $\widetilde w_{LNN'}^{\ell nn'}$ is typically a rectangular sparse matrix
that takes the average of neighboring modes $(\ell nn')\sim(LNN')$. The
operation \cref{eq:sfb_bandpowers} is a type of compression, where the
compression matrix $\widetilde w$ must satisfy the normalization
\ba
\label{eq:sfb_bandpowers_binning_condition}
\sum_{\ell nn'} \widetilde w_{LNN'}^{\ell nn'} &= 1\,.
\ea

In matrix notation, we write the compression operation \cref{eq:sfb_bandpowers}
and the corresponding decompression operation
\ba
\label{eq:wvobs_compression}
B^W &= \widetilde w C^W\,, &
C^W &\simeq \widetilde v B^W\,,\\
\label{eq:wv_compression}
B &= w C\,, &
C &\simeq v B\,,
\ea
where $\widetilde w$ and $\widetilde v$ are rectangular matrices operating on
window-convolved power spectra, and $w$ and $v$ are rectangular matrices
operating on cleaned power spectra. We use the index `W' to indicate that we
are only considering the window convolution. That is,
\ba
\label{eq:binned_matrix_mixing}
C^W &= \mathcal{M} C\,,&
B^W &= \mathcal{N} B\,,
\ea
where $\mathcal{M}$ is given by \cref{eq:sfb_cmix}, and we can use the first of
\cref{eq:wvobs_compression}, the first of \cref{eq:binned_matrix_mixing}, and
the last of \cref{eq:wv_compression} to get
\ba
\label{eq:sfb_N_mixing}
\mathcal{N} &= \widetilde w \mathcal{M} v\,.
\ea

The compression matrix $w$ is obtained by inverting the second of
\cref{eq:binned_matrix_mixing}, using the first of \cref{eq:wvobs_compression},
and the first of \cref{eq:binned_matrix_mixing} to get
\ba
B
&= \mathcal{N}^{-1} B^W
= \mathcal{N}^{-1} \widetilde w C^W
= \mathcal{N}^{-1} \widetilde w \mathcal{M} C\,,
\ea
or \citep{Alonso+:2019MNRAS.484.4127A},
\ba
\label{eq:binning_w}
w &= \mathcal{N}^{-1} \widetilde w \mathcal{M} \,.
\ea
Similarly, we find
\ba
\label{eq:binning_v}
\widetilde v &= \mathcal{M} v \mathcal{N}^{-1}\,.
\ea
\cref{eq:sfb_N_mixing} then implies that
\ba
\label{eq:wv_eq_I}
wv = \widetilde w \, \widetilde v &= I\,.
\ea
\cref{eq:wv_eq_I} is equivalent to assuming that a
decompression-then-recompression cycle is lossless. That is, the compressed
representation is unaffected by decompression. The opposite,
compression-then-decompression, however, will in general incur losses in the
compression step, so that $vw \neq I$ except in special cases.

Furthermore, once the information is lost, repeated
compression-then-decompression cycles do not lose more information. That is,
$(vw)^n=vw$ for integer $n\geq1$.

Note that $w$ and $\widetilde v$ must be calculated via
\cref{eq:binning_w,eq:binning_v}, because in the general case we have
$(vw)^\dagger \neq vw$, and, therefore, they are not unique in satisfying
\cref{eq:wv_eq_I}. That is, they are not the unique
Moore-Penrose inverses of the matrices $v$ and $\widetilde w$
\citep{Dresden:1920Sci....52..393D,Penrose:1955PCPS...51..406P}.

Since $w$ and $\widetilde v$ can be expressed in terms of $\widetilde w$, $v$,
and the window mixing matrix, our procedure consists of choosing a compression
matrix $\widetilde w$ and a decompression matrix $v$.

How to choose $\widetilde w$ and $v$? For $\widetilde w$ we have already
suggested that its operation on a power spectrum shall weigh neighboring modes
equally and satisfy the normalization
\cref{eq:sfb_bandpowers_binning_condition}. Since the modes $n$ and $n'$ refer
to modes $k_{n\ell}$, their spacing is not independent, and we need to bin four
modes, nine modes, or similar at a time. In this paper, our binning strategy is
completely specified by the two numbers $\Delta\ell$ and $\Delta n$.

A natural choice for the decompression $v$ is, then, as the Moore-Penrose
inverse of $\widetilde w$. Indeed, for the aforementioned choice for
$\widetilde w$ that takes the average of neighboring modes, this means that $v$
is a step function that assigns the same value or a value proportional to
$\ell\(\ell+1\)$ to all modes within a bin
\citep{Hivon+:2002ApJ...567....2H,Alonso+:2019MNRAS.484.4127A}.

Finally, to compare the binned power spectrum with a theoretical estimate the
first of \cref{eq:wv_compression} must be applied to the theoretical
prediction.

\subsection{Local Average Effect}
\label{sec:sfb_local_average_effect}
In this section we recognize that the average number density $\nbar$ in
\cref{eq:density_constrast_discrete_points} must in practice be measured from
the survey itself. This is often called the \emph{integral constraint}
\citep{Beutler+:2014MNRAS.443.1065B, deMattia+:2019JCAP...08..036D} or the
\emph{local average effect} \citep{dePutter+:2012JCAP...04..019D,
Wadekar+:2020PhRvD.102l3521W}, and here we show that it suppresses the largest
measured SFB mode in the survey.

Measuring the average number density is accomplished by dividing the total
number of galaxies in the survey by the effective volume. However, the total
number of galaxies in the survey is a stochastic quantity such that the average
number density is given by
\ba
\nbar &= \left(1 + \bar\delta\,\right) \nbar^\true\,,
\ea
where $\nbar^\true$ is the underlying density contrast in the whole universe,
and the average density contrast in the survey volume is
\ba
\label{eq:deltabar}
\bar\delta &= \frac{1}{V_\mathrm{eff}}\int\dd^3\vr\,W(\vr)\,\delta(\vr)\,,
\ea
with the effective volume defined in \cref{eq:Veff}.
Therefore, with our model in \cref{eq:density_constrast_discrete_points}, the
measured density contrast is \citep{Taruya+:2021PhRvD.103b3501T}
\ba
\label{eq:delta_estimated}
\delta^\obs(\vr) \equiv \delta^{W,A}(\vr)
&= W(\vr)\,\frac{\delta(\vr) - \bar\delta}{1+\bar\delta}\,,
\ea
where $\delta(\vr)$ is the true density contrast, the superscript `A' refers to
the local average effect, and the superscript `W' refers to the effect of the
window convolution.

The SFB transform of \cref{eq:delta_estimated} is
\ba
\label{eq:delta_W_sfb}
\delta^{W,A}_{n \ell m}
&=
\sum_{n'\ell'm'}
W_{n \ell m}^{n'\ell'm'}
\,\frac{\delta_{n'\ell'm'} - d_{n'\ell'm'}\,\bar\delta}
{1+\bar\delta}\,,
\ea
where we used \cref{eq:delta_mixing,eq:sfb_discrete_fourier_pair_b}, and we
defined
\ba
\label{eq:d_nlm}
d_{n'\ell'm'}
&=
\sqrt{4\pi}\,\delta^K_{\ell'0}\delta^K_{m'0}
\int\dd r\,r^2\,g_{n'0}(r)\,.
\ea
Using
\cref{eq:delta_mixing,eq:sfb_discrete_fourier_pair_a,eq:sfb_discrete_fourier_pair_b},
\cref{eq:deltabar} can be written
\ba
\bar\delta
&=
\frac{1}{V_\mathrm{eff}}
\sum_{n'\ell'm'}
d^{W,*}_{n'\ell'm'}
\,\delta_{n'\ell'm'}
\,,
\ea
where we used \cref{eq:sfb_window_symmetry} to define $d^W_{n'\ell'm'}$ as
\ba
\label{eq:dW_nlm}
d^{W}_{n'\ell'm'}
&=
\sum_{n\ell m}
W^{n\ell m}_{n'\ell'm'}
\,d_{n\ell m}
\,.
\ea
Then, expanding \cref{eq:delta_W_sfb} for small $\bar\delta$ we get
\ba
\delta^{W,A}_{n \ell m}
&=
\sum_{n'\ell'm'}
W_{n \ell m}^{n'\ell'm'}
\Big[
\delta_{n'\ell'm'} - d_{n'\ell'm'}\,\bar\delta
- \delta_{n'\ell'm'}\bar\delta
\vs&\qquad
+ d_{n'\ell'm'}\,\bar\delta^2
+ \delta_{n'\ell'm'}\bar\delta^2
+ \orderof(\bar\delta^3)
\Big].
\ea
Since $d_{n\ell m}\sim \sqrt{V}$ and $\bar\delta\sim 1/\sqrt{V}$ we expand in
the volume $V$. We also assume $C_{\ell nn'} \ll V_\mathrm{eff}$. Then, the
correlation function becomes
\ba
\label{eq:deltaWAdeltaWA}
&\<\delta^{W,A}_{NLM}\,\delta^{W,A,*}_{N'L'M'}\>
\vs
&=
\sum_{n\ell m}
W_{NLM}^{n\ell m}
\sum_{n'\ell'm'}
W_{N'L'M'}^{n'\ell'm',*}
\vs&\quad\times
\Big[
\<\delta_{n \ell m}\,\delta^*_{n'\ell'm'}\>
-\<\delta_{n \ell m}\,\bar\delta\> d_{n'\ell'm'}
- d_{n \ell m} \<\delta^*_{n'\ell'm'}\,\bar\delta\>
\vs&\qquad
+ d_{n\ell m}d_{n'\ell'm'}\<\bar\delta^2\>
+ \orderof\!\(V^{-\frac12}\)
\Big],
\ea
where we assume the field $\delta$ to be Gaussian.
The 2-point terms entering this expression are
\ba
\<\delta_{n \ell m}\,\delta^{*}_{n'\ell'm'}\>
&=
\delta^K_{\ell\ell'}\delta^K_{mm'}\,C_{\ell nn'}
+ \frac{1}{\nbar}\, (W^{-1})_{n \ell m}^{n' \ell' m'}
\,,\displaybreak[0]\\
\<\delta_{n \ell m}\,\bar\delta\>
&=
\frac{1}{V_\mathrm{eff}}
\sum_{n''}
d^{W}_{n''\ell m}
\,C_{\ell nn''}
+
\frac{1}{\nbar V_\mathrm{eff}}
\, d_{n \ell m}
\,,\displaybreak[0]\\
\label{eq:bardelta_squared}
\<\bar\delta^2\>
&=
\frac{1}{V^2_\mathrm{eff}}
\sum_{\ell_1 n_1 n_2}
D^W_{\ell_1 n_1 n_2}
\,C_{\ell_1 n_1 n_2}
\vs&\quad
+
\frac{1}{\nbar V^2_\mathrm{eff}}
\sum_{n_1\ell_1m_1}
\, d_{n_1\ell_1m_1}
\, d^W_{n_1\ell_1m_1}
\,,
\ea
where we included the shot noise term \cref{eq:shotnoise}, and we defined the
unnormalized power spectrum of a constant field
\ba
\label{eq:DW_lnn}
D^W_{\ell_1 n_1 n_2}
&=
\sum_{m_1}
d^{W}_{n_1\ell_1 m_1}
\,d^{W,*}_{n_2\ell_1 m_1}
\\
&=
\(2\ell_1+1\)
\sum_{n'_1}
\sum_{n'_2}
\mathcal{M}^{0 n'_1 n'_2}_{\ell_1 n_1 n_2}
\,d_{n'_100}
\,d_{n'_200}\,,
\ea
and \cref{eq:sfb_cmix} was used in the last line.
Now the correlation function becomes
\begin{widetext}
\ba
\label{eq:deltaAdeltaA}
\<\delta^{A}_{n\ell m}\,\delta^{A,*}_{n'\ell'm'}\>
&=
\delta^K_{\ell\ell'}\delta^K_{mm'}\,C_{\ell nn'}
+
\frac{1}{\nbar}\, (W^{-1})_{n \ell m}^{n' \ell' m'}
-
d_{n'\ell'm'}
\frac{1}{V_\mathrm{eff}}
\sum_{n''}
d^W_{n''\ell m}
\,C_{\ell nn''}
-
d_{n \ell m}
\frac{1}{V_\mathrm{eff}}
\sum_{n''}
d^{W,*}_{n''\ell'm'}
\,C_{\ell'n'n''}
\vs&\quad
+
\left[\frac{-2}{\nbar V_\mathrm{eff}} + \<\bar\delta^2\> \right]
\, d_{n \ell m}
\, d_{n'\ell'm'}
+ \orderof\!\(V^{-\frac12}\),
\ea
where we corrected for the window function. Only the first and fifth terms are
proportional to $\delta^K_{\ell\ell'}\delta^K_{mm'}$, and so we cannot take the
pseudo-$C_{\ell nn'}$ power spectrum at this stage.\footnote{If $d^W_{n\ell
  m}\,\propto\, \delta^K_{\ell0} \delta^K_{m0}$ would be a good approximation,
  which is the case in the absence of a window function, then the
  pseudo-$C_\ell$ approach works well. This is an argument for using
  eigenfunctions tailored to the survey geometry. That is, if the window
effects are already captured by the eigenfunctions, then the calculation here
would simplify dramatically.} To do so, we now add back the two window
functions in \cref{eq:deltaWAdeltaWA} to get a prediction for the observed
pseudo power spectrum
\ba
\label{eq:CWAlnn_direct}
C^{W,A}_{LNN'}
&=
\sum_{\ell nn'}
\mathcal{M}^{\ell nn'}_{LNN'}
\,C_{\ell nn'}
+
N^\obs_{LNN'}
-
\frac{1}{V_\mathrm{eff}}
\,\frac{1}{2L+1}
\sum_M
d^{W,*}_{N'LM}
\sum_{n\ell m}
W_{NLM}^{n\ell m}
\sum_{n''}
d^W_{n''\ell m}
\,C_{\ell nn''}
\vs&\quad
-
\frac{1}{V_\mathrm{eff}}
\,\frac{1}{2L+1}\sum_M
d^W_{NLM}
\sum_{n'\ell'm'}
W_{N'LM}^{n'\ell'm',*}
\sum_{n''}
d^{W,*}_{n''\ell'm'}
\,C_{\ell'n'n''}
+
\sum_{\ell nn'}
\mathcal{M}_{LNN'}^{\ell nn'}
\,\delta^K_{\ell0}d_{n00}d_{n'00}
\left[\frac{-2}{\nbar V_\mathrm{eff}} + \<\bar\delta^2\>\right]
\vs&\quad
+ \orderof\!\(V^{-\frac12}\),
\ea
where we used \cref{eq:Nshot_lnn}. The third and fourth terms are the same
except for $N \leftrightarrow N'$. To simplify these two terms, we express them
in terms of chains of window functions $W_k$ that we define in \cref{eq:sfb_Wk}
and study in \cref{sec:w_chains}. We use \cref{eq:dW_nlm} and get
\ba
\label{eq:sfb_WWW}
\sum_M
d^{W,*}_{N'LM}
\sum_{n\ell m}
W_{NLM}^{n\ell m}
\sum_{n''}
d^W_{n''\ell m}
\,C_{\ell nn''}
&=
\sum_M
\sum_{n'\ell'm'}
\sum_{n\ell m}
\sum_{n''}
\sum_{n_1 \ell_1 m_1}
W_{N'LM}^{n'\ell'm',*}
W_{NLM}^{n\ell m}
W_{n''\ell m}^{n_1 \ell_1 m_1}
\,d_{n'\ell'm'}
\,d_{n_1 \ell_1 m_1}
\,C_{\ell nn''}
\vs
&=
\sum_{\ell n n''}
\sum_{\ell' n_1 n'}
\sum_{Mmm'}
W^{N'LM}_{n'\ell'm'}
W_{NLM}^{n\ell m}
W_{n''\ell m}^{n_1 \ell' m'}
\,
\delta^K_{\ell'0}
\,d_{n'00}
\,d_{n_1 00}
\,C_{\ell nn''}
\vs
&=
\sum_{\ell' n_1 n'}
\delta^K_{\ell'0}
\,d_{n'00}
\,d_{n_1 00}
\sum_{\ell n n''}
C_{\ell nn''}
\,
W_3\!\!\begin{pmatrix}
  L  & \ell & \ell' \\
  N' & n    & n_1 \\
  N  & n''  & n'
\end{pmatrix},
\ea
where $W_3$ is defined in \cref{eq:sfb_Wk}.
\cref{eq:CWAlnn_direct} now becomes
\ba
\label{eq:CWAlnn}
C^{W,A}_{LNN'}
&=
\sum_{\ell nn'}
\mathcal{M}^{\ell nn'}_{LNN'}
\left[
  C_{\ell nn'}
  +
  \delta^K_{\ell0}d_{n00}d_{n'00}
  \left(\frac{-2}{\nbar V_\mathrm{eff}} + \<\bar\delta^2\>\right)
\right]
+
N^\obs_{LNN'}
\vs&\quad
-
\frac{1}{V_\mathrm{eff}}
\,\frac{1}{2L+1}
\sum_{\ell' n_1 n'}
\delta^K_{\ell'0}
\,d_{n_1 00}
\,d_{n'00}
\sum_{\ell n n''}
C_{\ell nn''}
\left[
W_3\!\!\begin{pmatrix}
  L  & \ell & \ell' \\
  N' & n    & n_1 \\
  N  & n''  & n'
\end{pmatrix}
+
\< N \leftrightarrow N' \>
\right].
\ea
\end{widetext}
The window de-convolved power spectrum is then $C^A=\mathcal{M}^{-1}C^{W,A}$.

In the absence of a window function and assuming $C_{\ell
nn'}\,\propto\,\delta^K_{nn'}$, as well as using \cref{eq:d_nlm}, we get
\ba
C^A_{\ell nn'}
&\simeq
\(1 + \frac{3 A\,C_{011}}{V_\mathrm{eff}}\)
C_{\ell nn'}
-
\delta^K_{\ell0}
\,\frac{d_{n00}\,d_{n'00}}{V_\mathrm{eff}}
\,B_{nn'}\,,
\ea
where we included further terms from the expansion \cref{eq:deltaWAdeltaWA},
and we defined
\ba
A &= 
\sum_{n_1}
\frac{d_{n_100}^2}{V_\mathrm{eff}}
\,\frac{C_{0n_1n_1}}{C_{011}}\,,
\\
B_{nn'}
&=
C_{0nn} + C_{0n'n'}
- A\,C_{011}
-\frac{6\, C_{0 nn} \,C_{0 n'n'} }{V_\mathrm{eff}}
\,.
\ea
To a good approximation $A\simeq1$, and the last term in $B_{nn'}$ can be
neglected if the effective volume is sufficiently large and the shot noise
sufficiently low.
Furthermore, a good approximation is $d_{n00}\,\propto\,\delta^K_{n1}$.
Therefore, only the $(\ell,n,n')=(0,1,1)$ mode is significantly affected. That
is, the main effect of super-sample variance on the measured power spectrum is
to reduce the power in the largest mode.

\subsection{Covariance matrix of power spectrum}
\label{sec:sfb_covariance}
\begin{figure*}
  \centering
  \incgraph{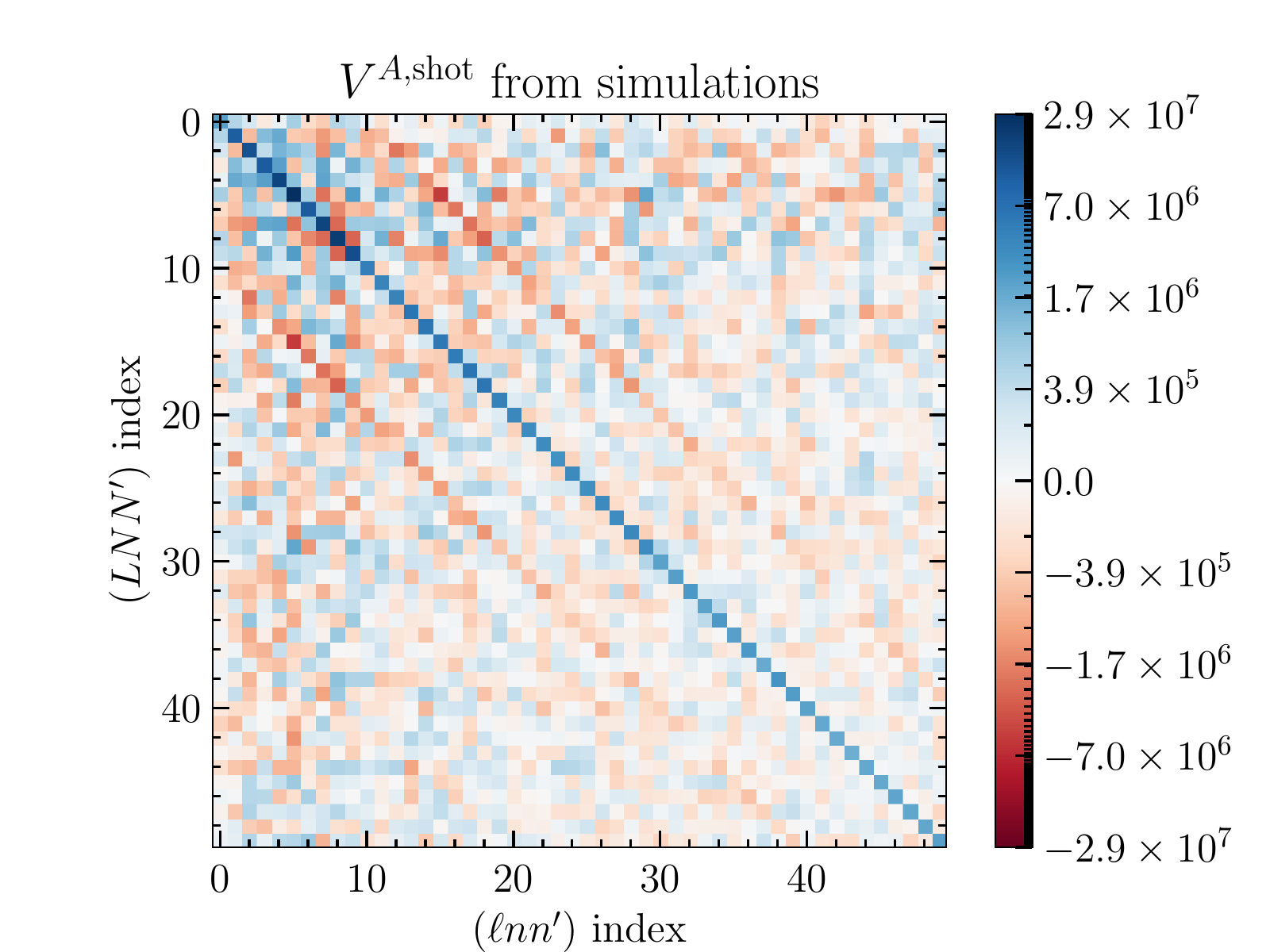}
  \incgraph{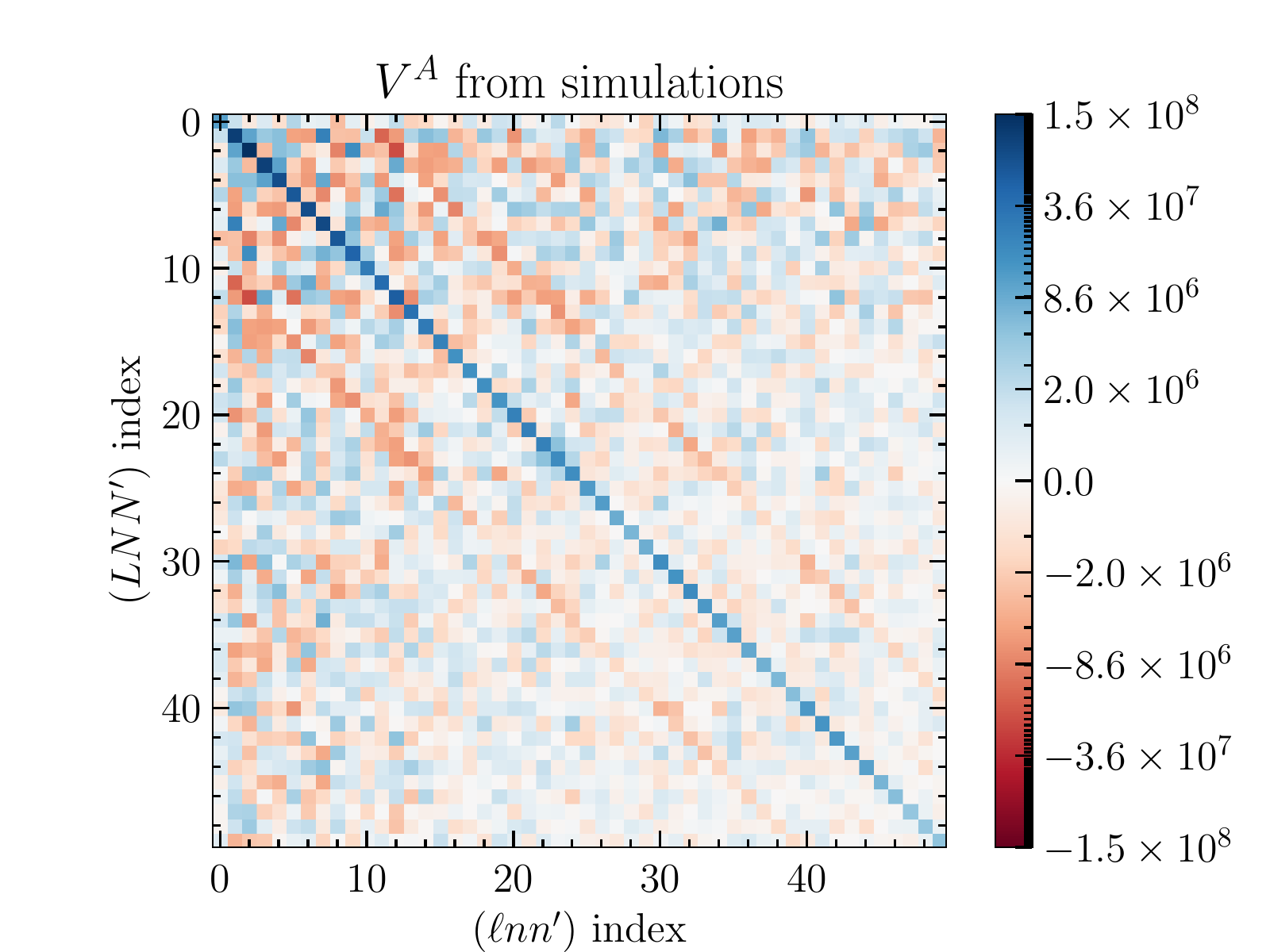}
  \incgraph{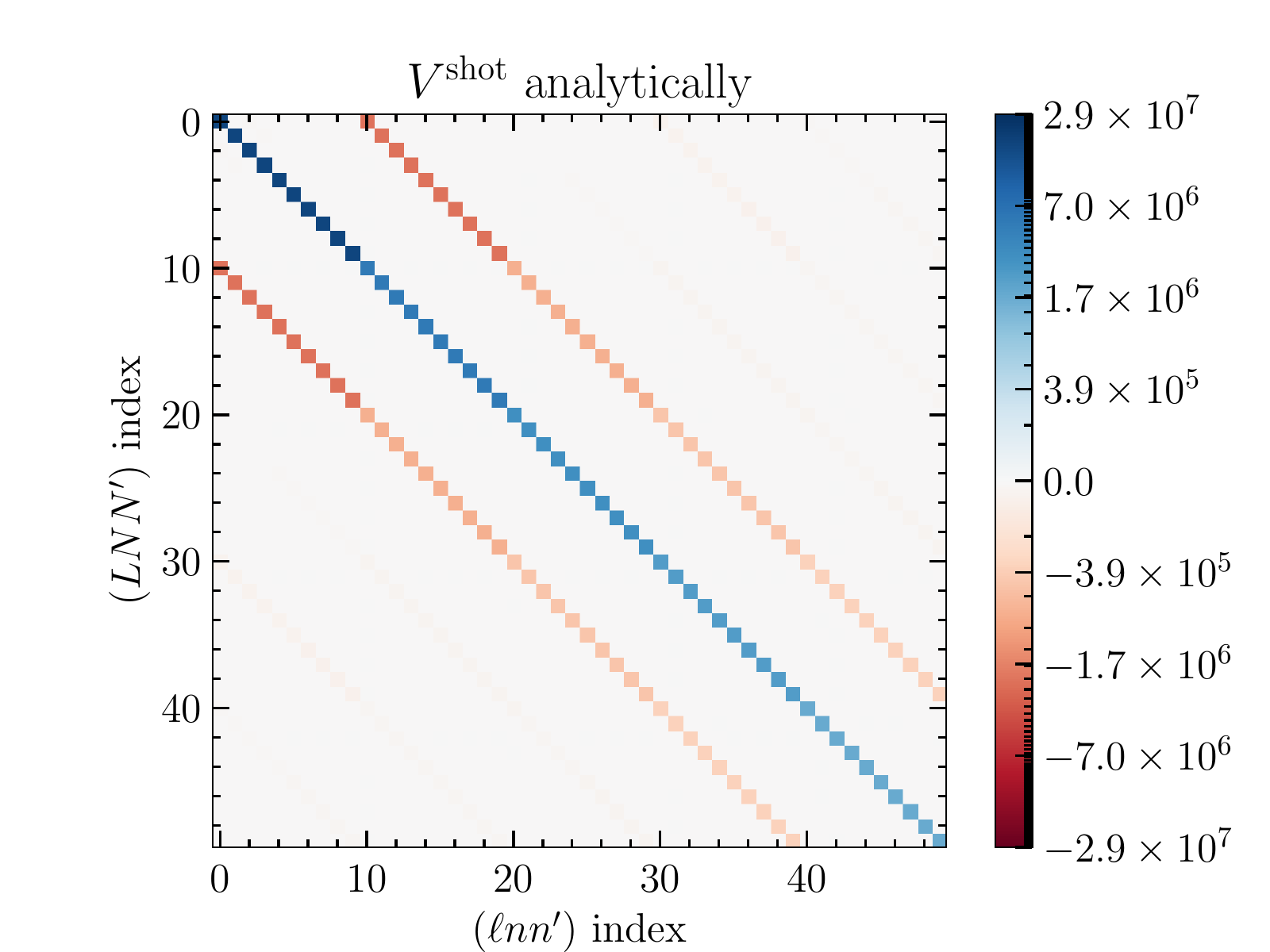}
  \incgraph{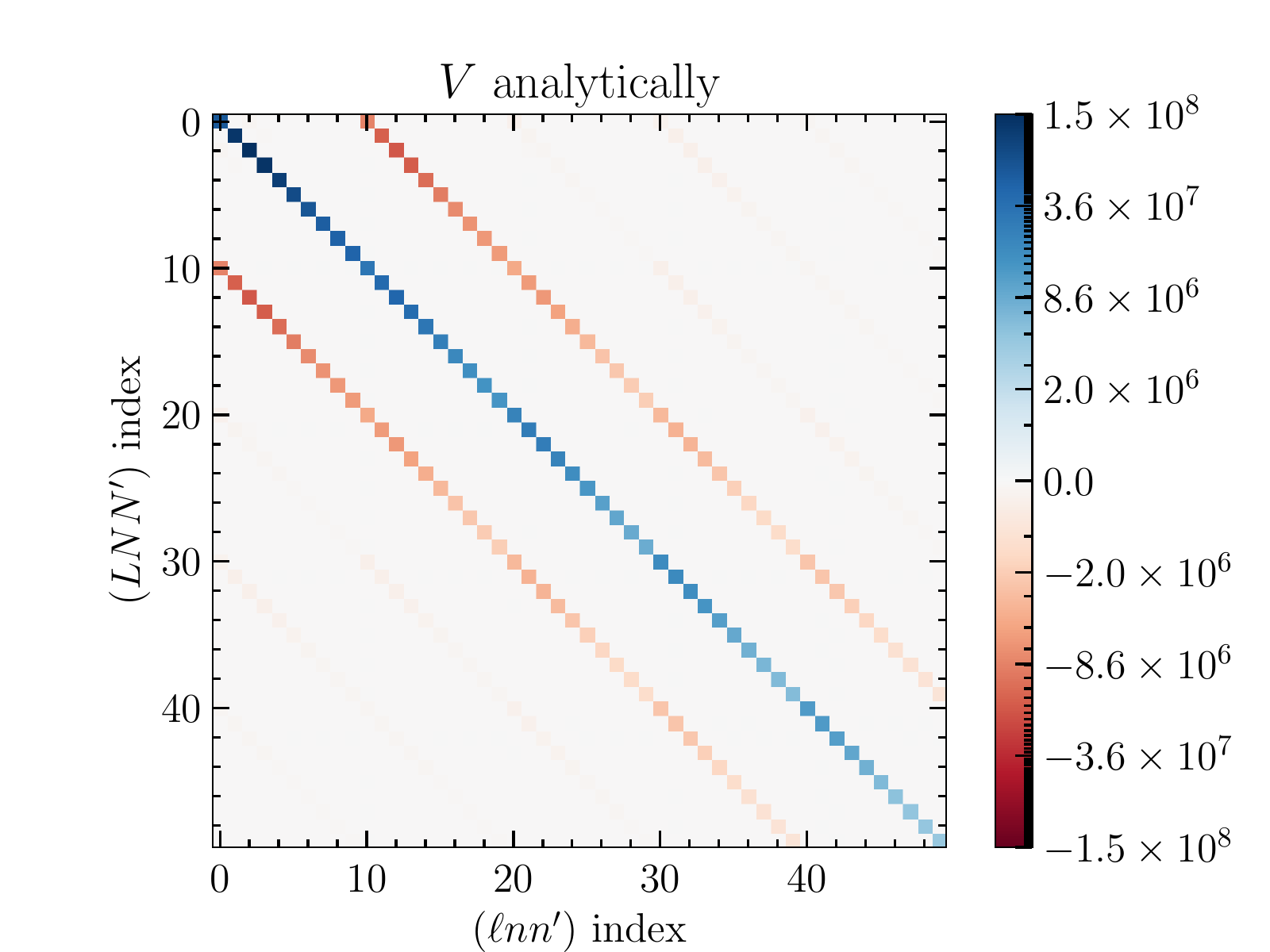}
  \caption{
    The panels show the covariance matrix $V$ defined in
    \cref{eq:covariance_matrix_window_deconvolution}.
    Top row: $V^A$ as measured from 100 simulations, including the local
    average effect.
    Bottom row: The analytical prediction, without local average effect.
    Left column: Shot noise only.
    Right column: Linear power spectrum with shot noise.
    Here we use a simulation with \SI{50}{\percent} sky coverage and bandpower
    binning with $\Delta\ell=2$. The order of the $(\ell nn')$ indices is such
    that each block of ten indices is for one $\ell$-bin starting with
    $\ell=0,1$ for the block starting with index 0 and ending with $\ell=8,9$
    for the block starting at index 40. Within each block $n=n'$ increases from
    one to 10.
    Note that the nonlinear color scheme amplifies small elements.
  }
  \label{fig:covariance_matrix}
\end{figure*}
\begin{figure*}
  \centering
  \incgraph{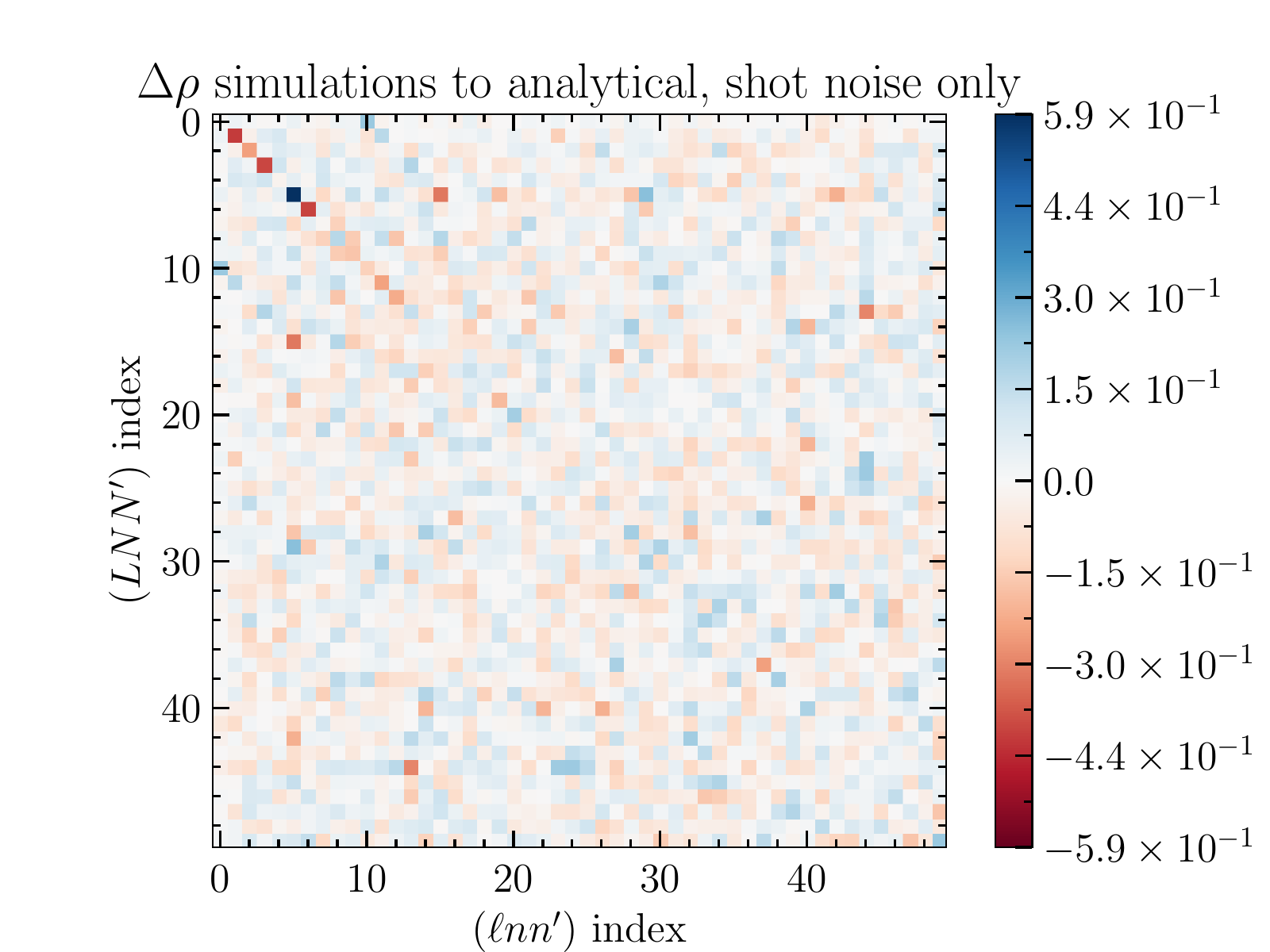}
  \incgraph{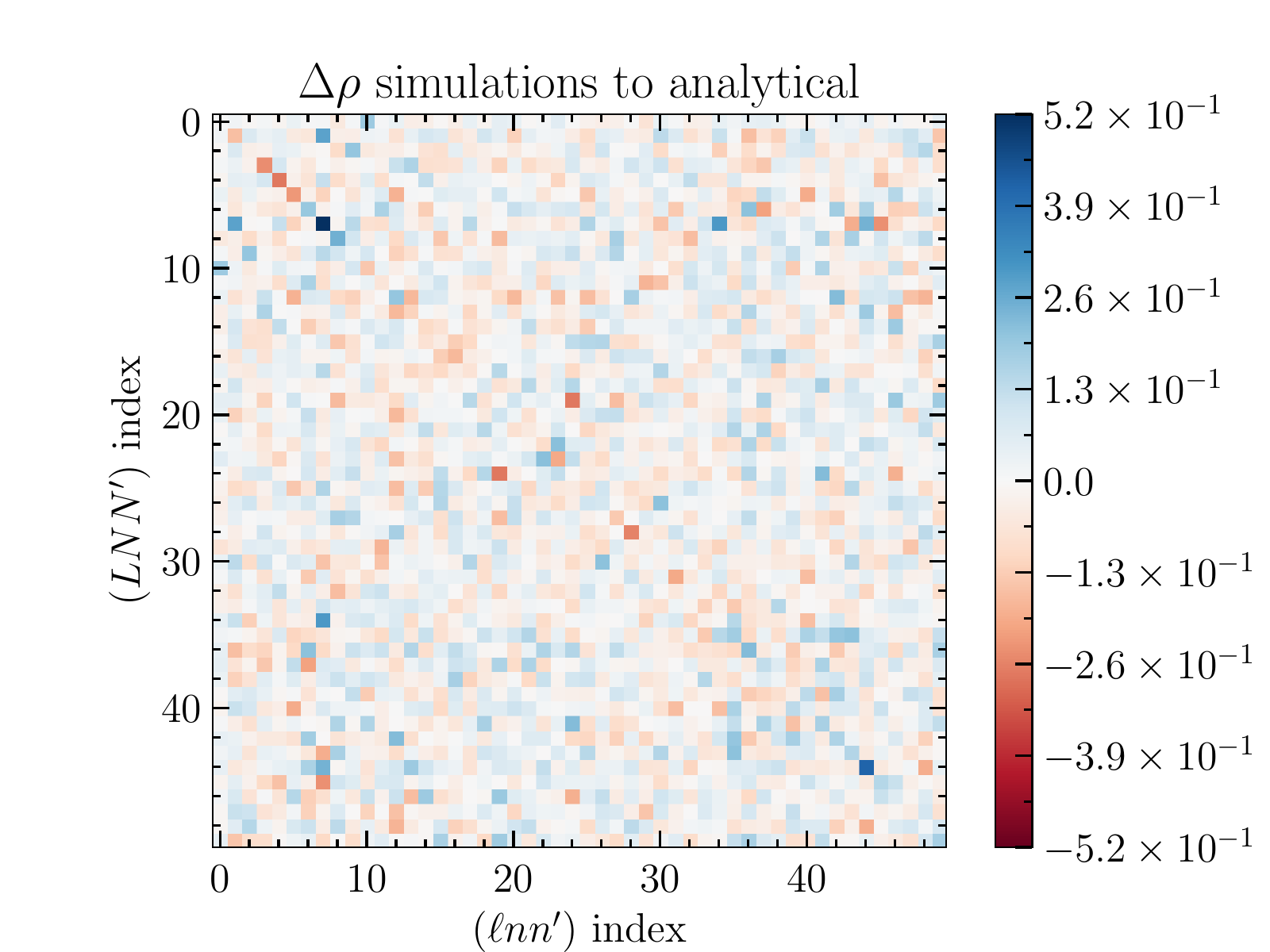}
  \caption{
    Relative comparison between the covariance matrices as in \cref{eq:drho}.
    This avoids amplifying small deviations far from the center diagonal, and
    it shows that the analytic result largely agrees with the simulations. The
    very largest mode is set to zero, because it is affected by the local
    average effect that is not included in the analytical result (see
    \cref{fig:covariance_matrix_diagonals}).
  }
  \label{fig:covariance_matrix_drho}
\end{figure*}
\begin{figure*}
  \centering
  \incgraph{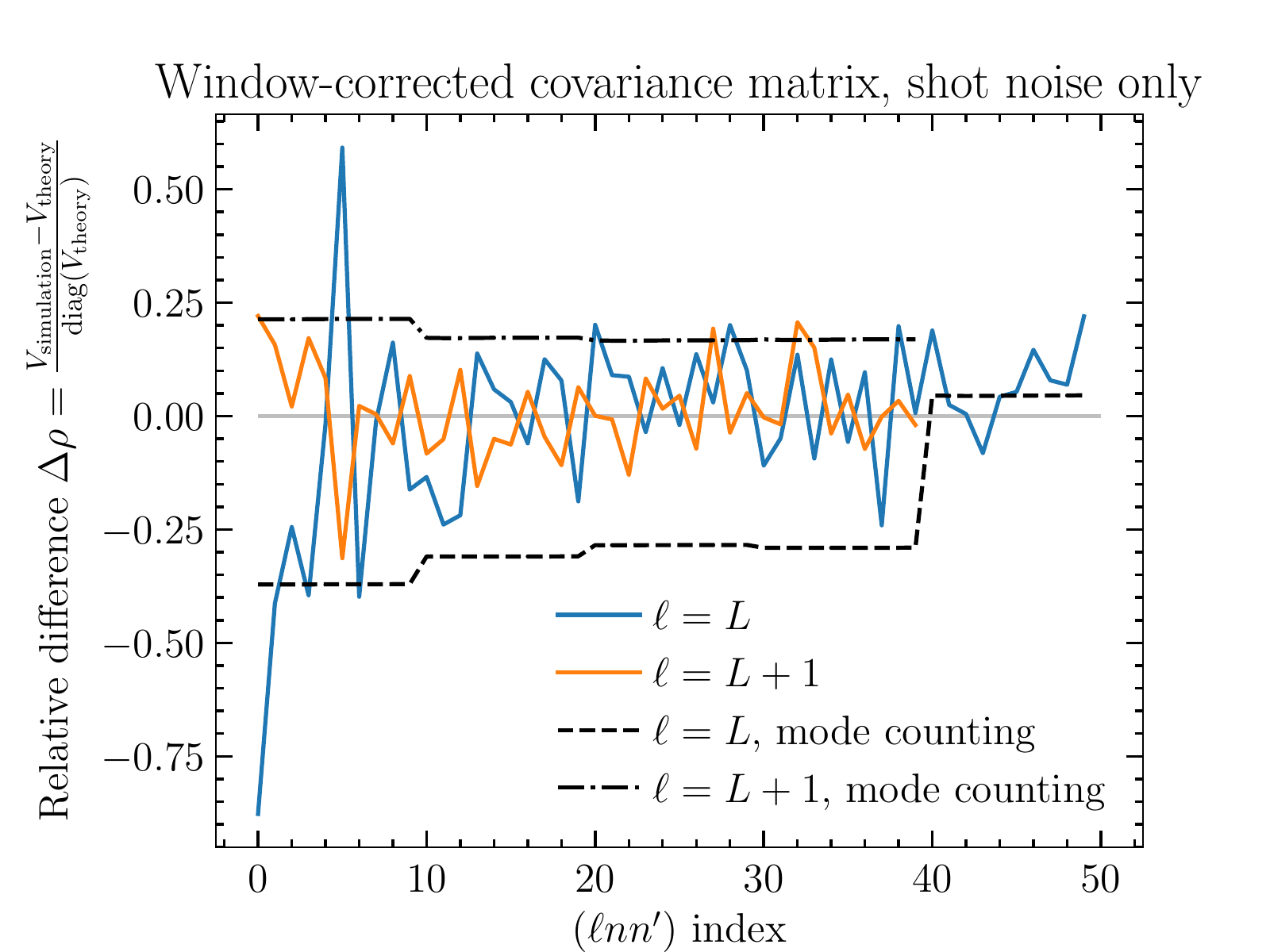}
  \incgraph{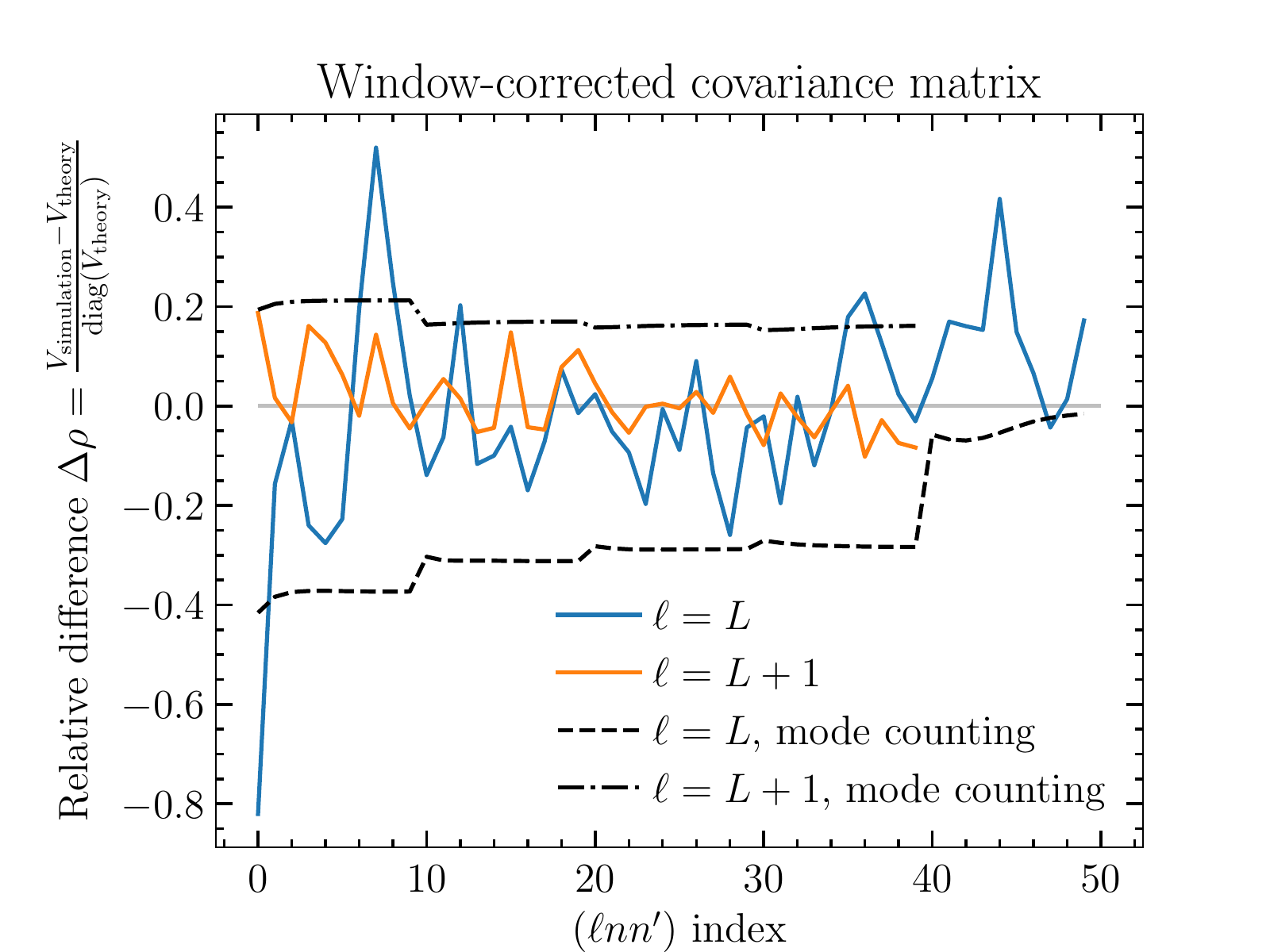}
  \caption{Comparison of the diagonal and the $\ell=L+1$ off-diagonal
    window-corrected covariance between simulations and analytical result. Left
    shows the shot noise only, and right including shot noise and a non-zero
    power spectrum. In the analytical result we do not include local average
    effect. Thus, the first mode in the simulations is suppressed compared to
    the analytical result.
    The black dashed and dash-dotted lines show the approximate result from
    mode-counting \cref{eq:sfb_covariance_modecounting}.
  }
  \label{fig:covariance_matrix_diagonals}
\end{figure*}
In this section we provide a covariance matrix for the SFB power spectrum.
Several approaches have been used previously.
\citet{Percival+:2004MNRAS.353.1201P,Wang+:2020JCAP...10..022W} trace the
likelihood function either on a grid or using Markov Chain Monte Carlo
techniques. \citet{Wang+:2020JCAP...10..022W}, e.g., use simulations to measure
the covariance matrix from suites of mock catalogues. An analytical approach
for the 3D power spectrum multipoles is presented in
\citet{Wadekar+:2020PhRvD.102l3517W}. In this paper, we get an analytical
estimate for the SFB power spectrum assuming that the density contrast is
Gaussian, and we compare to 100 log-normal simulations. Non-Gaussian terms in
the form of the disconnected trispectrum could be included similarly to
\citet{Taruya+:2021PhRvD.103b3501T,Sugiyama+:2020MNRAS.497.1684S}.

Super-sample variance \citep[e.g.,][]{dePutter+:2012JCAP...04..019D,
Lacasa+:2019A&A...624A..61L, Li+:2018JCAP...02..022L} can have a significant
impact on the covariance matrix. \emph{Beat coupling} is mode mixing due to the
window function with correlation between pairs of non-linear modes and one
large mode. The \emph{local average effect} is due to the large-scale mode
modulating the average number density inside the survey volume. Both these
effects can be treated in the manner of
\cref{sec:sfb_local_average_effect,eq:delta_estimated}.

The covariance matrix on the observed SFB power spectrum is
\ba
\label{eq:Clnnobs_covariance_raw}
&V_{\ell nn'}^{LNN',\obs}
\equiv
\<\hat C^\obs_{\ell nn'} \, \hat C^\obs_{LNN'}\> - C^\obs_{\ell nn'} \, C^\obs_{LNN'}
\vs
&=
\frac{1}{(2\ell+1)(2L+1)}\sum_{mM}\Big[
  \<\delta^{W,A}_{n\ell m}\delta^{W,A}_{NLM}\> \<\delta^{{W,A},*}_{n'\ell m}\delta^{{W,A},*}_{N'LM}\>
\vs&\quad
+\<\delta^{W,A}_{n\ell m}\delta^{{W,A},*}_{N'LM}\> \<\delta^{W,A}_{NLM}\delta^{{W,A},*}_{n'\ell m}\>
\Big],
\ea
where we used Wick's theorem for a Gaussian density contrast. We simplify
\cref{eq:Clnnobs_covariance_raw} in \cref{sec:sfb_covariance_simplification}.
However, an analytical calculation remains computationally expensive.

To get the covariance matrix for the window-corrected power spectrum,
we write the matrix equation
\ba
\label{eq:covariance_matrix_window_deconvolution}
V
&=
\mathcal{N}^{-1}\,V^{\obs}\,\mathcal{N}^{-1,T}\,.
\ea
where $\mathcal{N}$ is the bandpower-binned window coupling matrix given in
\cref{eq:sfb_N_mixing}, and the binning of the covariance matrix is implied.

A reasonably precise estimate can be obtained by counting modes and assuming
the covariance matrix is diagonal. That is,
\ba
\label{eq:sfb_covariance_modecounting}
V_{\ell nn'}^{LNN'}
&\simeq
\frac{\delta^K_{\ell L}}{N_\mathrm{modes}}
\left[
  C_{\ell nN}^\mathrm{binned}
  C_{L n'N'}^\mathrm{binned}
  +
  C_{\ell nN'}^\mathrm{binned}
  C_{L n'N}^\mathrm{binned}
\right],
\ea
where the power spectrum includes the shot noise, $C_{\ell nn'}^\mathrm{binned} =
C^\mathrm{signal}_{\ell nn'}+N^\mathrm{shot}_{\ell nn'}$, and
\ba
\label{eq:sfb_Nmodes}
N_\mathrm{modes} &= f_\mathrm{vol} \,(2\ell+1)\,\Delta\ell\,\Delta n
\,,
\ea
where $\Delta\ell$ and $\Delta n$ are the bin widths for modes $k_{n\ell}$, and
$f_\mathrm{vol}$ is the fraction of the SFB transform-volume that is occupied
by the survey, defined by
\ba
f_\mathrm{vol}
&\equiv
\frac{1}{V_\mathrm{SFB}}\int\dd^3\vr\,\tau\!\left[W(\vr) - W_\mathrm{threshold}\right],
\ea
where $\tau(x)$ is a step function and $W_\mathrm{threshold}$ is a threshold of
the window function.

The shot noise takes into account the variation of the number density across
the survey, and it enters in \cref{eq:sfb_covariance_modecounting} as part of
the power spectrum. The incomplete volume coverage enters as a reduction in the
number of modes, and it is needed for the stability of the window-deconvolution
when there are large unobserved regions in the SFB volume.

In \cref{fig:covariance_matrix} we show the covariance matrices for a set of
simulations that contain only shot noise (top left) as well as for a set of
simulations with a physical galaxy power spectrum with bias $b=1.5$ at
effective redshift $z_\mathrm{eff}=2$ (top right). In the figure we also show
the analytical result from \cref{eq:Clnnobs_covariance_raw} (bottom panels).

The colorbar in the figure is nonlinear. As a result, small elements appear
amplified. To provide a more useful comparison, we introduce the difference
between two covariance matrices, scaled to the center diagonal. That is, we
introduce the relative difference
\ba
\label{eq:drho}
\Delta\rho_{ij} &= \frac{C^A_{ij} - C^B_{ij}}{\sqrt{C^B_{ii}\,C^B_{jj}}}\,,
\ea
and we choose $C^B$ to refer to the analytic result. $\Delta\rho$ does not
suffer from amplification of small differences far from the diagonal.

Therefore, in \cref{fig:covariance_matrix_drho} we show the relative difference
between the covariance matrix as obtained from simulations and the analytical
result. However, in the figure we remove the largest mode, since we have not
included the local average effect in the analytical calculation. All other
modes are statistically essentially equal between simulation result and
analytics.

To show this more clearly, we present \cref{fig:covariance_matrix_diagonals},
where we compare the main diagonal and the $\ell=L+1$ diagonal of the
covariance matrices using the same statistic \cref{eq:drho}. Within the
noise, we find good agreement between simulations and analytical result.

\subsection{Performance scaling}
\begin{figure}
  \centering
  \incgraph{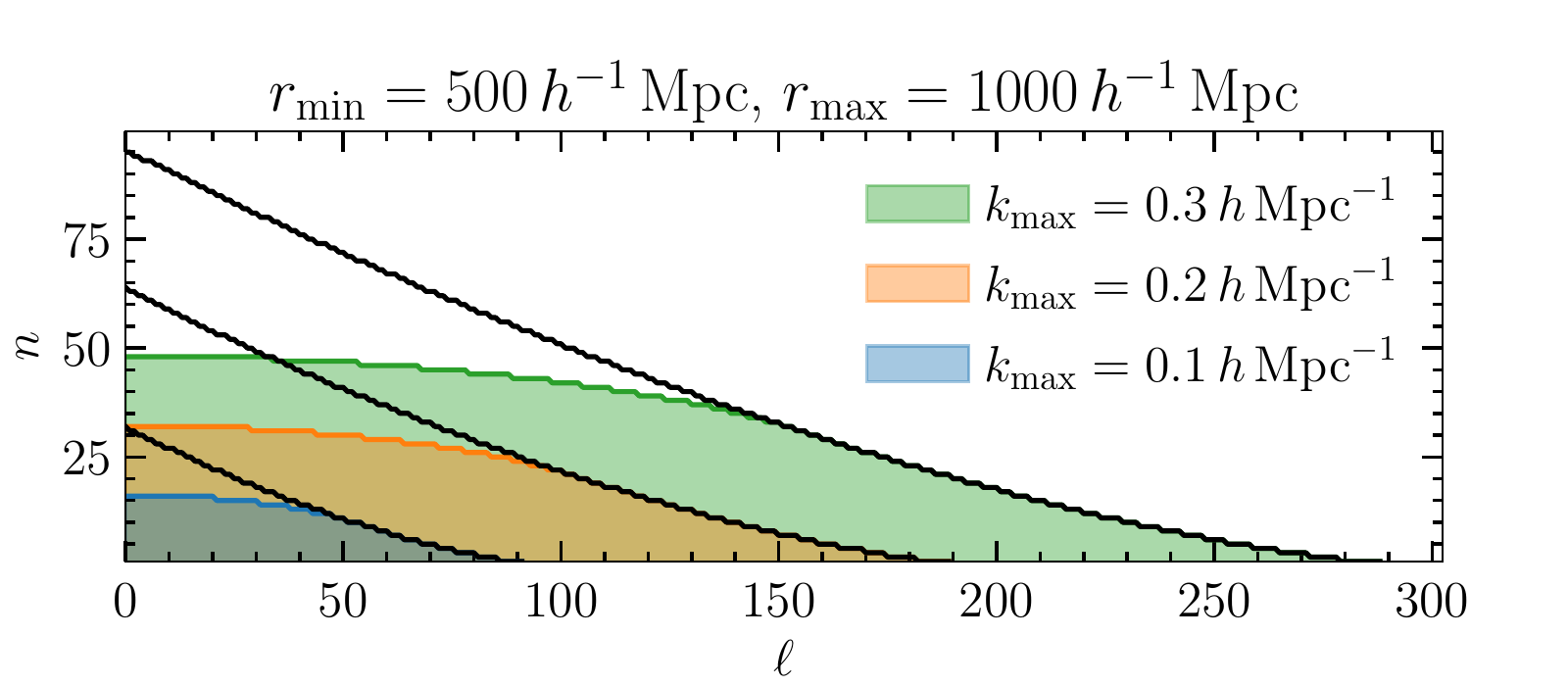}
  \caption{
    Here we show the modes that are required to achieve a given $k_\max$ for
    the given radial boundaries. The solid black curves are constant-$k$
    contours for $r_\min=0$.
  }
  \label{fig:modes_nl}
\end{figure}
In this section we give a brief overview of the performance behavior of the
code. We consider the scaling of the SFB decomposition, the coupling matrix,
and the number of modes with the parameters of the SFB power spectrum
estimation.

Typically, it is desirable to calculate all modes up to some $k_\max$. The
total number of modes can be estimated in the same way as for a standard
Fourier transform by estimating the fundamental frequency from the volume that
is being transformed, i.e.,
$V_\mathrm{SFB}\simeq\frac{4\pi}{3}\(r_\max^3-r_\min^3\)$ and
$k_F\simeq2\pi/V_\mathrm{SFB}^{1/3}$. Then, the total number of modes is
approximately $N_\mathrm{modes}\simeq\frac{4\pi}{3}\,k_\max^3 /
k_F^3\simeq\frac{1}{18\pi}\,k_\max^3\(r_\max^3-r_\min^3\)$. These are the modes
that need to be calculated for transform.

However, as shown in \cref{fig:modes_nl}, the boundary at $r_\min$ changes the
structure about which modes need to be calculated. In the figure, modes with
$k\leq k_\max$ are in the shaded region when, for illustration,
$r_\min=\SI{500}{\per\h\mega\parsec}$, and all the modes below the solid black
curves need to be calculated if $r_\min=0$. The figure shows that fewer
low-$\ell$ modes are needed when $r_\min$ is finite. However, most modes are at
large $\ell$, and, therefore, this is only a small computational reduction.

The algorithm now scales as follows. First, the sum in
\cref{eq:delta_nl_theta_phi_exact} is performed, and then for each
$(n,\ell)$-combination the spherical harmonic transform
\cref{eq:sfb_spherical_harmonic_transform} is performed. Hence, the execution
time of the transform scales as
\ba
T &\sim\orderof\!\left[
  n_\max \, \ell_\max \( N_\mathrm{gal}
  + N^\mathrm{Healpix}_{\ell} \)
\right],
\ea
where $N^\mathrm{Healpix}_\ell\sim N_\mathrm{pix}^{3/2}$ is the number of
operations needed for the spherical harmonic transform, and $N_\mathrm{pix}$ is
the number of HEALPix pixels.

To estimate $N_\mathrm{pix}=12 n_\mathrm{side}^2$, we need to estimate
$n_\mathrm{side}$, which we do by considering the angular resolution.
Recommended\footnote{\url{https://healpix.jpl.nasa.gov/}} is $\ell_\max \simeq
2 n_\mathrm{side}$. However, our algorithms dealing with the window function
will need to go to $L=2\ell_\max$. Hence, we estimate
\ba
\label{eq:nside}
n_\mathrm{side} &= 2^{\mathrm{ceil}\(\log_2\(\ell_\max + \frac12\)\)},
\ea
where $\mathrm{ceil}(x)$ is the smallest integer greater than $x$.
\cref{eq:nside} guarantees that $n_\mathrm{side}$ is a power of two.

The angular resolution is determined by $\ell_\max$, which is determined by
$k_\max$ and $r_\max$ by Limber's relation \cref{eq:sfb_limber_ratio}. However,
we note that the actual number needed for $\ell_\max$ tends to be smaller by a
few percent.

For the three surveys in \cref{sec:sfb_applications} below, the SFB
analysis takes \num{\sim10}~CPU-min per simulation for the \emph{Roman}- and
\emph{SPHEREx}-like surveys, and \num{\sim1.5}~CPU-hour for the
\emph{Euclid}-like survey.
Calculation of the mixing matrix $\mathcal{M}$ for the three surveys is on the
order of a few CPU-minutes, exploiting the angular/radial split, and would take
several CPU-hours without that split.
At present, the analytical covariance matrix is only feasible for the largest
scales.

\section{Future Applications}
\label{sec:sfb_applications}
\begin{figure*}
  \centering
  \incgraph{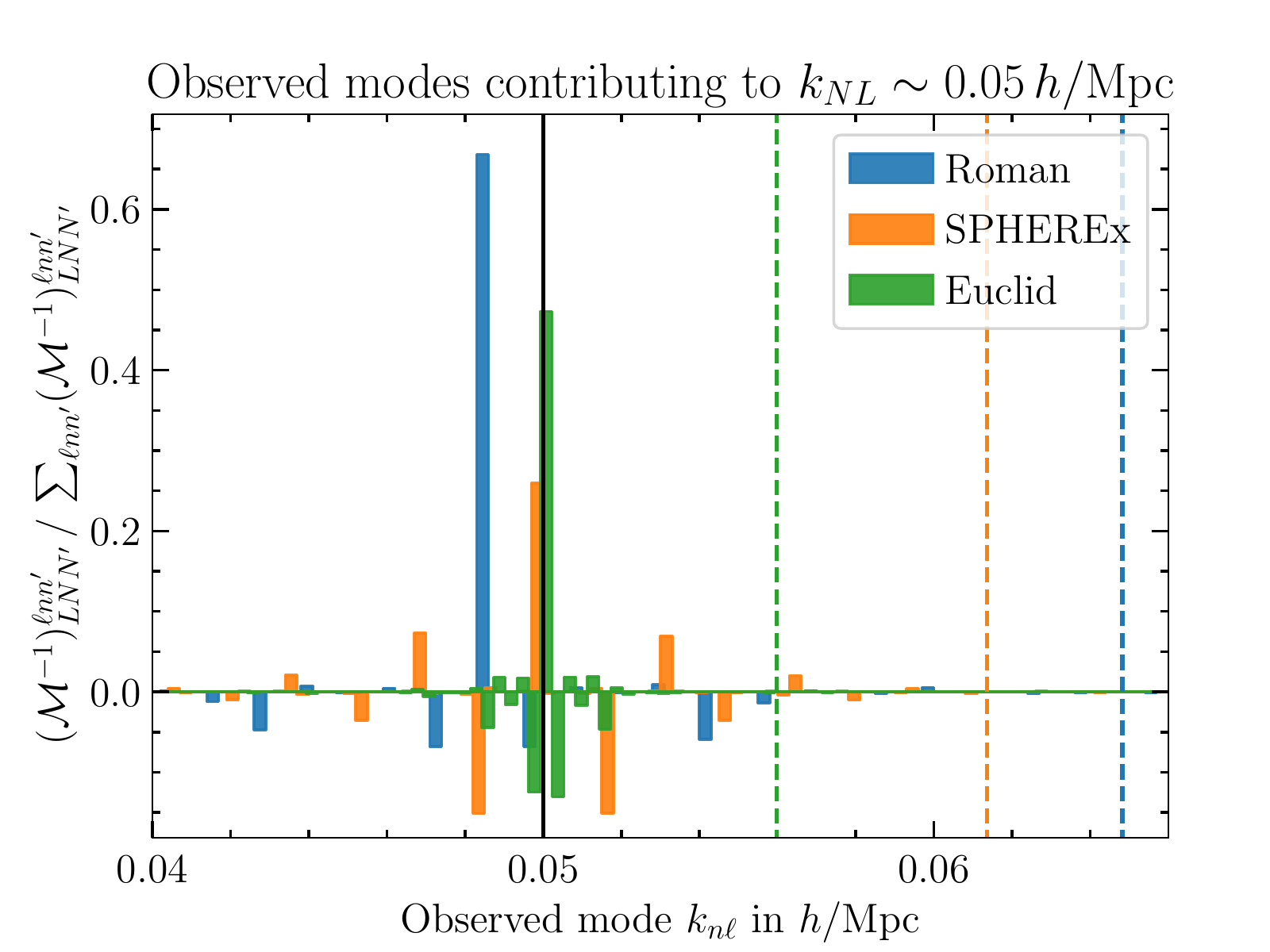}
  \incgraph{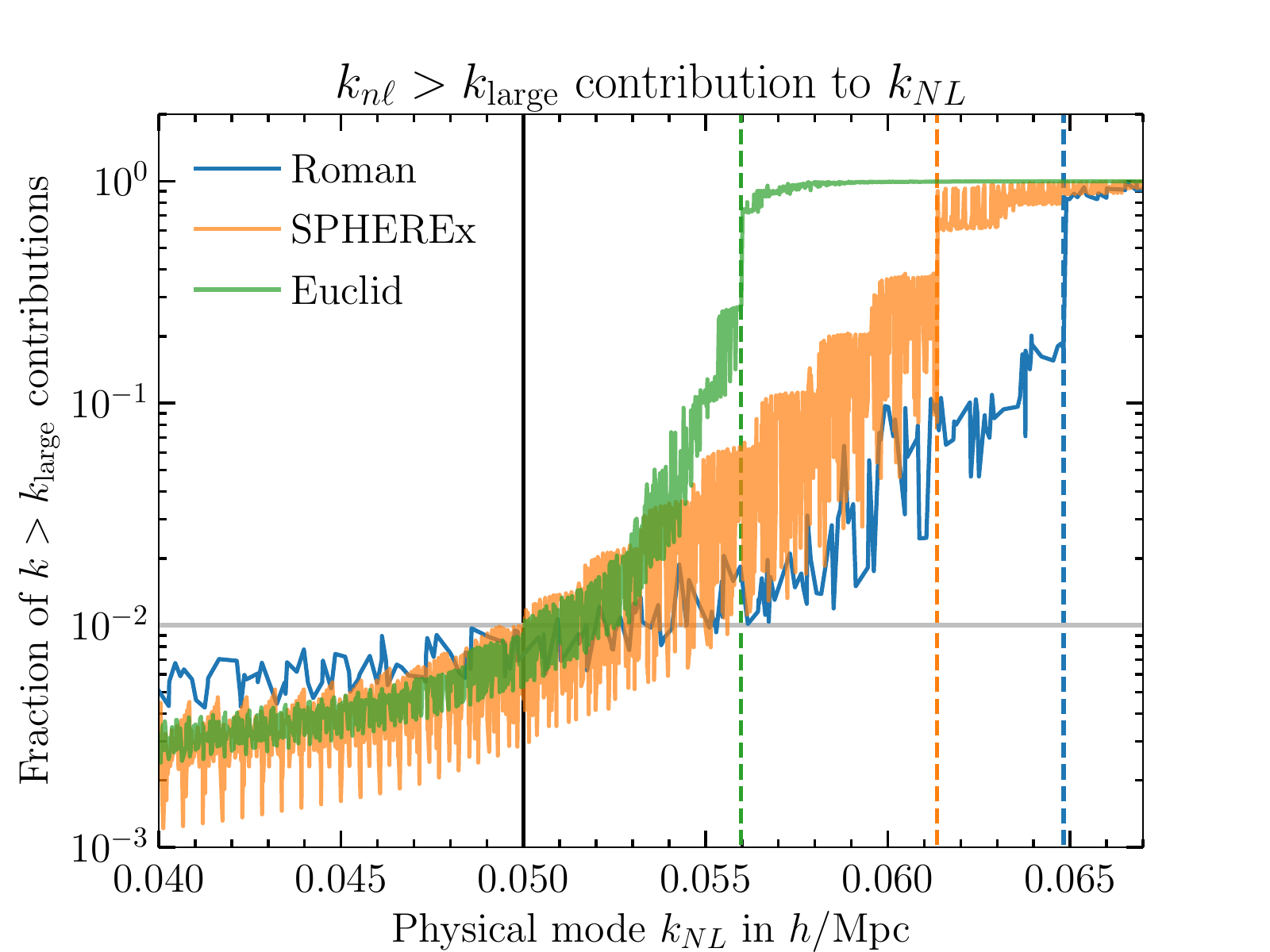}
  \caption{
    This figure shows the relative contributions from hi-$k$ observed
    (window-convolved) modes to physical modes close to
    $k_\max=\SI{0.05}{\h\per\mega\parsec}$ (vertical black line).
    The left panel shows a histogram of the contributions to a single physical
    mode. Since there are modes above $k_\max$ that are contributing in the
    deconvolution of the window (the inversion of \cref{eq:cell_mixing_matrix})
    we include all the modes up to the dashed lines at $k_\mathrm{large}$,
    which are colored by survey. The specific physical mode chosen is the one
    that has the most contribution from modes above $k_\mathrm{large}$ (the
    dashed line), and the dashed lines are chosen so that their cumulative
    contribution is less than \SI{1}{\percent}.
    The right panel shows the cumulative contribution from observed modes
    $k>k_\mathrm{large}$ for each physical mode $k_{NL}$. The vertical lines
    are the same as in the left plot, the horizontal line marks our target
    maximum contribution of \SI{1}{\percent}.
  }
  \label{fig:sfb_contributing_modes}
\end{figure*}
In this section we will test the SFB estimator presented in this paper for
several use cases. First, we will consider simplistic simulations of surveys similar to the
High-Latitude Spectroscopic Survey (HLSS) of the \emph{Nancy Grace Roman Space
Telescope} which will benefit from large radial modes, and we will consider
\emph{SPHEREx} and \emph{Euclid} for wide-angle surveys.

We bin into bandpowers by selecting
\ba
\label{eq:sfb_bin_dl}
\Delta\ell &\simeq \frac{1}{f_\mathrm{sky}}\,,
\ea
and then we round to the nearest integer. \cref{eq:sfb_Nmodes} then suggests
\ba
\label{eq:sfb_bin_dn}
\Delta n &\simeq \frac{f_\mathrm{sky}}{f_\mathrm{vol}}\,.
\ea
For all cases in this paper, this results in $\Delta n=1$.

To do the window deconvolution, it is important to ensure that all modes are
complete.
For the angular power spectrum,
\citet{Leistedt+:2013MNRAS.435.1857L,Alonso+:2019MNRAS.484.4127A} suggest
estimating up to $2\ell_\max$, and then discarding all the modes above
$\ell_\max$. \citet{Wang+:2020JCAP...10..022W} argue that (in our notation) the
sum \cref{eq:cell_mixing_matrix} only converges with the inclusion of modes
$k_{n\ell}>k_\max$, and they do numerical experiments to estimate the maximum $k$
needed.

We take a similar approach, which is demonstrated in
\cref{fig:sfb_contributing_modes}, where in the left panel the relative
contribution of window-convolved modes to a physical mode near $k_\max$
is shown. That is, we plot the relative contributions of all observed modes
$(\ell nn')$ that contribute to the physical mode $(LNN')$ using
\cref{eq:cell_mixing_matrix}. Assuming a flat power spectrum and summing the
absolute values of the coupling matrix $\mathcal{M}^{-1}$, then, allows us to
estimate the contribution from all modes above some $k_\mathrm{large}$. This is
shown in the right panel of \cref{fig:sfb_contributing_modes}.

Next, we iteratively increase $k_\mathrm{large}$ until the contribution from
$k_{n\ell}>k_\mathrm{large}$ to the most affected mode $k_{NL}\leq{}k_\max$ is less
than \SI{1}{\percent}. The results for $k_\mathrm{large}$ are the dashed
vertical lines in either panel of \cref{fig:sfb_contributing_modes}.

That is, by including all modes up to $k_\mathrm{large}>k_\max$ in the SFB
power spectrum estimation, we get reasonable confidence that all modes
$k<k_\max$ can be fully deconvolved by the inversion of
\cref{eq:cell_mixing_matrix}.

\subsection{Roman}
\begin{figure*}
  \centering
  \incgraph{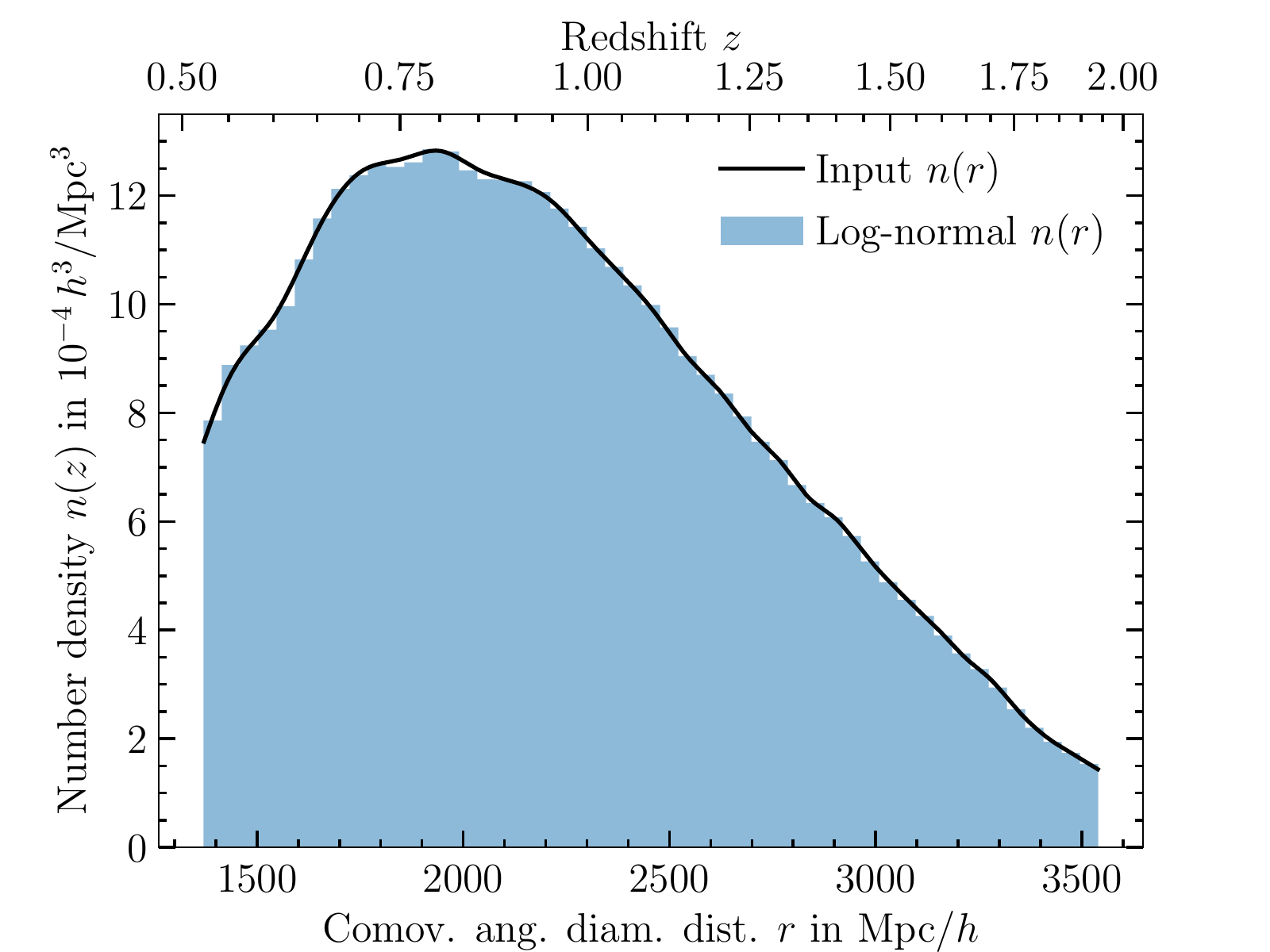}
  \incgraph{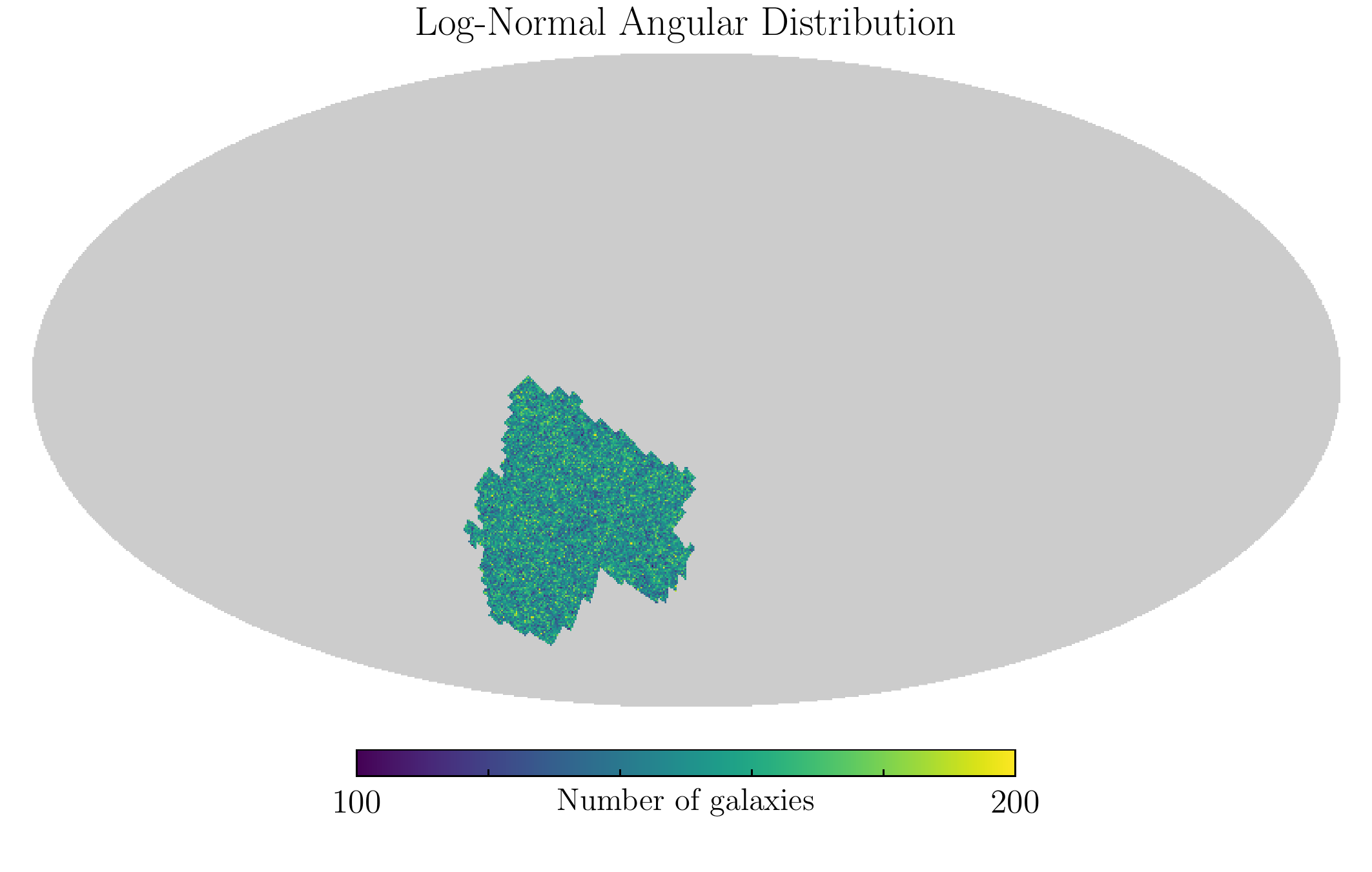}
  \caption{The plots show the approximate Roman radial selection function and
    angular mask. In addition one realization of a log-normal simulation is
    shown.
  }
  \label{fig:roman_selection_and_mask}
\end{figure*}
\begin{figure*}
  \centering
  \incgraph{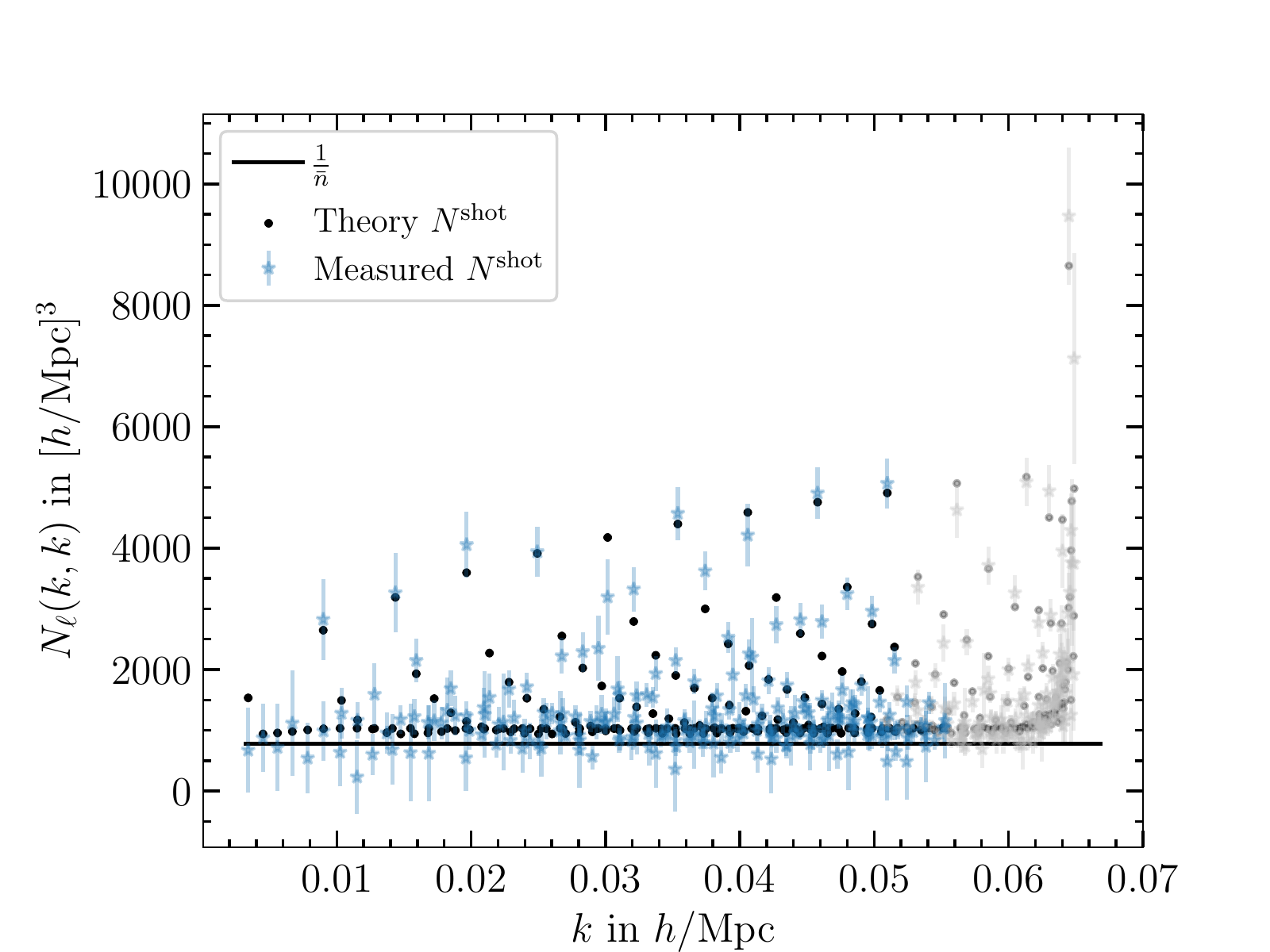}
  \incgraph{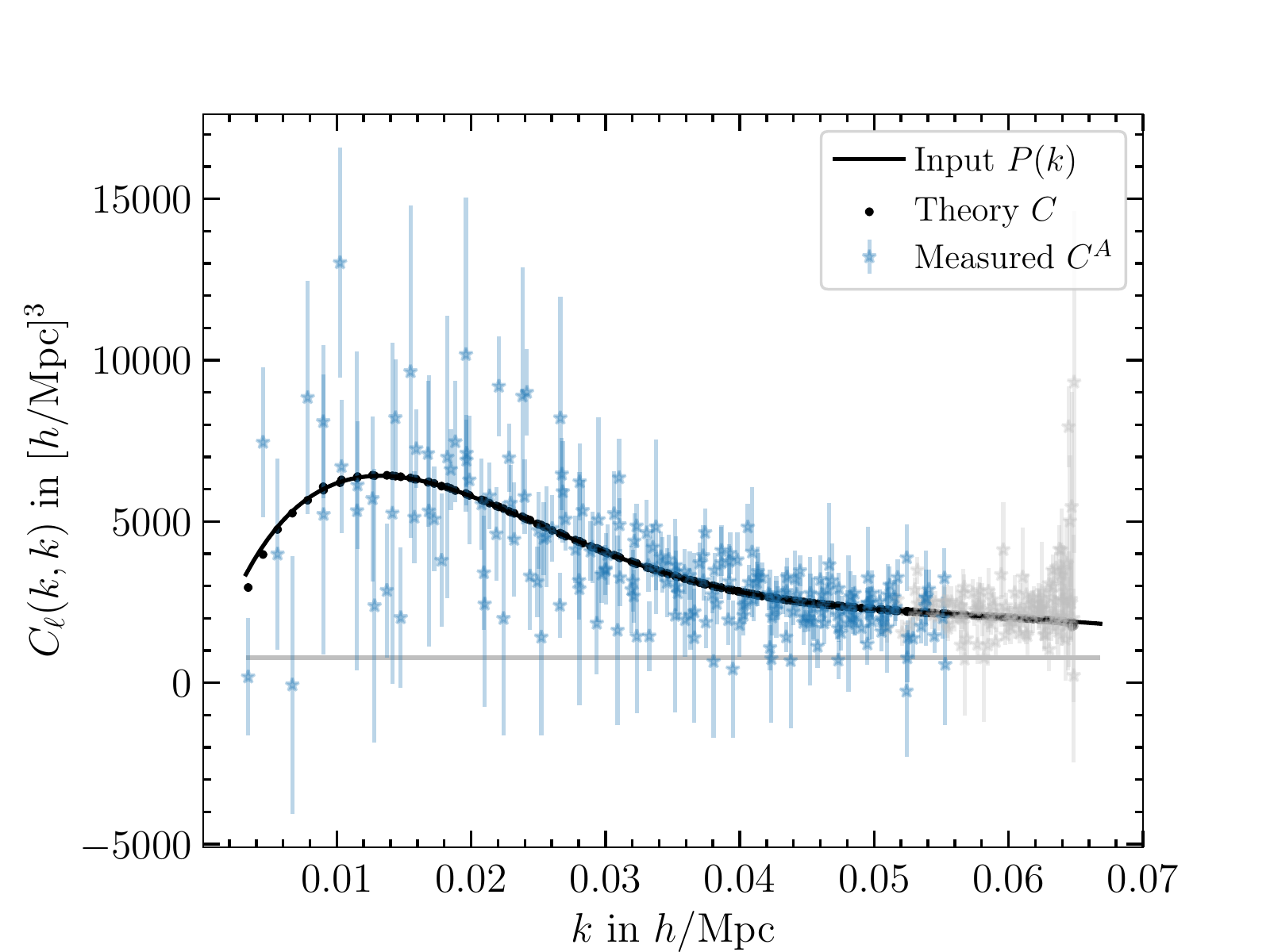}
  \incgraph{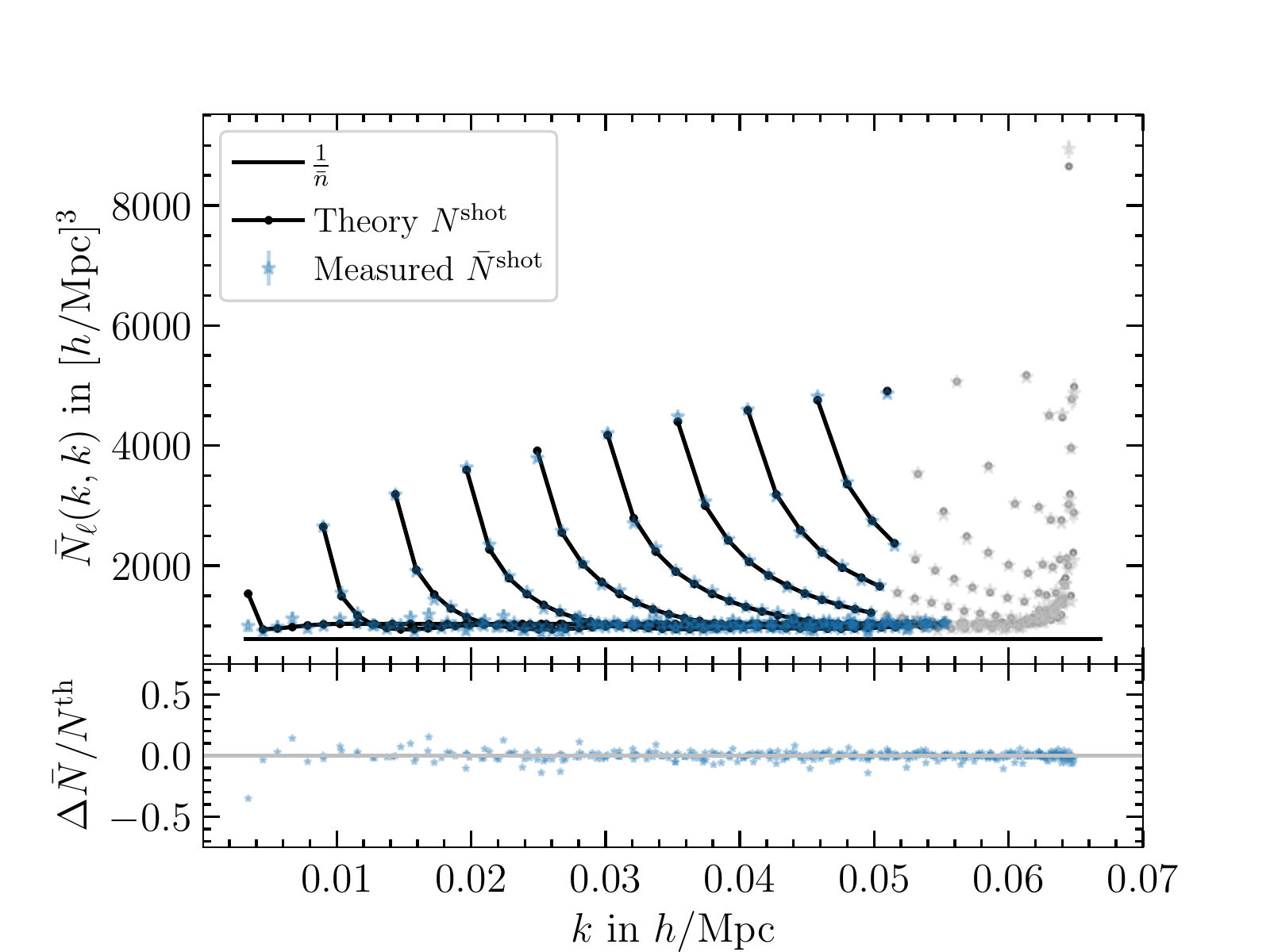}
  \incgraph{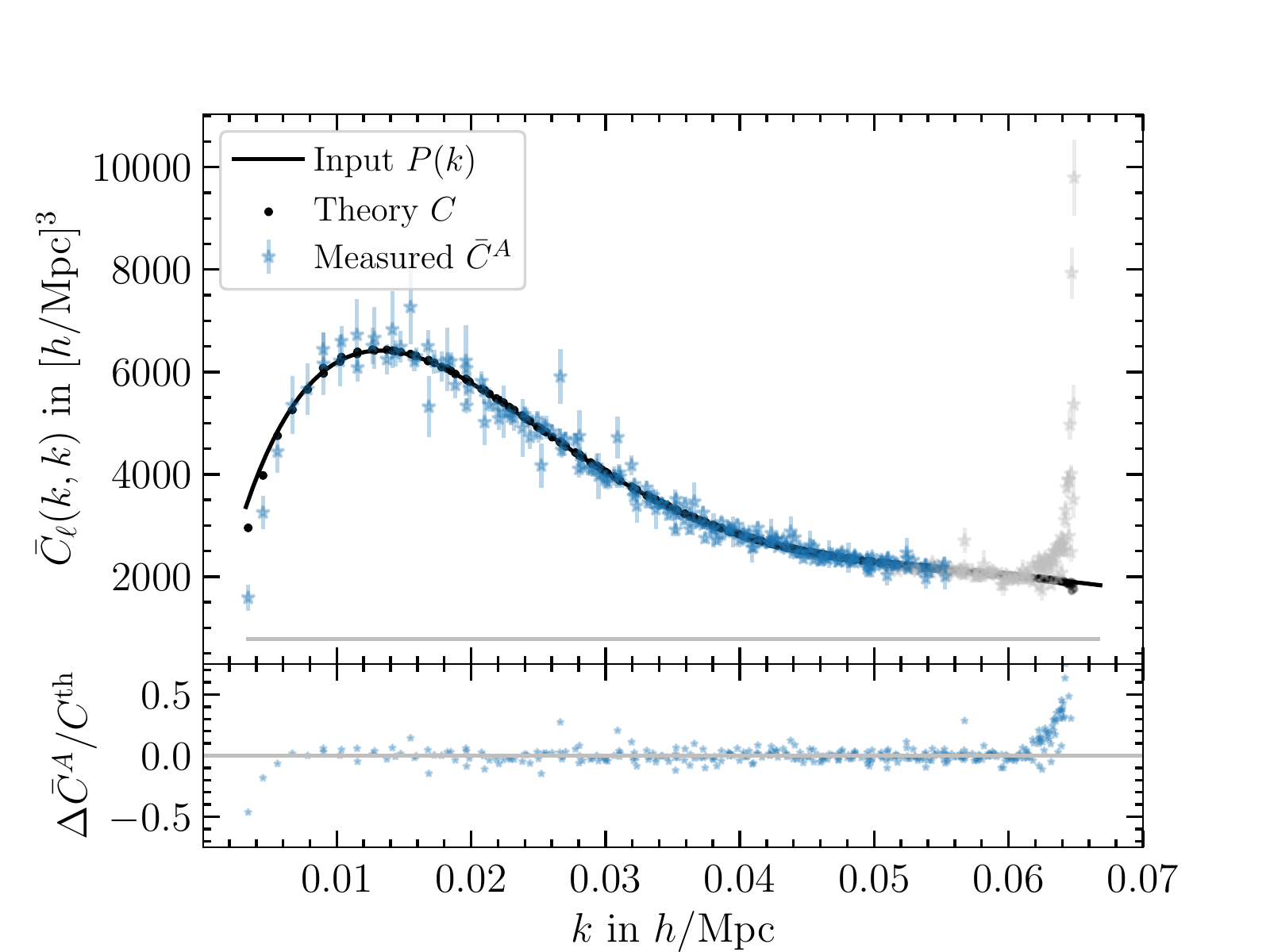}
  \caption{SFB power spectrum measurement from uniform shot-noise-only (left column) and
    log-normal (right column) simulations with Roman window function.
    The top row is for a single simulation, the bottom shows the average over
    50 simulations, and in the top panels the error bars are for a single
    simulation, while in the bottom panels the error bars are divided by
  $\sqrt{50}$. In each panel, the grey-painted modes near $k_\max$ are
    incomplete bandpower bins, and we show them for illustration only.
    On the right panels, the theory points take into account the bandpower
    binning with $\Delta\ell=18$.
    The horizontal line in each plot is $\frac{1}{\nbar}$.
    The local average effect suppresses the largest-scale mode in the
    simulations.
    For better comparison between the average measured power spectrum
    and the theory points, we show the fractional difference of each mode at
    the bottom of the lower panels.
    Note that here we only consider an H$\alpha$ sample of \emph{Roman}.
  }
  \label{fig:roman_power_spectra}
\end{figure*}
In this section we apply the SFB power spectrum estimator to a log-normal
simulation for the High-Latitude Spectroscopy Survey (HLSS) of the \emph{Nancy
Grace Roman Space Telescope} \citep{Spergel+:2015arXiv150303757S}. The notional survey
area is currently planned as \SI{\sim6}{\percent} of the sky. Our main
objective here, however, is to exploit the large radial selection that
\emph{Roman} will provide.

The grism spectroscopy of the HLSS will yield observed wavelengths
\SIrange{1}{1.93}{\micro\meter}\footnote{\url{https://roman.ipac.caltech.edu/sims/Param_db.html\#wfi_grism}}, 
which for the H$\alpha$ line at \SI{6562.8}{\angstrom} results in a redshift
range $0.523\leq z \leq 1.94$, and for the simulations we round this to the
range \SIrange{1370}{3540}{\per\h\mega\parsec}. The radial selection for our
simulation is shown in the left panel of \cref{fig:roman_selection_and_mask} \citep{Eifler+:2021MNRAS.507.1746E}.

The right panel of \cref{fig:roman_selection_and_mask} shows the HEALPix
projection of our log-normal simulation \citep{Agrawal+:2017JCAP...10..003A},
where we use an approximate binary Roman mask. The window function is then
constructed as the multiplication of the radial selection and mask, normalized
so that the maximum is unity.

Our log-normal simulation assumes a non-evolving linear power spectrum at
redshift 1.5 and linear galaxy bias $b=1.5$. The number of galaxies is
\num{\sim6.6e6}. We use a flat $\Lambda$CDM Planck cosmology.

The results of the SFB analysis are shown in \cref{fig:roman_power_spectra}.
The top left panel shows the shot noise from a simulation with vanishing power
spectrum as well as the theoretical shot noise prediction from
\cref{eq:Nshot_lnn}. The bottom left shows the average over 50
simulations.

For a simulation with signal, we show the SFB measurement from our log-normal
simulation in the top right of \cref{fig:roman_power_spectra}. Here, we have
subtracted the theoretical shot noise. The theory points and the input $P(k)$
differ due to application of \cref{eq:binning_w} to account for the bandpower
binning with $\Delta\ell=18$. The bottom right panel shows the same as an
average over our ensemble of simulations.

We get good agreement between the ensemble measurement and theory power spectra
if we restrict ourselves to the blue modes in \cref{fig:roman_power_spectra}.
There are 213 modes with a $\chi^2_\nu=1.033$ in one simulation.
The greyed-out modes cannot be fully deconvolved due to the possibility of
high-$k$ contributions, as explained at the beginning of
\cref{sec:sfb_applications}. Clearly, our estimations there were conservative,
because high-$k$ contributions vary in sign and can cancel each other.

The black curves in the lower left panel of \cref{fig:roman_power_spectra}
connect theory points with the same $\ell$. For a given $\ell$, then, the
limber ratio \cref{eq:sfb_limber_ratio} suggests that higher $k$ corresponds to
lower redshift. Since the number density tends to be higher at lower redshift,
we expect the shot noise to decrease with $k$ given constant $\ell$. This
effect is much more pronounced for \emph{SPHEREx} below.

\subsection{SPHEREx}
\begin{figure*}
  \centering
  \incgraph{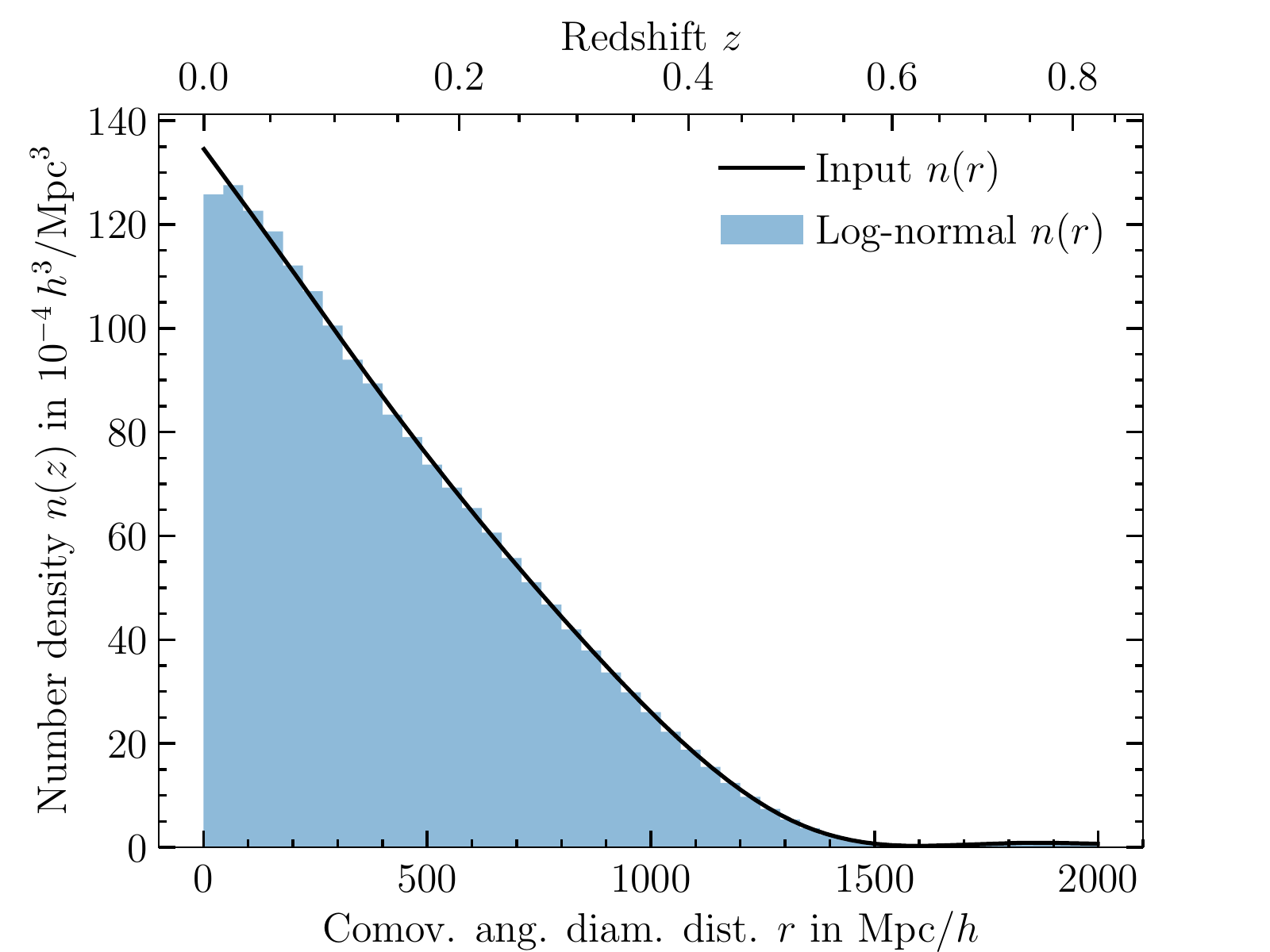}
  \incgraph{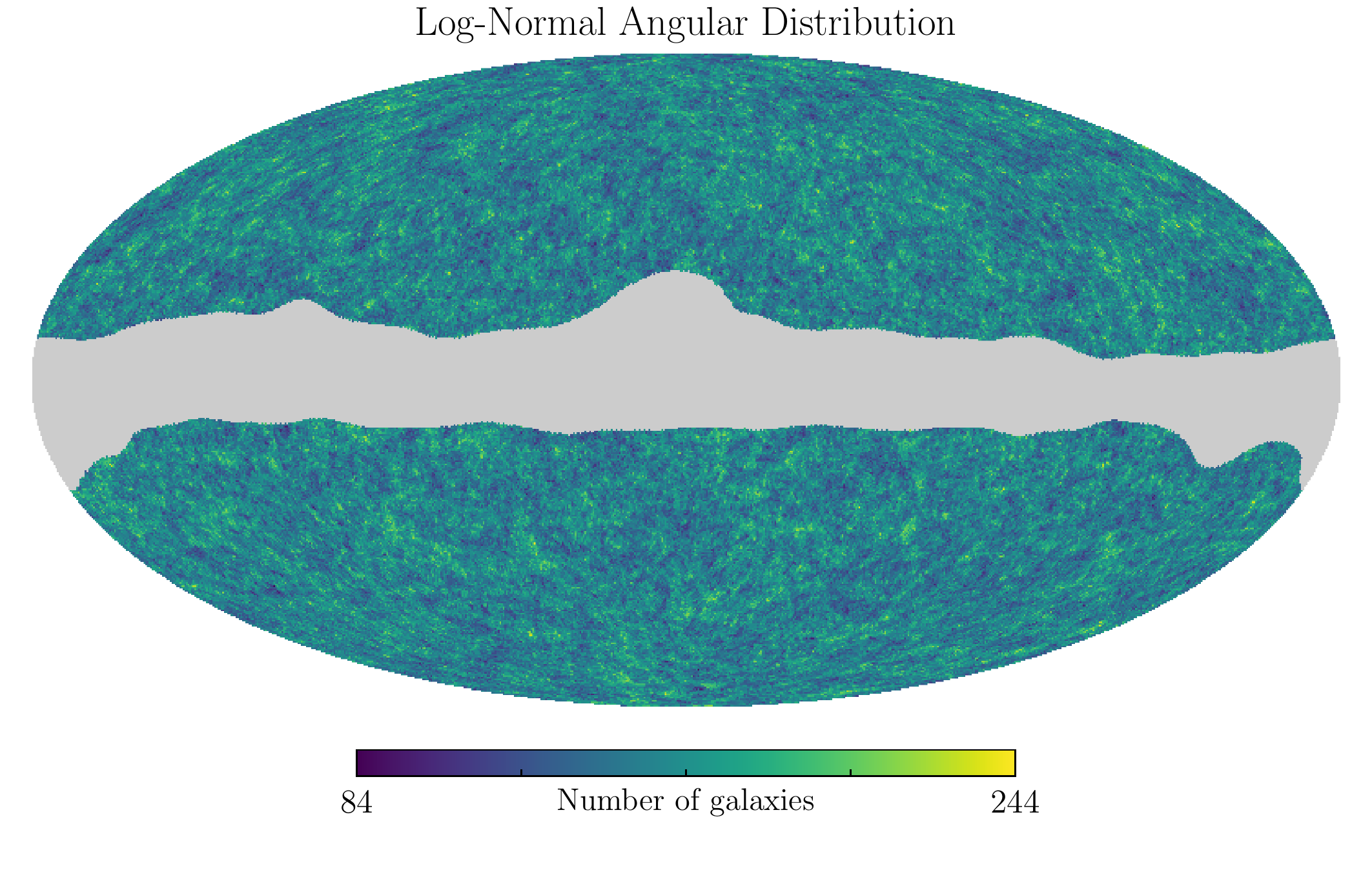}
  \caption{The plots show the approximate radial selection function and
    angular mask for our \emph{SPHEREx}-like survey.
  }
  \label{fig:spherex_selection}
\end{figure*}
\begin{figure*}
  \centering
  \incgraph{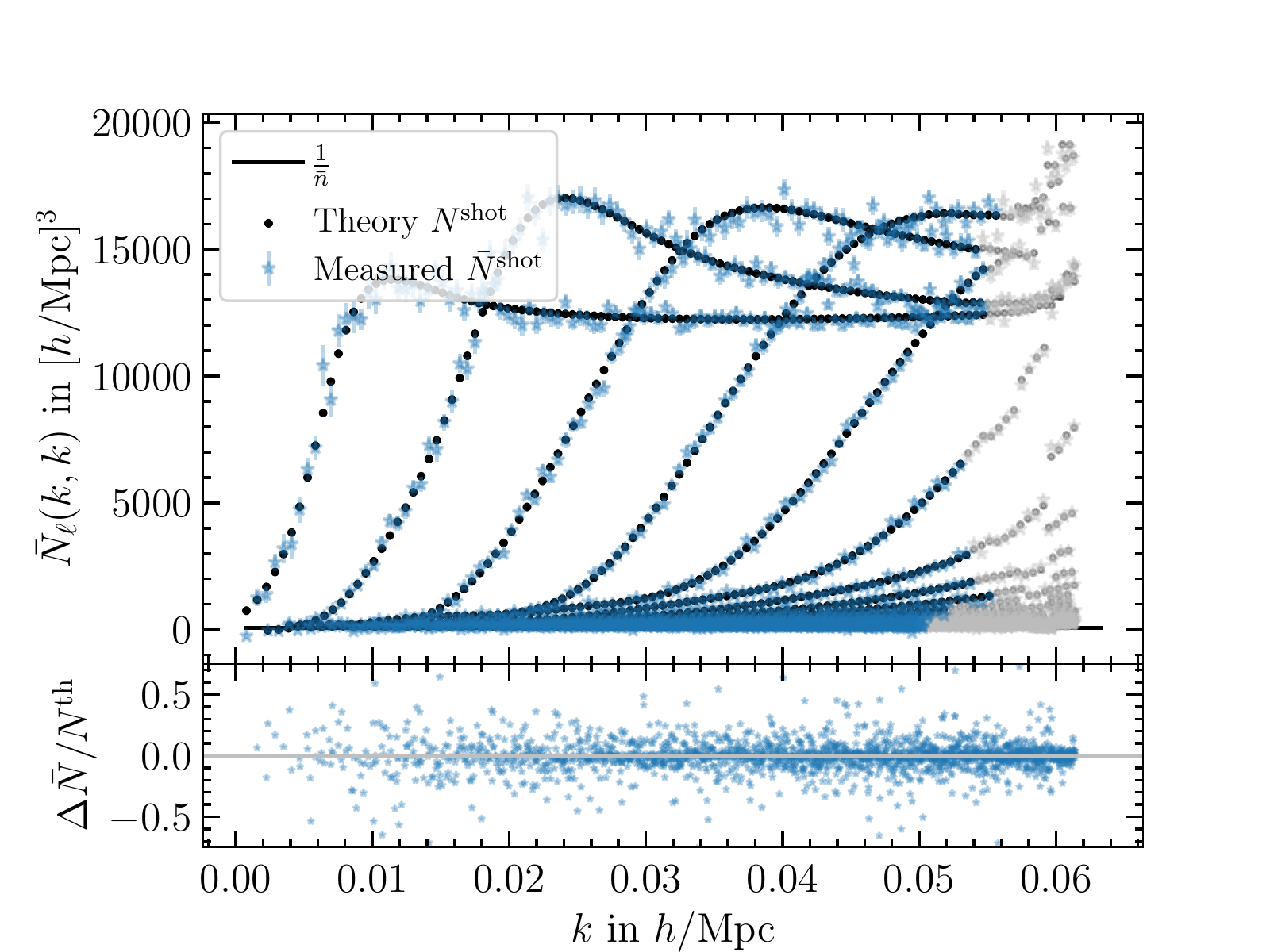}
  \incgraph{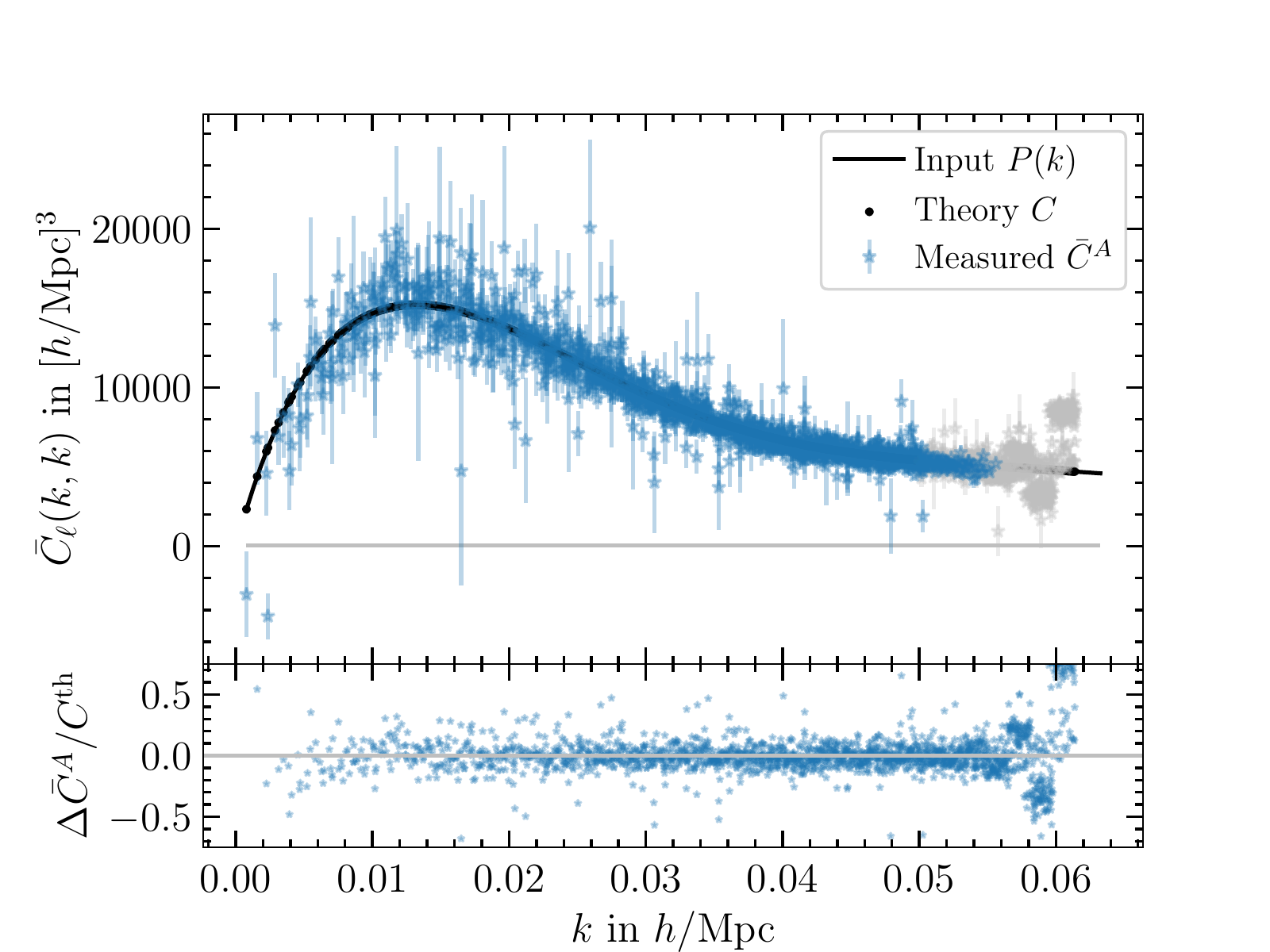}
  \caption{The plot on the left shows the shot noise and the plot on the right
    the SFB power spectrum similar to \cref{fig:roman_power_spectra}, but now
  for a full-sky mission like \emph{SPHEREx}, averaged over 50 simulations with error bars divided by $\sqrt{50}$.
    Note that while the mask and selection are realistic, we only consider a
    small part of the full \emph{SPHEREx} volume due to limitations of our
    mocks.
  }
  \label{fig:spherex_power_spectra}
\end{figure*}
In this section we aim to show the feasibility of applying our estimator to
\emph{SPHEREx}\footnote{\url{http://spherex.caltech.edu}}
\citep{Dore+:2014arXiv1412.4872D}.

For the radial selection, we use the \emph{SPHEREx} public
products\footnote{\url{https://github.com/SPHEREx/Public-products}}, and for
demonstration we limit ourselves to the range $0\leq r \leq
\SI{2000}{\per\h\mega\parsec}$, corresponding to a maximum redshift of 0.83. We
impose this limit due to our current lack of a large number of better
simulations. The radial selection is shown in \cref{fig:spherex_selection}.
\emph{SPHEREx} is able to go down to essentially $z=0$ since it is not limited
by the detection of a single emission line but measures galaxy redshifts with 102 narrow photometric bands.

For the mask we use the HFI GAL080 mask with no apodization from the \emph{Planck}
Collaboration\footnote{\url{https://irsa.ipac.caltech.edu/data/Planck/release_2/ancillary-data/previews/HFI_Mask_GalPlane-apo0_2048_R2.00/index.html}}.
This cuts out the galactic plane, as shown in \cref{fig:spherex_selection}.
The binning strategy in \cref{eq:sfb_bin_dl,eq:sfb_bin_dn} yields no binning,
or $\Delta\ell=\Delta n=1$. The number of galaxies per simulation is \num{\sim24}~million.

Since our log-normal simulations do not take into account redshift-evolution
effects, we choose a fixed effective redshift $z_\mathrm{eff}=0.5$ and galaxy
bias $b=1.5$.

The estimation of the SFB shot noise and power spectrum is shown in
\cref{fig:spherex_power_spectra}. The ``dotted curves'' of the theory shot
noise that rise quickly and then settle on an approximately constant value are
curves of constant $n$, and each dot along a curve signifies the increase of
$\ell$ by one. That is, lines of constant $\ell$ start on the first of these
curves, and then decrease rapidly, as expected from the Limber ratio
\cref{eq:sfb_limber_ratio} in conjuction with a high number density at low
redshifts.

\cref{fig:spherex_power_spectra} shows that we get good agreement between our
measured SFB power spectrum with the theory power spectrum.
There are 1345 modes with a $\chi^2_\nu=1.020$ in one simulation.
It is only in the
greyed-out modes that are not fully deconvolved that a spurious oscillatory
pattern is introduced, and measuring those modes accurately is simply a matter
of increasing $k_\mathrm{large}$.

Furthermore, since in this paper we are primarily interested in testing the SFB
estimator, \cref{fig:spherex_power_spectra} shows every mode by itself. A more
intuitive visualization of the constraining power of the survey should bin the
information from neighboring modes, and this would bring the error bars down
significantly. We leave such visualization to a future paper. For now, every
mode for itself!

\subsection{Euclid}
\begin{figure*}
  \centering
  \incgraph{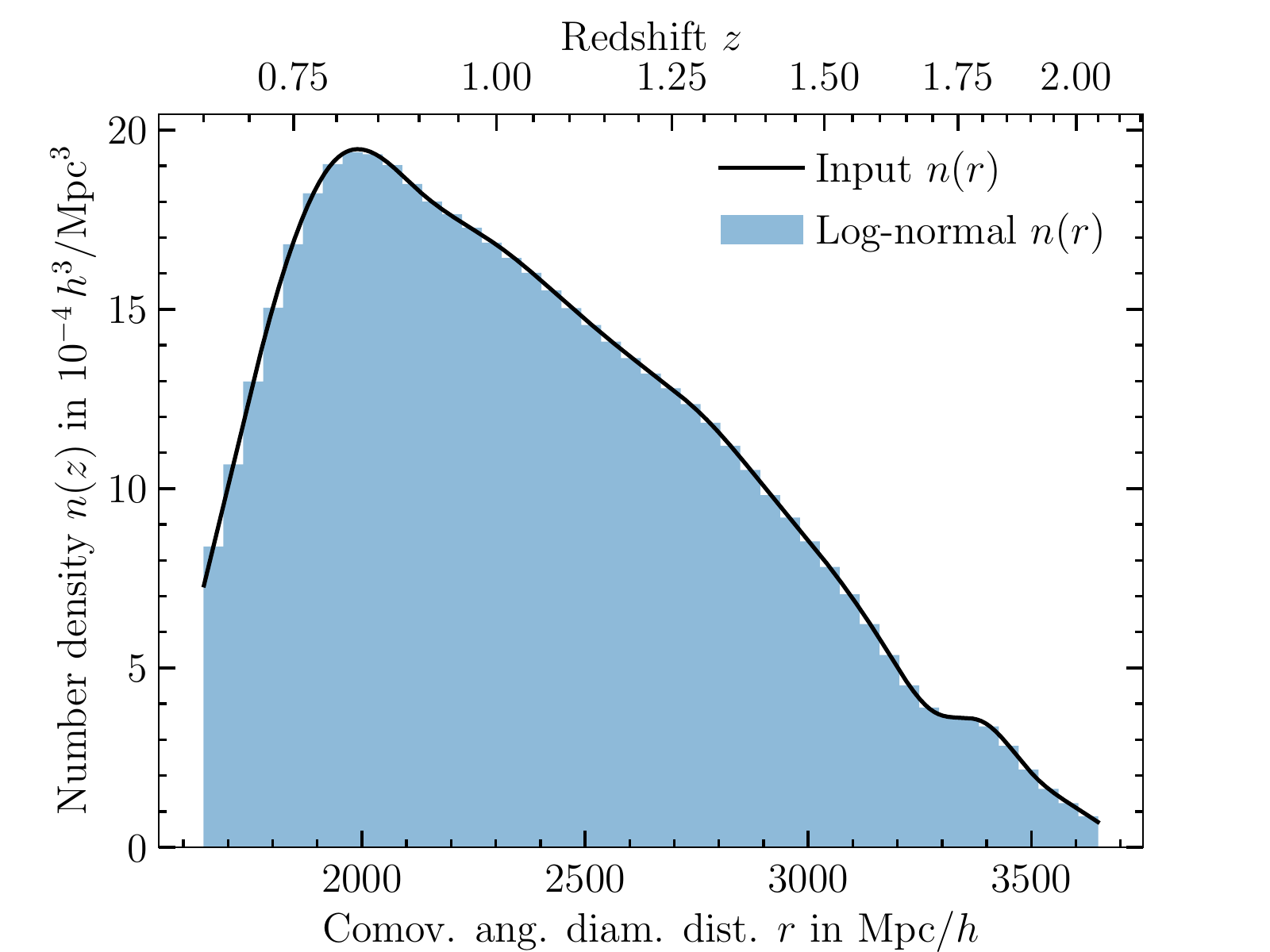}
  \incgraph{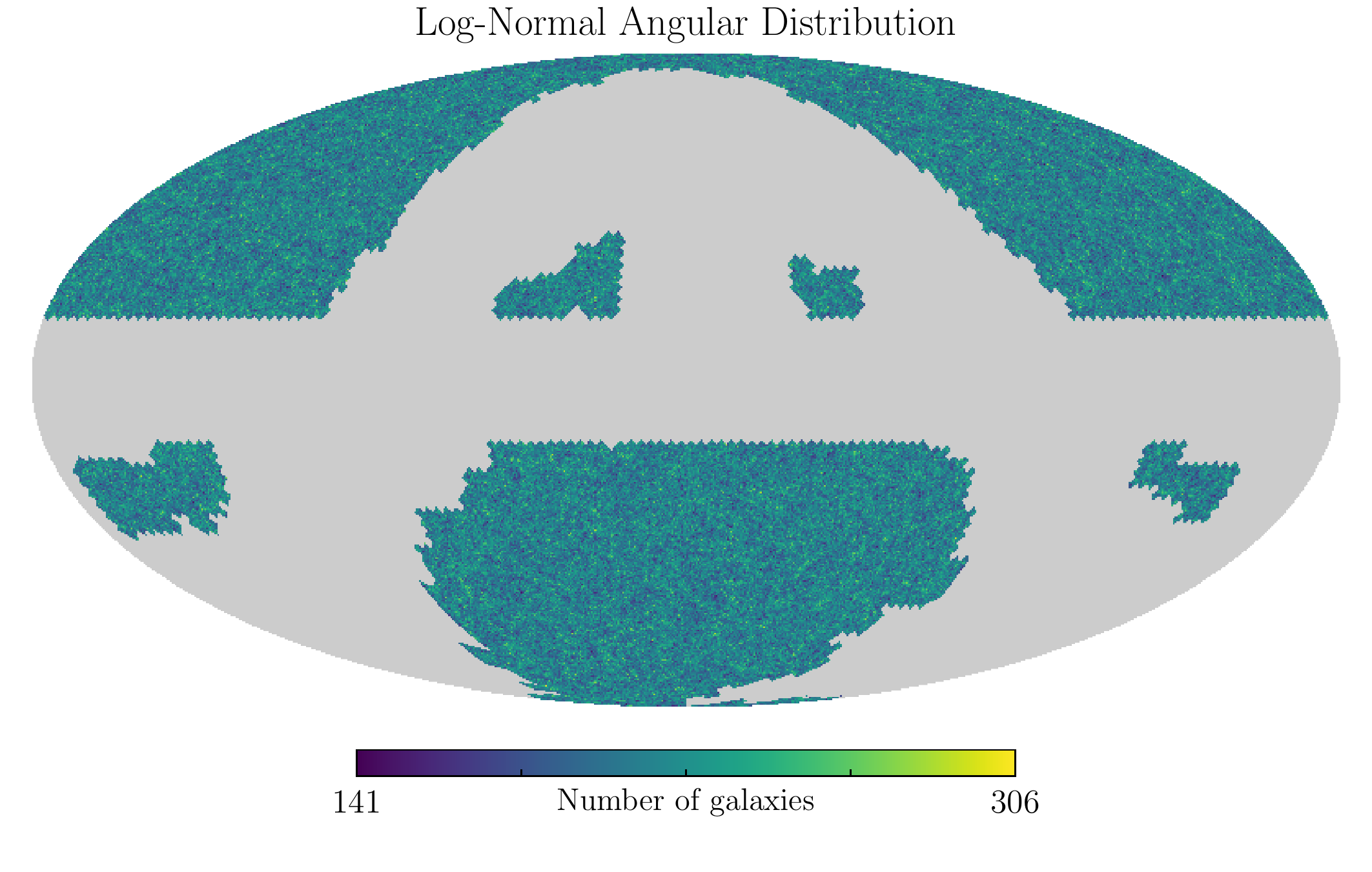}
  \caption{The plots show the approximate radial selection function and
    angular mask for our \emph{Euclid}-like survey with one
    log-normal simulation.
  }
  \label{fig:euclid_selection}
\end{figure*}
\begin{figure*}
  \centering
  \incgraph{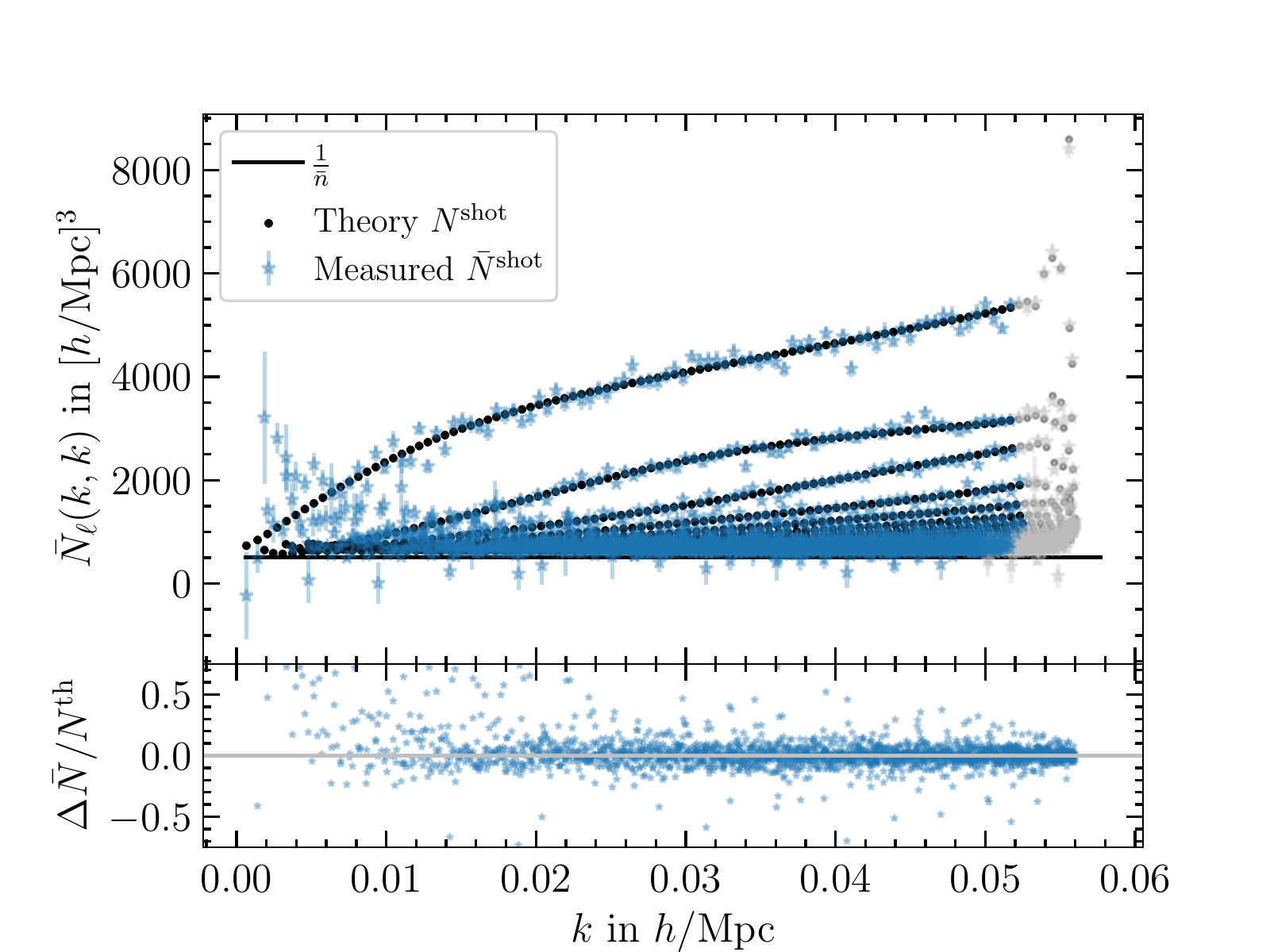}
  \incgraph{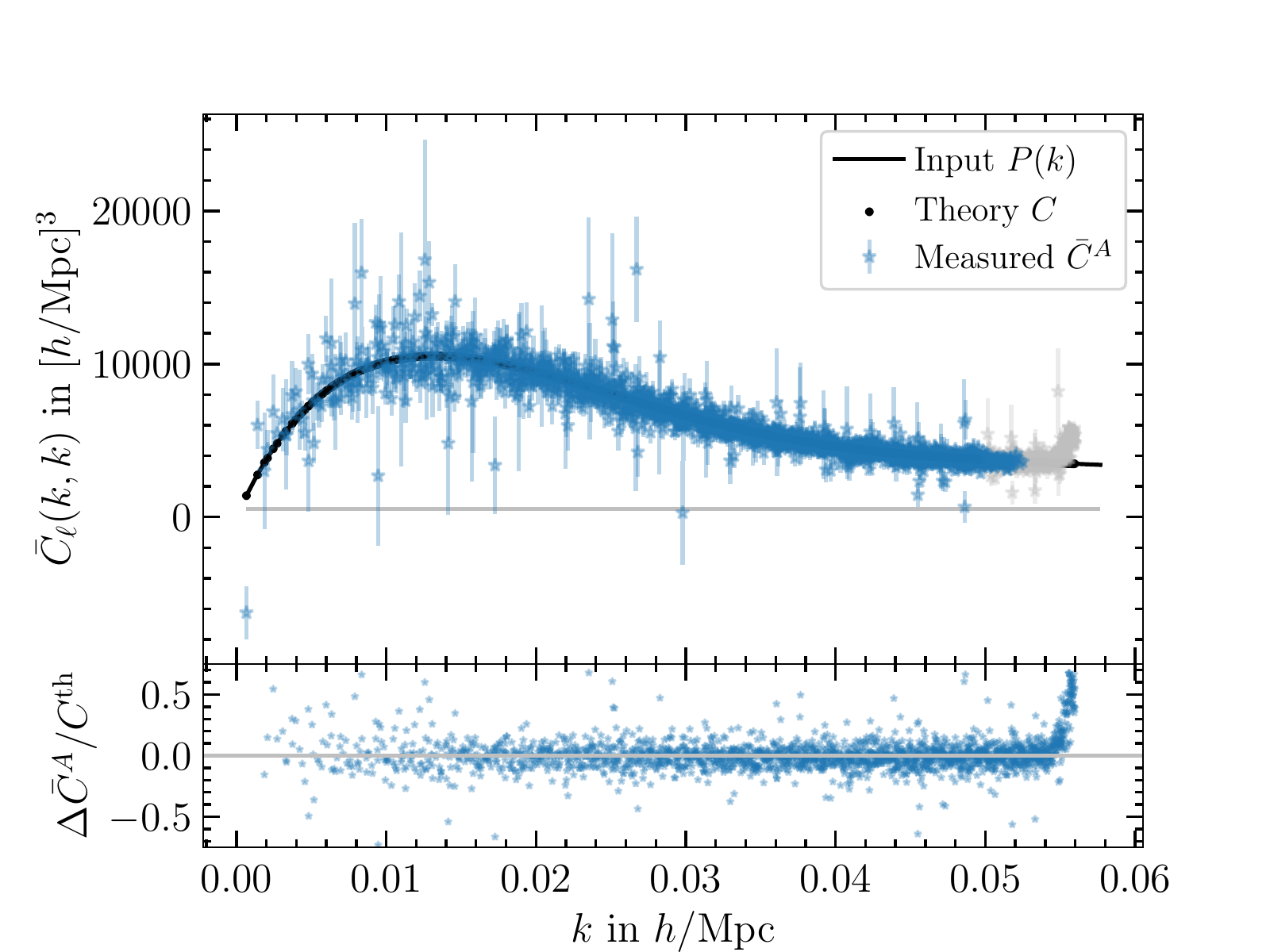}
  \caption{The plot on the left shows the shot noise and the plot on the right
    the SFB power spectrum similar to \cref{fig:roman_power_spectra}, but now
  for a full-sky mission like \emph{Euclid}, averaged over 20 simulations, and error bars divided by $\sqrt{20}$.
  }
  \label{fig:euclid_power_spectra}
\end{figure*}
As a final test for a realistic mask and selection, in this section we apply
the SFB estimator to make a forecast for the spectroscopic survey of
\emph{Euclid}, which is an all-sky mission that covers approximately
\SI{40}{\percent} of the sky. For the number density, we adopt the reference
case given in \citet{Amendola+:2018LRR....21....2A}. We use the radial range
$1645 \leq \frac{r}{\si{\per\h\mega\parsec}} \leq 3650$, shown in
\cref{fig:euclid_selection}. Also shown in the figure is the mask that cuts the
galactic and ecliptic planes \citep{euclid_mask}.
The number of galaxies per simulation is \num{\sim68}~million.

As our primary purpose is to show the applicability of the SFB
estimator, we have not updated our simulations with the parameters in
\citet{EuclidCollaboration:2020A&A...642A.191E}.

Our simulation results are shown in \cref{fig:euclid_power_spectra}, for both
shot noise-only and with a power spectrum signal with and galaxy bias $b=1.6$.

As for the other surveys, once we ignore the modes that are not fully
deconvolved, we get good agreement for both cases,
with $\chi^2_\nu=0.983$ for 1744 modes of one simulation.

\section{Conclusion}
\label{sec:sfb_conclusion}
In this paper we present a new pseudo-SFB power spectrum estimator, \codename.
The estimator analytically accounts for shot noise, mask, and selection
effects. We also investigate the impact of the local average effect and the
covariance matrix.

\codename{} works by performing the radial transform before the angular
transform, similar to \citet{Leistedt+:2012A&A...540A..60L}. In the radial
direction the galaxies are treated as point-particles so that no radial pixel
window needs correction. The angular transform is performed using HEALPix
\citep{Gorski+:2005ApJ...622..759G,Zonca+:2019JOSS....4.1298Z}.

Furthermore, we derive the radial eigenmodes with potential boundary conditions
at $r_\min$ and $r_\max$, as suggested by \citet{Samushia:2019arXiv190605866S}.
The boundary at $r_\min\neq0$ eliminates the need for bandpower-binning in the
radial direction, the boundary at $r_\max$ discretizes the measured modes
$k_{n\ell}$ for integer $n$ and $\ell$.

We demonstrate that \codename{} will be able to analyze all large-scale modes
of upcoming wide and deep galaxy surveys such as \emph{Roman}, \emph{SPHEREx},
and \emph{Euclid}.

We also review the SFB power spectrum theory and provide intuition using the
Limber approximation. Notably, redshift-dependence of the power spectrum and
bias factors primarily enter as the interplay between $k$ and $\ell$ modes,
such that $r\sim(\ell+0.5)/k$  is an approximation for the angular diameter
distance. We leave a more precise and detailed analysis for the projection of the 3D power
spectrum to SFB space and the connection with cosmological parameters for a
future paper.

We also leave for a future paper the extension of the estimator to
cross-correlations between samples with differing selection functions.

Since the SFB power spectrum is uniquely suited for all-sky surveys, we expect
a particularly intriguing application of \codename{} will be intensity mapping
at high redshift, and we look forward to this possibility.

For surveys covering a compact area on the sky we can introduce boundary
conditions to the angular basis functions as well, as pointed out by
\citet{Samushia:2019arXiv190605866S}. This would significantly reduce the size
of the computational problem in these cases.

\acknowledgments
\textcopyright 2020. All rights reserved.
The authors thank Katarina Markovi\v{c}, Christopher Hirata, Lado
Samushia, Chen Heinrich, and the Nancy Grace Roman Telescope \textit{Cosmology with the High Latitude Survey} Science Investigation Team (SIT) for discussions and providing
selection functions and masks.
We also thank the anonymous referee for comments that improved the paper.
Part of this work was done at Jet Propulsion Laboratory, California Institute of Technology, 
under a contract with the National Aeronautics and Space Administration. This work was 
supported by NASA grant 15-WFIRST15-0008 \textit{Cosmology with the High Latitude Survey} Roman Science Investigation Team (SIT).
Henry S. G. Gebhardt's research was supported by an appointment to the NASA Postdoctoral Program at the Jet Propulsion Laboratory, administered by Universities Space Research Association under contract with NASA.

\bibliography{../references}

\appendix

\section{Useful formulae}
\label{app:sfb_useful_formulae}
For any function $f(\vk)$
\ba
&\int k^2\dd{k}\,\dd^2\khat\,\delta^D(\vk-\vk')\,f(\vk)
\vs
&= \int\dd{k}\,\delta^D(k-k')\int\dd^2\khat\,\delta^D(\khat-\khat')\,f(\vk)\,.
\ea
Therefore,
\ba
\label{eq:dirac3D}
\delta^D(\vk-\vk')
&= k^{-2} \, \delta^D(k-k')\,\delta^D(\khat-\khat')\,.
\ea
Furthermore,
\ba
\frac{1}{r}\,\delta^D\!\(\frac{1}{r} - \frac{1}{r_0}\)
&= r\,\delta^D\!\(r - r_0\)\,.
\ea

Spherical Bessel functions and spherical harmonics satisfy orthogonality
relations
\ba
\label{eq:jljlDelta}
\delta^D(k-k')
&= \frac{2kk'}{\pi}\int_0^\infty\dd{r}\,r^2\,j_\ell(kr)\,j_\ell(k'r)\,, \\
\label{eq:YlmYlmDelta}
\delta^K_{\ell\ell'}\delta^K_{mm'}
&= \int\dd{\Omega}_{\rhat}\,Y_{\ell m}(\rhat)\,Y^*_{\ell'm'}(\rhat)\,.
\ea
Spherical harmonics can be expressed in terms of a complex exponential and real
associated Legendre functions $\mathrm{P}_\ell^m(x)$ as
\ba
\label{eq:spherical_harmonics}
Y_{\ell m}(\rhat)
&=
e^{im\phi}
\(\frac{(\ell - m)! (2\ell+1)}{4\pi\(\ell+m\)!}\)^\frac12
\mathrm{P}_\ell^m\!\(\cos\theta\).
\ea
The completeness relation is
\ba
\label{eq:Ylm_completeness}
\sum_{\ell m} Y_{\ell m}(\rhat)\,Y^*_{\ell m}(\rhat')
&= \delta^D(\rhat - \rhat')\,.
\ea
Rayleigh's formula decomposes the plane waves into spherical Bessels and
spherical harmonics,
\ba
\label{eq:rayleigh}
e^{i\vq\cdot\vr} &=
4\pi\sum_{\ell',m'} i^{\ell'} j_{\ell'}(qr)\,
Y^*_{\ell'm'}(\qhat)\,Y_{\ell'm'}(\rhat)\,.
\ea
Legendre polynomials can be expressed as a sum over spherical harmonics as
\ba
\label{eq:legendre_spherical_harmonics}
\mathcal{P}_\ell(\khat\cdot\rhat)
&= \frac{4\pi}{2\ell+1}\sum_m Y_{\ell m}(\khat)\,Y_{\ell m}^*(\rhat)\,.
\ea
Flipping the sign of the component angular momentum or the direction of the
argument to spherical harmonics gives
\ba
\label{eq:Ylm_conjugate}
Y^*_{\ell m}(\rhat)
&= (-1)^m Y_{\ell,-m}(\rhat)
\,,\\
\label{eq:Ylm_parity}
Y_{\ell m}(-\rhat)
&= (-1)^\ell Y_{\ell,m}(\rhat)\,.
\ea
The Gaunt factor is
\ba
\label{eq:gaunt_factor}
\mathcal{G}^{\ell L L_1}_{mMM_1}
&=
\int\dd^2\rhat
\,Y_{\ell m}(\rhat)
\,Y_{LM}(\rhat)
\,Y_{L_1M_1}(\rhat)\,,
\ea
and it can be expressed in terms of Wigner-$3j$ symbols,
\ba
\mathcal{G}^{\ell L L_1}_{mMM_1}
&=
\(\frac{(2\ell+1)(2L+1)(2L_1+1)}{4\pi}\)^\frac12
\begin{pmatrix}
  \ell & L & L_1 \\
  0 & 0 & 0
\end{pmatrix}
\vs&\quad\times
\begin{pmatrix}
  \ell & L & L_1 \\
  m & M & M_1
\end{pmatrix}\,.
\label{eq:gaunt_3j}
\ea
The Wigner $3j$ symbols obey an orthogonality relation
\ba
\sum_{mM}
\begin{pmatrix}
  \ell & L & L_1 \\
  m & M & M_1
\end{pmatrix}
\begin{pmatrix}
  \ell & L & L_2 \\
  m & M & M_2
\end{pmatrix}
&=
\frac{\delta^K_{L_1L_2}
\delta^K_{M_1M_2}
\delta^T(\ell,L,L_1)}
{2L_1+1}\,,
\label{eq:gaunt_orthogonality}
\ea
where $\delta^T(\ell,L,L_1)$ enforces the triangle relation that is also obeyed
by the 3$j$-symbols. That is, the Gaunt factor is only nonzero when
\ba
\label{eq:triangle_mmm}
m + M + M_1 = 0\,,
\\
\label{eq:triangle_lll}
|\ell - L| \leq L_1 \leq \ell + L\,.
\ea
Assuming the triangle condition is satisfied, for even $J=\ell+L+L_1$ we have
\ba
\begin{pmatrix}
  \ell & L & L_1 \\
  0 & 0 & 0
\end{pmatrix}
&=
(-1)^{\frac12 J}
\(\frac{(J-2\ell)!(J-2L)!(J-2L_1)!}{(J+1)!}\)^\frac12
\vs&\quad\times
\frac{\(\frac12J\)!}{\(\frac12J-\ell\)!\(\frac12J-L\)!\(\frac12J-L_1\)!}
\,,
\ea
for odd $J=\ell+L+L_1$, those $3j$'s vanish when $m=M=M_1=0$.

\section{The Laplacian in an expanding universe}
\label{app:metric}
We use the SFB decomposition because that correlates radial and angular modes
of the same scale. Here we show that the Laplacian in a flat expanding universe
takes the form \cref{eq:laplacian_spherical}.

The flat Robertson-Walker metric is
\ba
ds^2 &= -dt^2 + a^2\left[dr^2 + r^2 d\theta^2 + r^2 \sin^2\!\theta\,d\phi^2\right]\,,
\ea
where $r$ is the comoving coordinate, and the metric has the non-zero
Christoffel symbols
\ba
\Gamma^0_{11} &= a^2 H  \,,&
\Gamma^0_{22} &= a^2r^2 H \,,&
\Gamma^0_{33} &= a^2r^2\sin^2\!\theta\,H \,,
\displaybreak[0]\vs
\Gamma^1_{01} &= \Gamma^1_{10} = H  \,,&
\Gamma^1_{22} &= -r \,,&
\Gamma^1_{33} &= -r\sin^2\!\theta \,,
\displaybreak[0]\vs
\Gamma^2_{02} &= \Gamma^2_{20} = H  \,,&
\Gamma^2_{12} &= \Gamma^2_{21} = \frac{1}{r} \,,&
\Gamma^2_{33} &= -\cos\theta\sin\theta \,,
\displaybreak[0]\vs
\Gamma^3_{03} &= \Gamma^3_{30} = H  \,,&
\Gamma^3_{13} &= \Gamma^3_{31} = \frac{1}{r} \,,&
\Gamma^3_{23} &= \Gamma^3_{32} = \frac{\cos\theta}{\sin\theta} \,. \nonumber
\ea
Therefore,
\ba
g^{\mu\nu}\Gamma^0_{\mu\nu} &= 3H
\,,\displaybreak[0]\\
g^{\mu\nu}\Gamma^1_{\mu\nu} &= -2\,a^{-2}\,r^{-1}
\,,\displaybreak[0]\\
g^{\mu\nu}\Gamma^2_{\mu\nu} &= -a^{-2}\,r^{-2}\,\frac{\cos\theta}{\sin\theta}
\,,\displaybreak[0]\\
g^{\mu\nu}\Gamma^3_{\mu\nu} &= 0\,.
\ea
The d'Alembertian operator for a scalar $f$ is
\ba
\Box f
&= g^{\mu\nu}\,f_{:\mu\nu}
\vs
&= g^{\mu\nu}\(\partial_\mu\partial_\nu - \Gamma^\sigma_{\mu\nu} \partial_\sigma\) f
\vs
&= \bigg[
-\partial_0^2
+a^{-2}\,\partial_1^2
+a^{-2}r^{-2}\,\partial_2^2
+a^{-2}r^{-2}\sin^{-2}\!\theta\,\partial_3^2
\vs&\quad
- 3H\, \partial_0
+ 2\,a^{-2}r^{-1}\, \partial_1
+ a^{-2}r^{-2}\,\frac{\cos\theta}{\sin\theta}\, \partial_2
\bigg] f
\vs
&= \bigg[
-\partial_0^2
- 3H\, \partial_0
+a^{-2}\bigg(
  \partial_1^2
+r^{-2}\,\partial_2^2
+r^{-2}\sin^{-2}\!\theta\,\partial_3^2
\vs&\quad
+ 2\,r^{-1}\, \partial_1
+ r^{-2}\,\frac{\cos\theta}{\sin\theta}\, \partial_2
\bigg)
\bigg] f\,.
\ea
Identifying the term in parentheses as given by \cref{eq:laplacian_spherical}, we get
\ba
\label{eq:dAlembertian}
\Box f
&= \left[ -\partial_0^2 - 3H\, \partial_0 +a^{-2}\nabla^2 \right] f
\ea
in a flat expanding universe.
We can also write \cref{eq:dAlembertian} as
\ba
\Box f
&= \left[ -a^{-3}\partial_t\!\(a^3\,\partial_t\) + a^{-2}\nabla^2 \right] f\,.
\ea
The eigenfunctions to the d'Alembertian are, therefore, separable.
If we write an eigenfunction
\ba
f(t,r,\rhat)=p(t)\,g(r)\,h(\rhat)
\ea
with
\ba
\nabla^2[g(r)\,h(\rhat)]=-k^2\,g(r)\,h(\rhat)\,,
\ea
then
\ba
0
&=
a^{-3}\,\partial_t\!\(a^3\,\partial_t p\) + \left[a^{-2}k^2 - \lambda^2\right]p\,,
\ea
where $-\lambda^2$ is the eigenvalue of the d'Alembertian. Since $ar$ is the
angular diameter distance and $r$ is the comoving distance (also comoving
angular diameter distance), we can call $k$ a comoving mode.

\section{Limber's Approximation}
\label{sec:sfb_limber}
\begin{figure*}
  \centering
  \incgraph{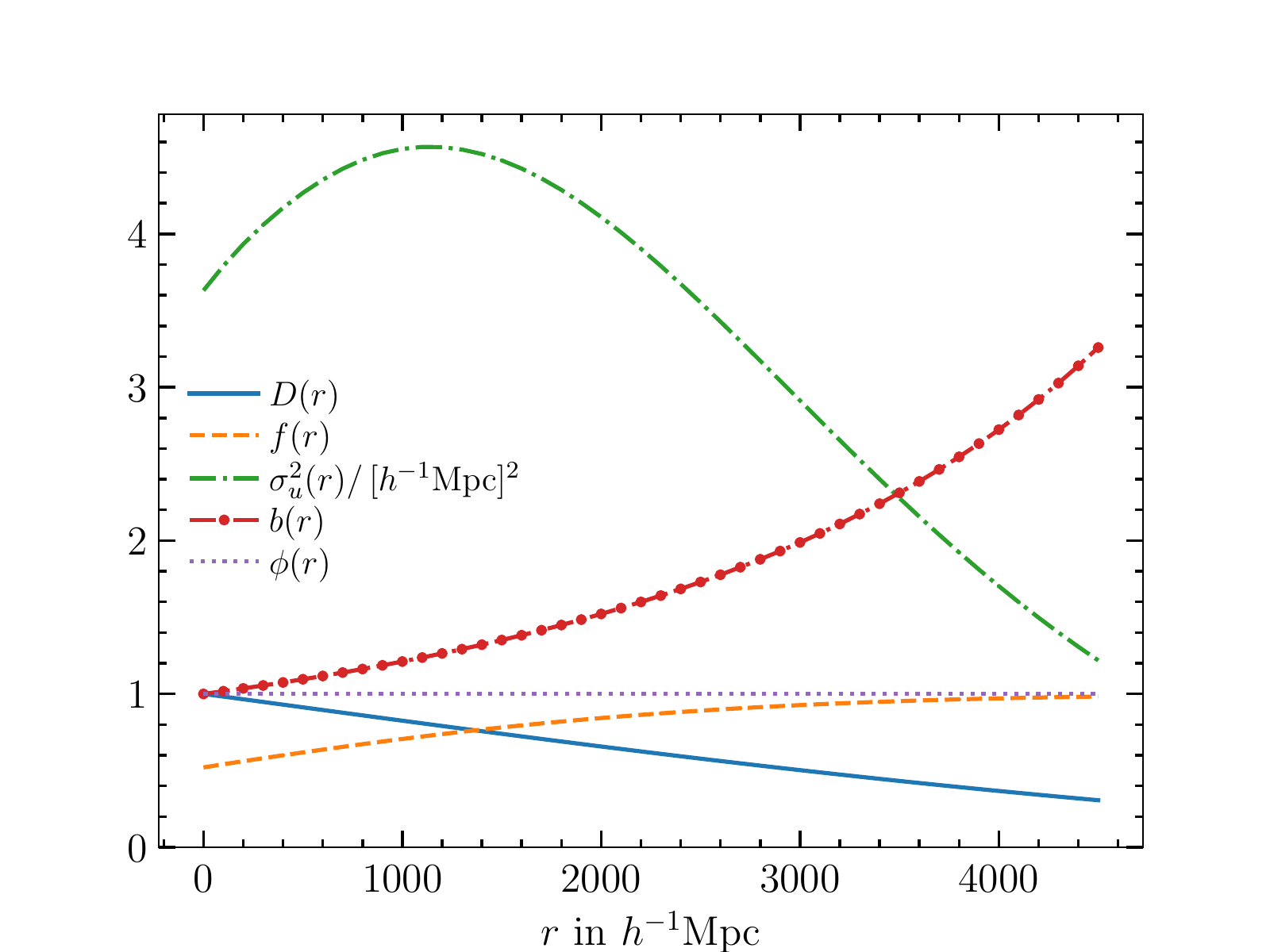}
  \incgraph{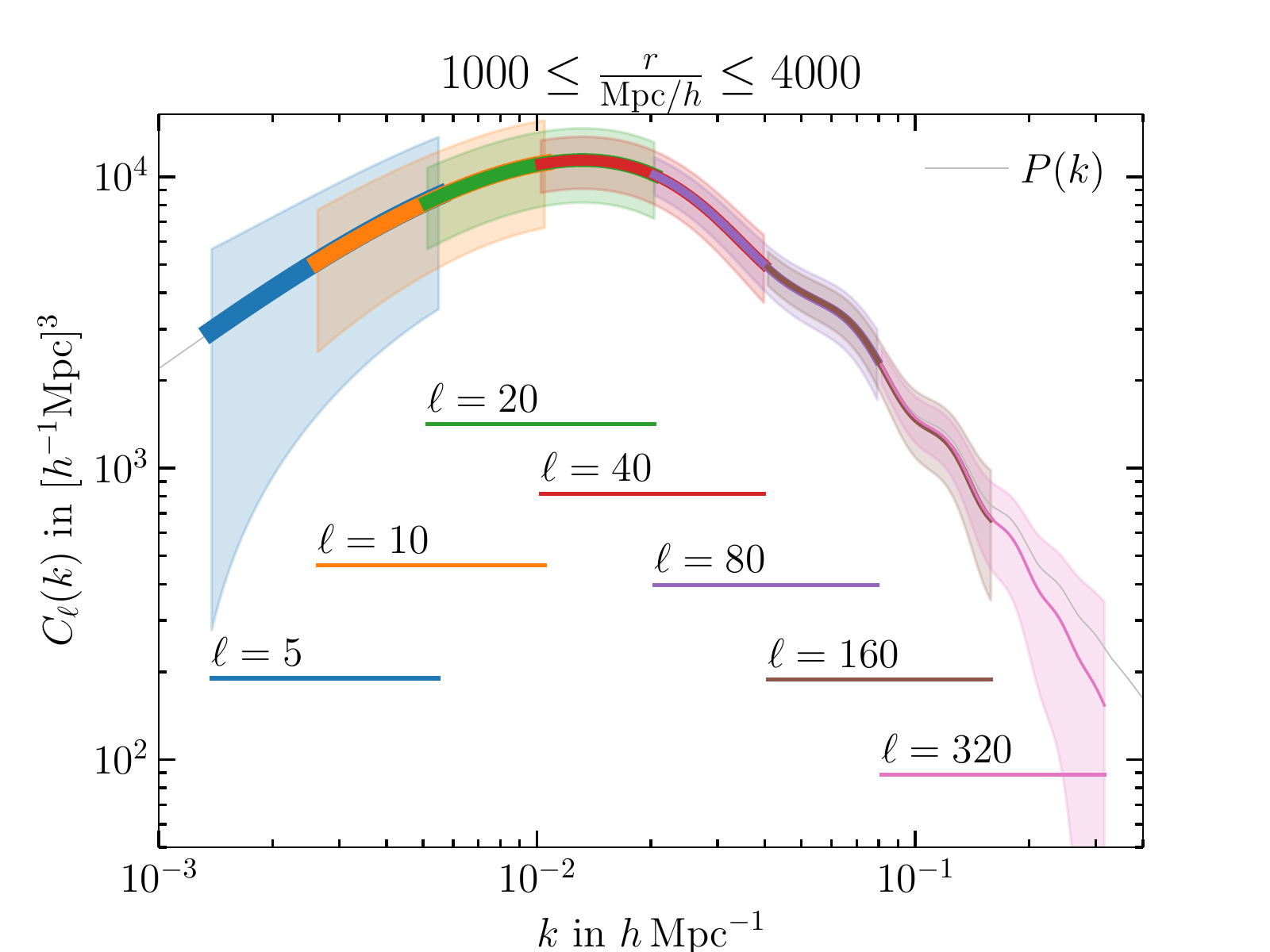}
  \incgraph{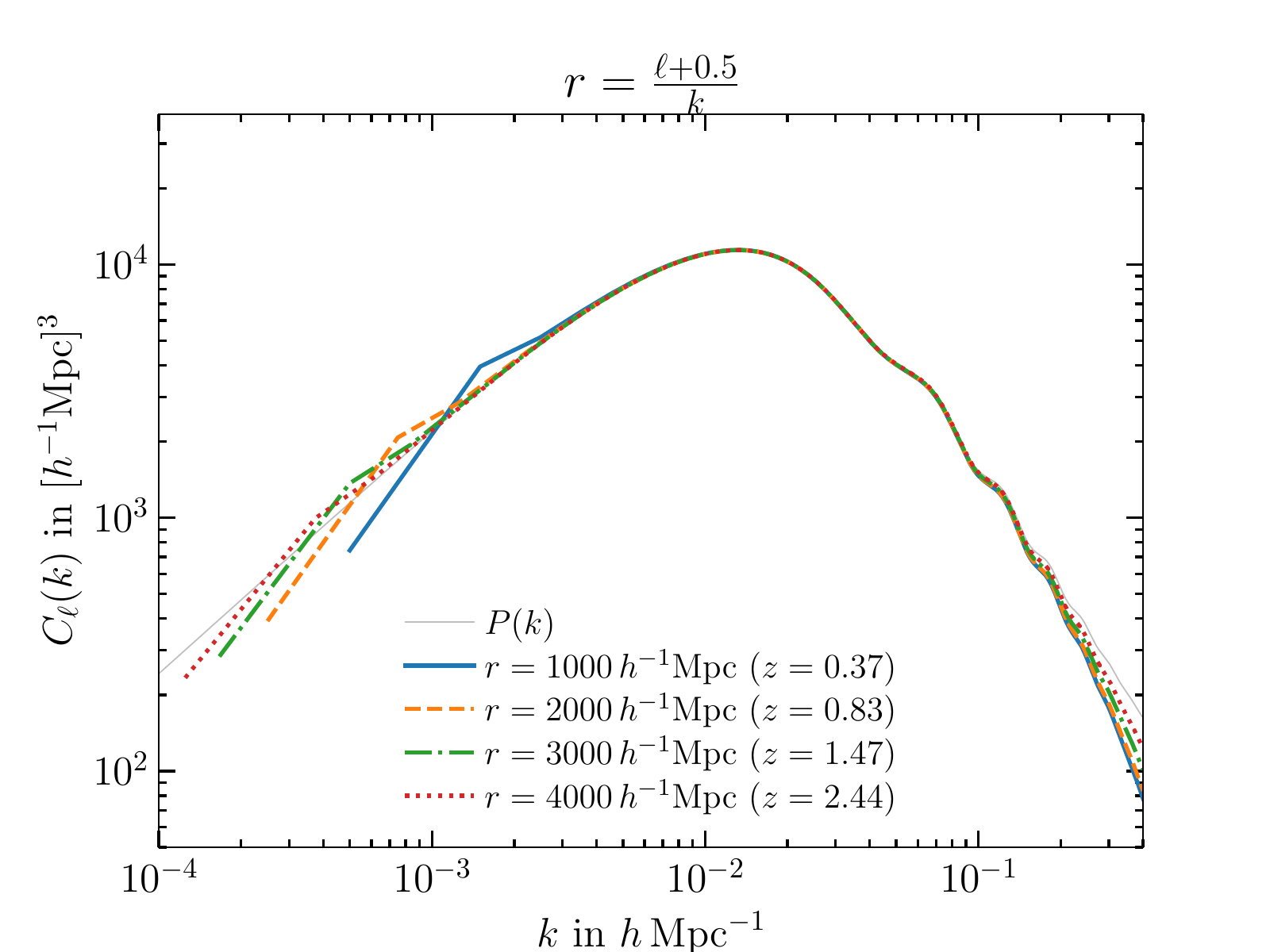}
  \incgraph{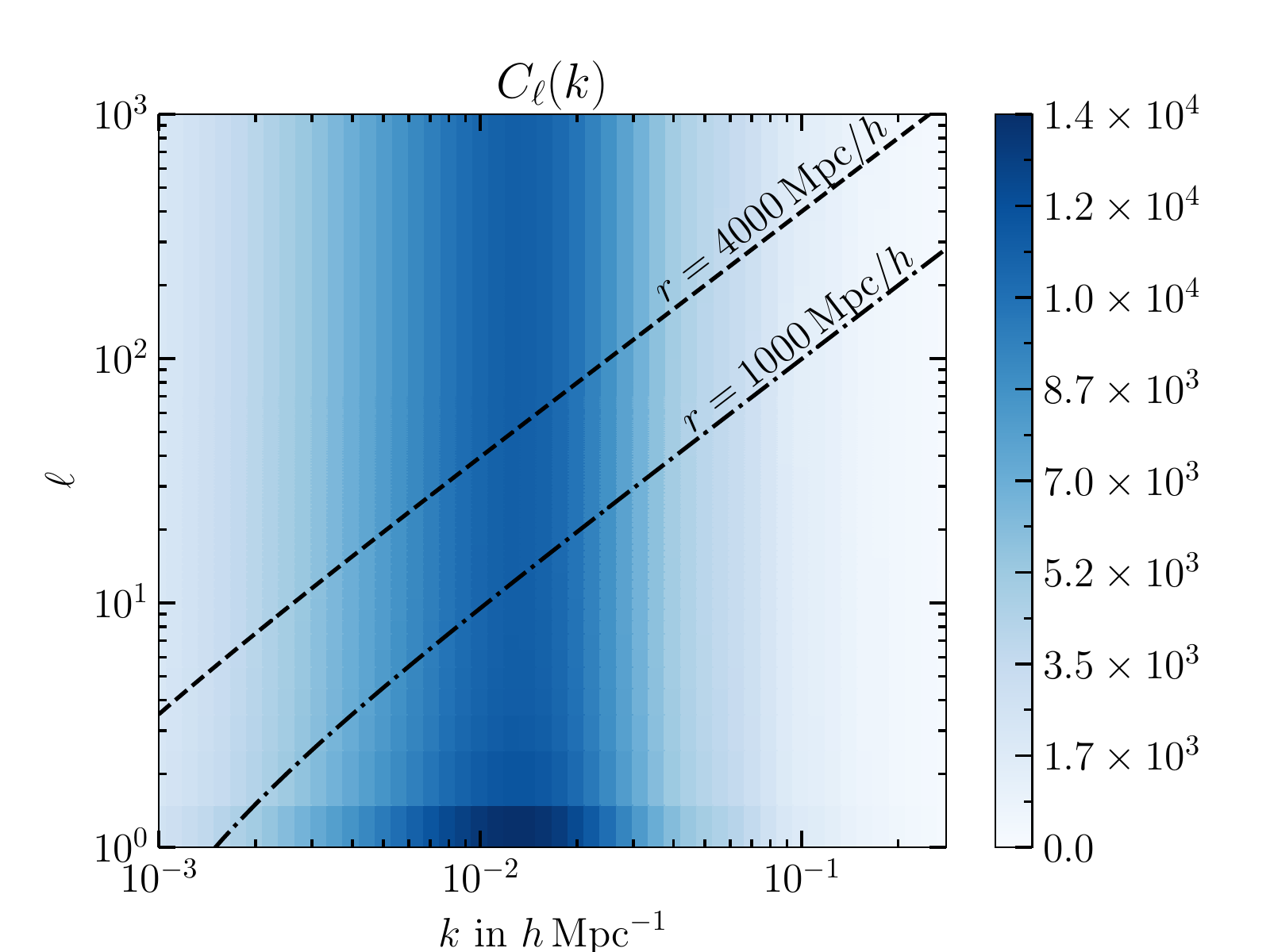}
  \caption{
    Top left: Linear growth factor $D(r)$, linear growth rate $f(r)$, velocity
    dispersion $\sigma_u^2=\sigma_v^2/(aH)^2$, galaxy bias $b(r)=b_0/D(r)$
    where $b_0=1$, and selection function $\phi(r)$ defined in \cref{eq:phi}.
    Top right: The SFB power spectrum in the Limber approximation closely
    traces the 3D power spectrum. However, for a given perpendicular mode
    $\ell$ and redshift range, only a part of the power spectrum is measured.
    The horizontal lines show the range of $k$ modes that a given $\ell$ mode
    is able to measure for a survey within $1000\leq \frac{r}{h^{-1}{\rm
    Mpc}}\leq 4000$, and the shaded bands show an estimate for the $1\sigma$
    measurement uncertainty for that particular $\ell$-mode.
    Bottom left: Here, each line fixes the redshift, and all $\ell$ modes are
    used. The Kaiser effect is not visible due to the Limber approximation
    becoming invalid on large scales.
    Bottom right: Here we show the SFB power spectrum on a grid of $\ell$-$k$
    modes. Within the Limber approximation, the SFB power spectrum
    can be measured within a band such that $r\simeq\(\ell+\frac12\)/\,k$ is
    within the survey. Outside this band we expect the Limber
    approximation to be too inaccurate even for
    the qualitative reasoning that is our objective here, and we leave a
    detailed treatment to a future paper.
  }
  \label{fig:sfb_limber}
\end{figure*}
\begin{figure*}
  \centering
  \incgraph{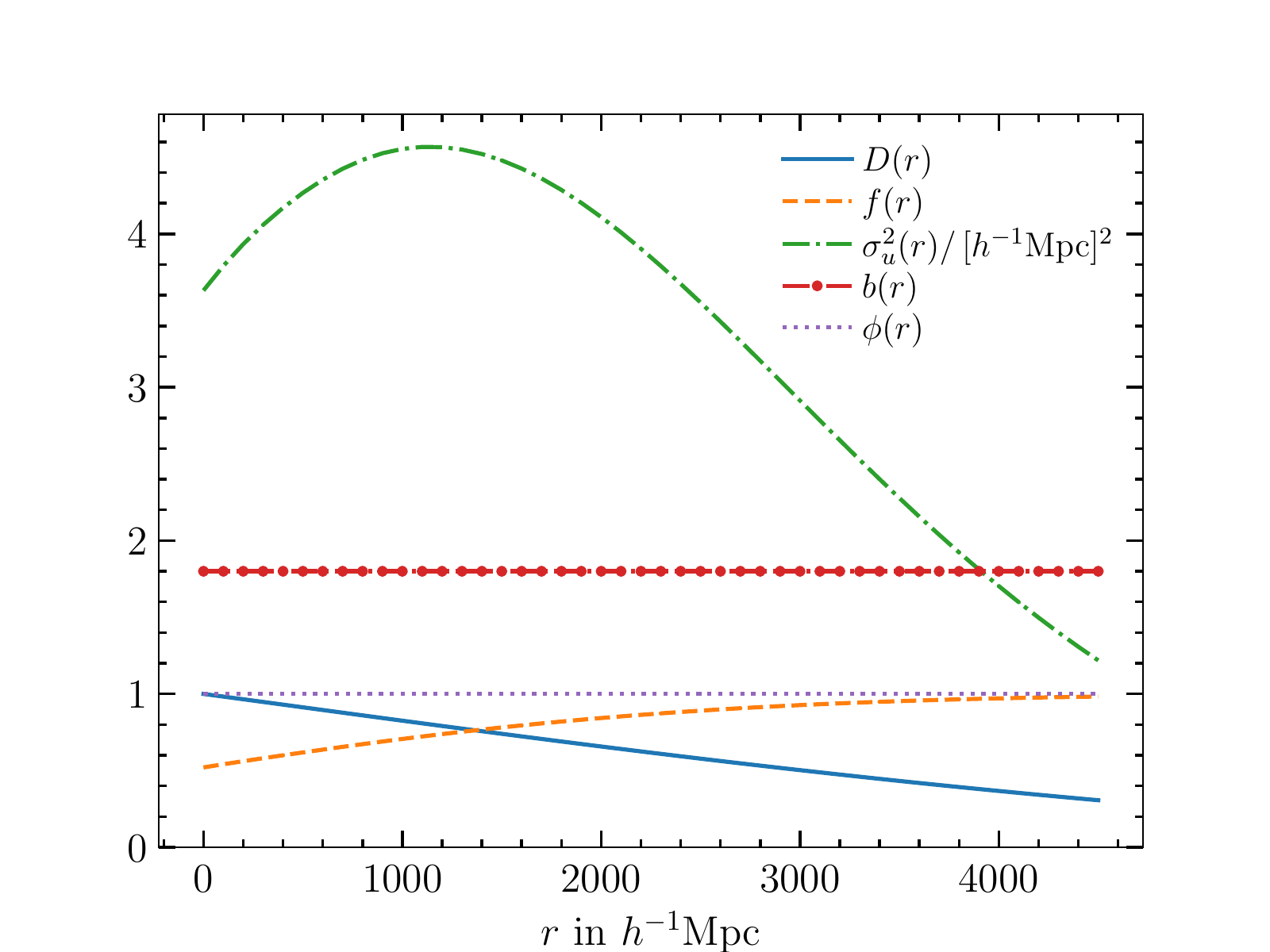}
  \incgraph{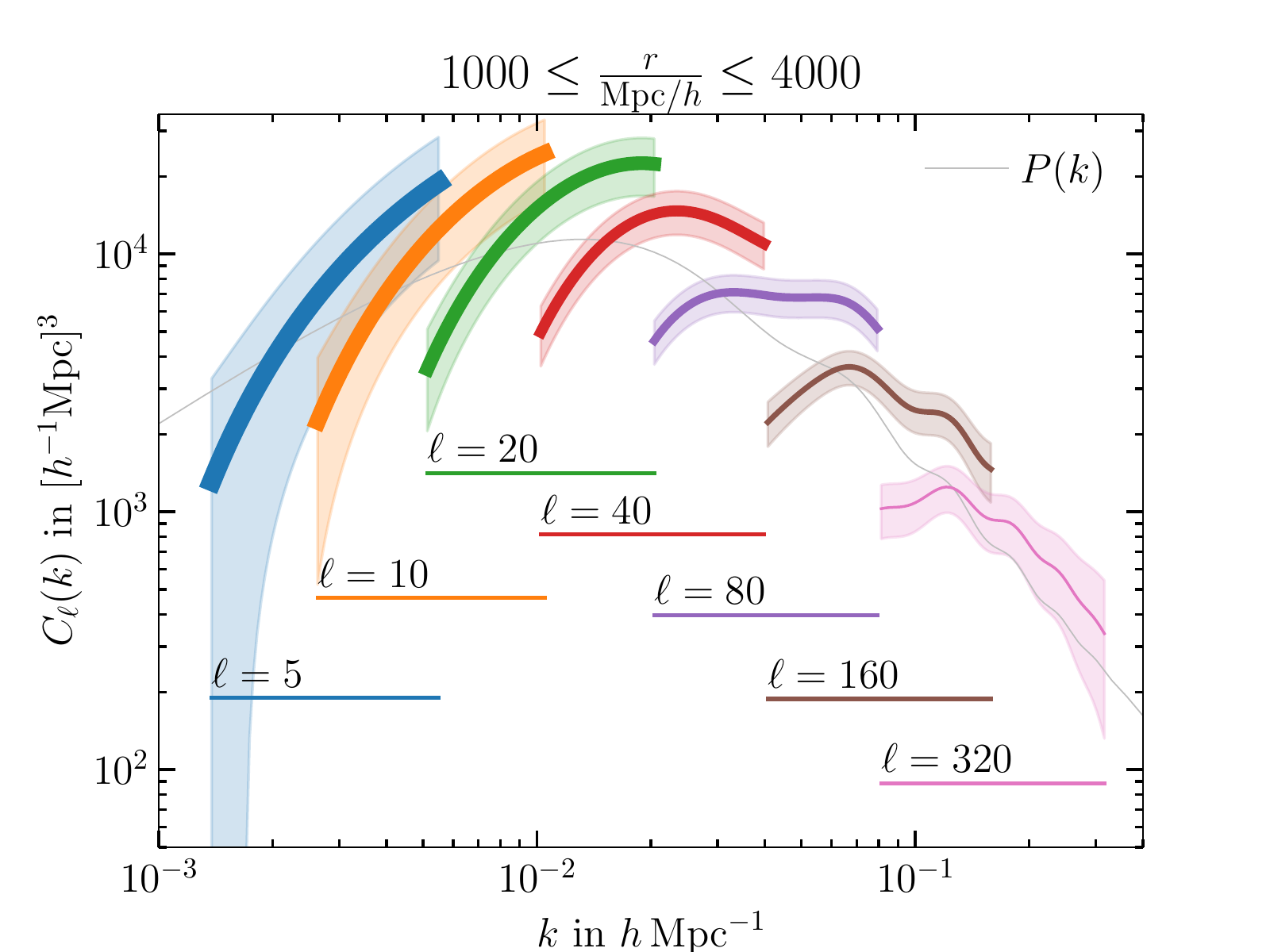}
  \incgraph{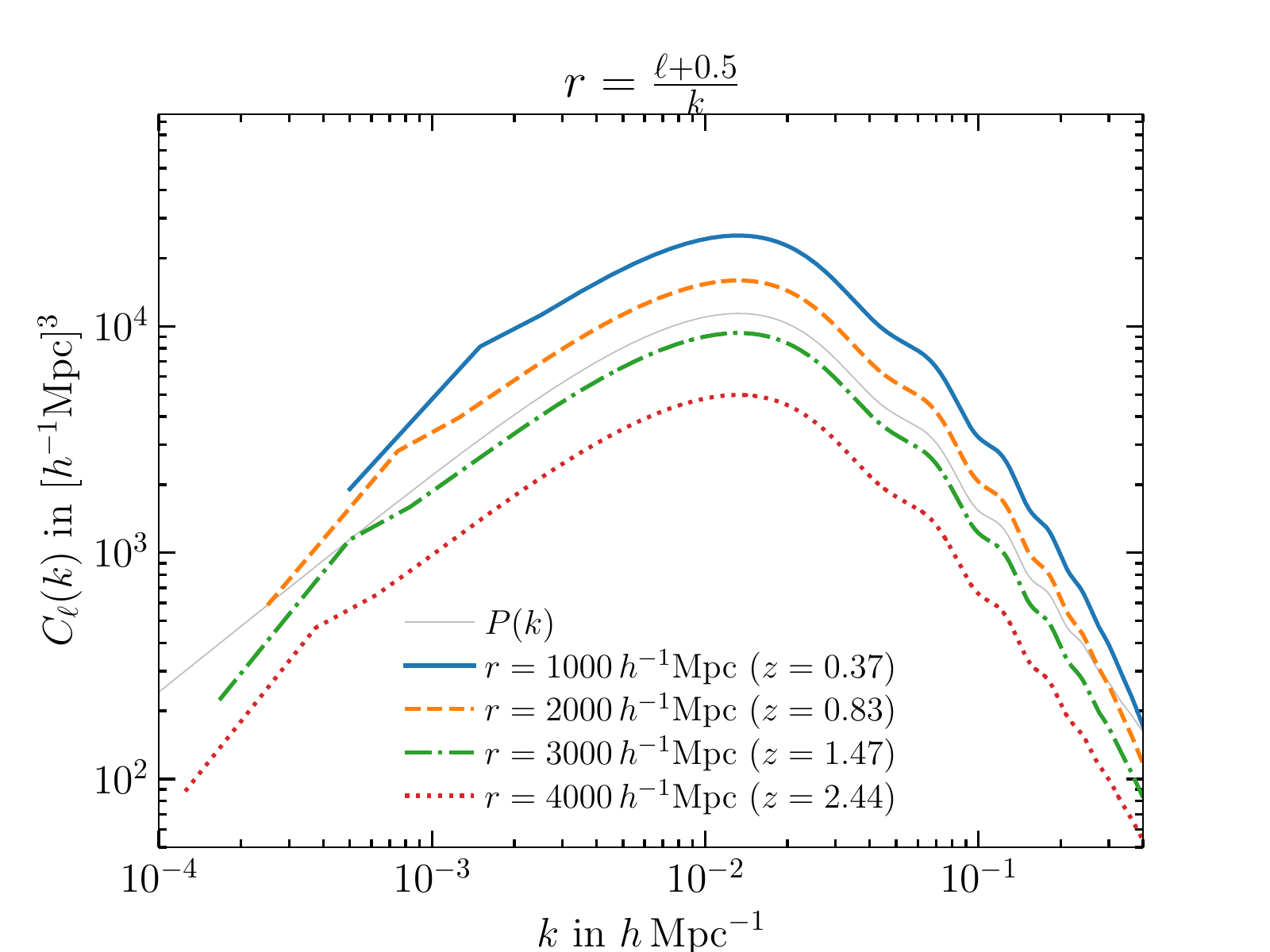}
  \incgraph{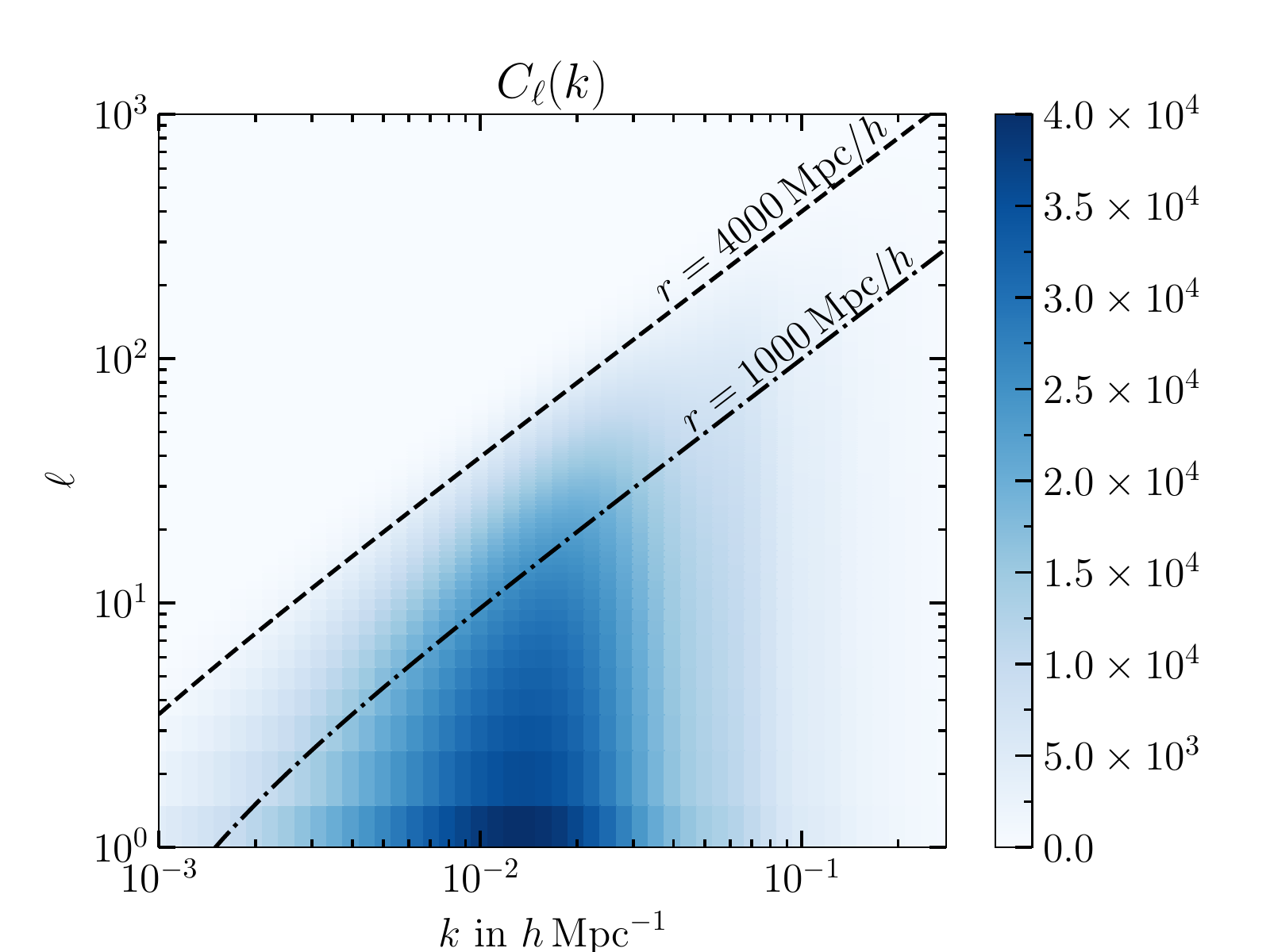}
  \caption{
    Same as \cref{fig:sfb_limber}, except that the linear galaxy bias is now
    constant $b(r)=1.8$. As a result, the redshift evolution of the linear
    growth factor $D(r)$ is no longer cancelled by the bias.
    Top right: At a fixed $\ell$, larger-scale modes are probed at higher
    redshift where the growth factor is smaller. Thus, compared to
    \cref{fig:sfb_limber}, each $\ell$-segment appears tilted.
    Bottom left: The redshift evolution of the linear growth factor $D(r)$
    causes a shift in the power spectrum amplitude with the Limber ratio
    $r=(\ell+\frac12)\,/\,k$.
    Bottom right: High-$\ell$ modes are suppressed because they primarily probe
    the high redshifts.
  }
  \label{fig:sfb_limber_bconst}
\end{figure*}
\begin{figure}
  \centering
  \incgraph{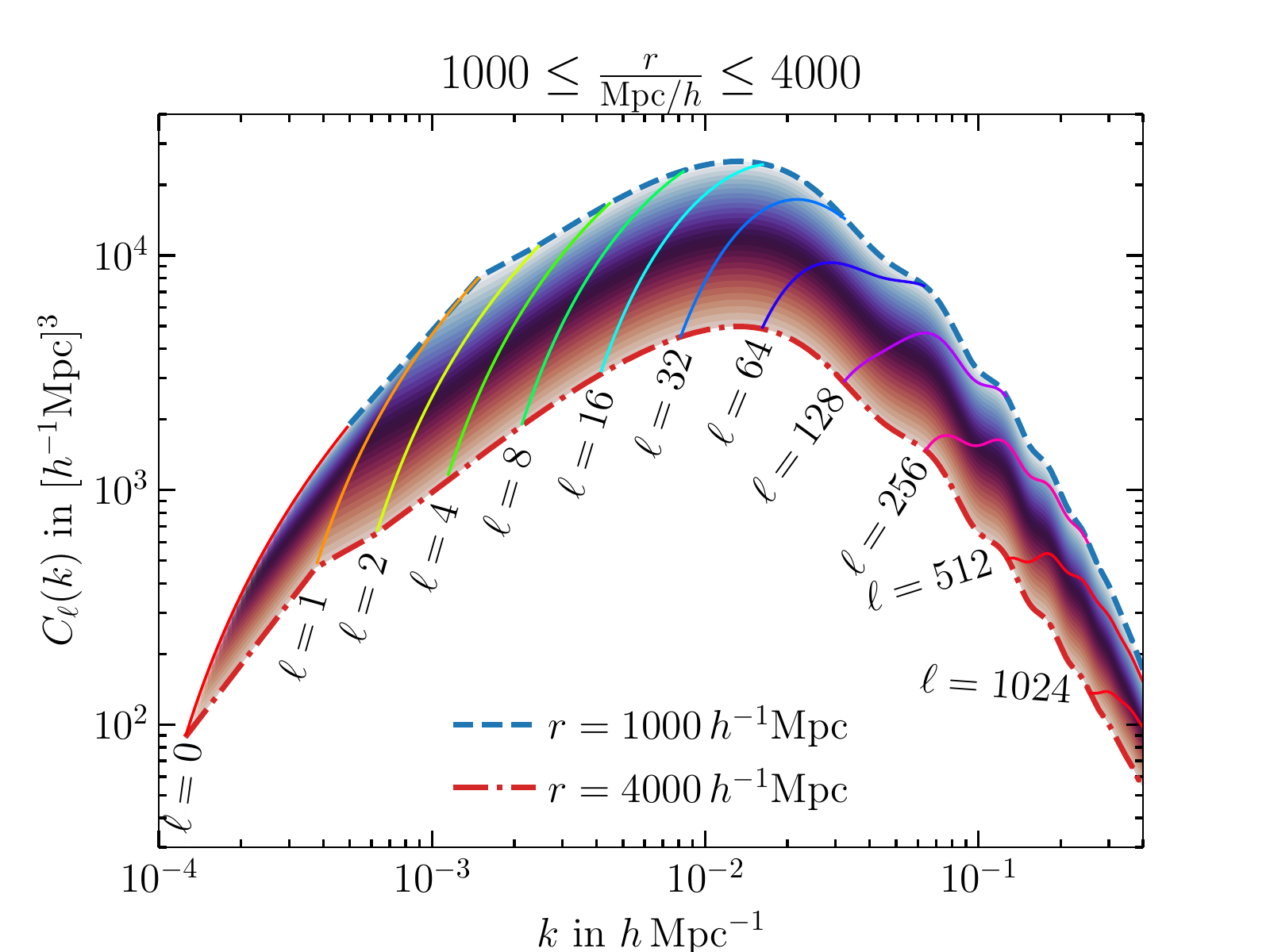}
  \caption{
    Here we combine the top right and bottom left plots in
    \cref{fig:sfb_limber_bconst} to show the region that the SFB power
    spectrum probes in the Limber approximation for a given redshift range and
    linear power spectrum evolution.
  }
  \label{fig:sfb_limber_bconst_mesh}
\end{figure}
In this section we aim to gain some intuition for the SFB power spectrum in
\cref{eq:sfb_power_spectrum} by applying a type of Limber approximation.
We stress that the approximation used here is inadequate as a precise model and is only intended for the purpose of gaining intuition, especially for how redshift evolution is encoded in the SFB power spectrum at high $\ell$.
To do so, we
will also approximate the effect of the FoG. We write \cref{eq:Arsd,eq:Afog}
acting on a spherical Bessel function as
\ba
&\widetilde A_\mathrm{RSD}(-iq\partial_{qr},-i\partial_{qr},r) \, j_\ell(qr)
\vs
&=
e^{\frac12\sigma_u^2q^2\partial_{qr}^2}
\(1 - \beta \partial_{qr}^2\)
j_\ell(qr)\,.
\ea
The second derivative is obtained exactly via a recursion relation for the
derivative of Spherical Bessel function,
\ba
\(1 - \beta \partial_{qr}^2\)
j_\ell(qr)
&=
\(1-\beta f^{\ell}_0\) j_{\ell}(qr)
- \beta f^{\ell}_{-2}\, j_{\ell-2}(qr)
\vs&\quad
- \beta f^{\ell}_{2}\, j_{\ell+2}(qr)
\\
&=
\sum_{\Delta\ell}
\(\delta^K_{\Delta\ell,0}-\beta f^{\ell}_{\Delta\ell}\) j_{\ell+\Delta\ell}(qr)\,,
\ea
where the only non-zero $f_{\Delta\ell}^\ell$ are
\ba
f_{-2}^\ell &= \frac{\ell (\ell-1)}{(2\ell-1)(2\ell+1)}\,, \\
f_0^\ell &= -\frac{2\ell^2 + 2\ell - 1}{(2\ell-1) (2\ell+3)}\,, \\
f_2^\ell &= \frac{(\ell+1) (\ell+2)}{(2\ell+1)(2\ell+3)}\,.
\ea
The FoG term acting on the spherical Bessel function is a convolution
\ba
\widetilde A_\mathrm{FoG}(-iq\partial_{qr}) \, j_\ell(qr)
&=
\int\frac{\dd k}{2\pi}
\, e^{ikqr}
\, \widetilde A_\mathrm{FoG}(qk)
\, \widetilde j_\ell(k)
\vs
&=
\int\dd y \, A_\mathrm{FoG}(r - y) \,j_\ell(qy)
\\
&=
\int\dd y
\, \frac{1}{\sqrt{2\pi}\,\sigma_u} \,e^{-\frac{(r-y)^2}{2\sigma_u^2}}
\, j_\ell(qy)
\,,
\ea
where the tilde signify Fourier transforms, and we took the inverse transform
of \cref{eq:Afog} \citep[see e.g.][]{GrasshornGebhardt+:2020PhRvD.102h3521G}. As a
first approximation, if the frequency $q$ is low, then the convolution will
have little effect. If the frequency is high, the convolution will erase the
oscillations to vanish. That is, we approximate
\ba
\widetilde A_\mathrm{FoG}(-iq\partial_{qr}) \, j_\ell(qr)
&\approx
e^{-\frac12\sigma_u^2q^2}\,j_\ell(qr)\,.
\ea

We are now in a position to apply a version of Limber's approximation.
The first-order result from \citet{LoVerde+:2008PhRvD..78l3506L} can be written
as
\ba
\label{eq:J_limber}
J_\nu(kr) &\simeq \delta^D\!\(kr-\nu\)\,,
\ea
where $J_\nu(x)$ is the Bessel function.
Therefore, for a spherical Bessel function
$j_\ell(x)=\sqrt{\pi/2x}\,J_{\ell+\frac12}(x)$ we get
\ba
\label{eq:jl_limber}
j_\ell(kr)
&\simeq \sqrt{\frac{\pi}{2rk}}\,\frac{1}{k}\,\delta^D\!\(r-\frac{\ell+\frac12}{k}\)
\ea
to first order.
\cref{eq:jl_limber} is valid only when all
other functions are slowly-varying compared to the frequency of the
spherical Bessel, and the integration should be over a wide interval. However,
in the special case that one has two spherical Bessel functions, applying
\cref{eq:jl_limber} reproduces \cref{eq:jljlDelta}, and that leads to the
Limber approximation in the context of smooth power spectra integrated over a redshift bin \citep[also see][their
Appendix C]{Jeong+:2009PhRvD..80l3527J}. For simplicity we will refer to this
as the Limber approximation. The
Limber approximation \cref{eq:jl_limber} needs to be used with care, and we will list some of the
caveats throughout this section.
Then, \cref{eq:sfb_wlLkqrhat,eq:sfb_wlkq} become
\ba
\mathcal{W}_\ell(k,q)
&=
\frac{2qk}{\pi}
\int\dd r\,r^2
\,\phi(r) \,D(r)\,b(r,q)
\,j_\ell(kr)
\vs&\quad\times
\,e^{-\frac12\sigma_u^2 q^2}
\sum_{\Delta\ell}
\(\delta^K_{\Delta\ell,0}-\beta f^{\ell}_{\Delta\ell}\) j_{\ell+\Delta\ell}(qr)
\displaybreak[0]\\
&=
\sqrt{\frac{q}{k}}
\,\phi\!\(\frac{\ell+\frac12}{k}\) D\!\(\frac{\ell+\frac12}{k}\) b\!\(\frac{\ell+\frac12}{k},q\)
\vs&\quad\times
\,e^{-\frac12\sigma_u^2 q^2}
\sum_{\Delta\ell}
\(\delta^K_{\Delta\ell,0}-\beta f^{\ell}_{\Delta\ell}\)
\vs&\quad\times
\,\delta^D\!\(q - \frac{\ell+\Delta\ell+\frac12}{\ell+\frac12}\,k\)\,.
\ea
Since the Limber approximation is only applicable for large $\ell$, we further
assume $\Delta\ell \ll \ell$. Then,
\ba
\mathcal{W}_\ell(k,q)
&=
\delta^D\!\(q - k\)
\phi\!\(\frac{\ell+\frac12}{k}\)
D\!\(\frac{\ell+\frac12}{k}\)
b\!\(\frac{\ell+\frac12}{k},k\)
\vs&\quad\times
e^{-\frac12\sigma_u^2 k^2}
\sum_{\Delta\ell}
\(\delta^K_{\Delta\ell,0}-\beta f^{\ell}_{\Delta\ell}\).
\ea
Therefore, the SFB power spectrum \cref{eq:sfb_power_spectrum} in the Limber
approximation is
\ba
\label{eq:sfb_power_spectrum_limber}
C_\ell(k,k')
&=
P(k)
\,e^{-\sigma_u^2 k^2}
\,\delta^D\!\(k - k'\)
\vs&\quad\times
\phi^2\!\(\frac{\ell+\frac12}{k}\)
D^2\!\(\frac{\ell+\frac12}{k}\)
b^2\!\(\frac{\ell+\frac12}{k},k\)
\vs&\quad\times
\left[1 - \beta \(f_{-2}^\ell + f_0^\ell + f_{2}^\ell\)\right]^2.
\ea
The exponential is the suppression due to the FoG. The Dirac-delta function
shows that even with redshift-evolution most of the power is on the diagonal
$k=k'$, as for a non-evolving universe. Redshift evolution manifests itself
mainly through the interplay between $\ell$ and $k$ such that in the Limber
approximation the ratio
\ba
\label{eq:sfb_limber_ratio}
r &= \frac{\ell+\frac12}{k}
\ea
is the comoving angular diameter distance. For example, if the scale $k$ is
fixed, then changing the angular scale $\ell$ corresponds to changing the
redshift. However, we caution the reader that
\cref{eq:sfb_limber_ratio} comes from the approximation \cref{eq:jl_limber},
and a detailed treatment especially at high $k$ is needed in general. We note
that primordial non-Gaussianity will lead to a
scale-dependent bias that will be absorbed directly in the bias term on the
second line. Finally, the last line in \cref{eq:sfb_power_spectrum_limber}
accounts for the linear Kaiser effect.

In \cref{fig:sfb_limber} we show the SFB power spectrum in the Limber
approximation. We define $C_\ell(k)$ such that $C_\ell(k,k')=\delta^D(k-k')\,C_\ell(k)$. For the galaxy bias we choose
$b(r,k)=b_0/D(r)$ with $b_0=1$ so that the bias and linear growth factor
cancel, and our selection function is a constant $\phi(r)=1$ for illustration.
The top left panel shows the inputs to our calculation.

The top right panel of \cref{fig:sfb_limber} shows the SFB power spectrum for
several fixed $\ell$ modes. The shaded areas correspond to the $1\sigma$
measurement uncertainty estimated via
\ba
\Delta C_\ell(k)
&=
\sqrt{\frac{2}{2\ell + 1}}\(C_\ell(k) + \frac{1}{\nbar}\),
\ea
\citep{Pratten+:2013MNRAS.436.3792P}. Since $\ell$ corresponds to perpendicular
wave modes, at small $r$ the corresponding $k$ modes are large, and at large
$r$ the corresponding $k$ modes are small. Therefore, at a constant $\ell$, the
SFB power spectrum as a function of $k$ sweeps through both redshift and $k$
modes, measuring a redshift corresponding to
$r\simeq\SI{4000}{\per\h\mega\parsec}$ at lower $k$, and a redshift
corresponding to $r\simeq\SI{1000}{\per\h\mega\parsec}$ at higher $k$.

Consequently, in the $\ell$-$k$ plane, only a band of modes can be measured, as
illustrated in the bottom right plot of \cref{fig:sfb_limber}. Fixing the
redshift, which is possible in the Limber approximation, results in a bona-fide
power spectrum measurement, as illustrated in the bottom left panel.

In the Limber approximation, the SFB power spectrum does not exhibit a strong
Kaiser effect. We attribute this to our approximations being inadequate for
such analysis, and we refer the reader to
\citet{Yoo+:2013PhRvD..88b3502Y,Munshi+:2016MNRAS.456.1627M} for further
details.

Because \cref{eq:sfb_limber_ratio} relates the $\ell$ and $k$ modes to a
definite redshift, we can only measure a band of modes. We illustrate this in
the bottom right panel of \cref{fig:sfb_limber}. More generally,
\cref{eq:sfb_limber_ratio} is valid only approximately, and the detailed
treatment outside this band is dependent on the exact choice of basis
functions.

The choice $b(r,q)\,\propto\,D^{-1}(r)$ results in linear bias and linear
growth cancelling each other. In general, this may not be the case, and we
illustrate redshift evolution by setting the bias constant, $b(r,q)=1.8$, in
\cref{fig:sfb_limber_bconst}. Because for fixed $\ell$ larger-scale modes $k$
correspond to higher redshift, the linear growth evolution tilts each
$\ell$-segment of the power spectrum in the top right panel. The bottom left
panel shows the power spectrum at fixed redshift according to the Limber ratio
\cref{eq:sfb_limber_ratio}, sweeping through $\ell$.

Each segment in the top right panel of \cref{fig:sfb_limber_bconst} crosses the
lines in the bottom left panel. We illustrate this further in
\cref{fig:sfb_limber_bconst_mesh}.

We hope that this appendix gives some insight into how the SFB power spectrum
works. However, we stress again that the approximations made here are
inadequate for a full cosmological analysis, and the reader should keep this
caveat in mind.

\section{Radial spherical Fourier-Bessel modes with potential boundary conditions}
\label{app:gnl}
In this appendix we derive the radial basis functions of the Laplacian with
potential boundary conditions at $r_\min$ and $r_\max$.

We first isolate the radial part of \cref{eq:laplacian_spherical}. Writing
\ba
f(\vr) &= g(r)\,h(\rhat)\,,
\ea
we require
\ba
-k^2 gh
&=
\frac{h}{r^2}\,\frac{\partial}{\partial r}\left(r^2\,\frac{\partial g}{\partial r}\right)
+ g\,\nabla^2h\,.
\ea
Given the spherical harmonic solution for the angular term $h$,
\ba
\nabla^2 h &= -\frac{\ell(\ell+1)}{r^2}\,h\,,
\ea
we get
\ba
0
&=
\frac{\dd}{\dd r}\left(r^2\,\frac{\dd g_{\ell}(kr)}{\dd r}\right)
+ \big[(kr)^2 - \ell(\ell+1)\big]\,g_{\ell}(kr)\,,
\label{eq:helmholtz}
\ea
where we now added that the function $g$ depends on $\ell$. Our first aim is to
derive the discrete spectrum of $k$ modes for a given $\ell$. We then use that
to derive the form of the $g_\ell$. Following
\citet{Fisher+:1995MNRAS.272..885F}, we demand that the orthogonality relation
\cref{eq:gnl_orthonormality} is satisfied. However, we modify the approach in
\citet{Fisher+:1995MNRAS.272..885F} to integrate from $r_\min$ to $r_\max$.
\cref{eq:helmholtz} multiplied by $g_{\ell}(kr)$ then yields
\ba
&\int_{r_\min}^{r_\max}\dd r\,
\frac{\dd}{\dd r}\left(r^2\,\frac{\dd g_{\ell}(kr)}{\dd r}\right)g_{\ell}(k'r)
\vs
&=
\int_{r_\min}^{r_\max}\dd r\,
\big[\ell(\ell+1) - (kr)^2\big]\,g_{\ell}(kr)\,g_{\ell}(k'r)\,.
\ea
Subtract from this equation the same equation with $k$ and $k'$ interchanged,
\ba
&\big[k'^2 - k^2\big]\int_{r_\min}^{r_\max}\dd r\,r^2
\,g_\ell(kr)\,g_{\ell}(k'r)
\vs
&=
\int_{r_\min}^{r_\max}\dd r\,
\bigg\{
\frac{\dd}{\dd r}\left(r^2\,\frac{\dd g_\ell(kr)}{\dd r}\right)g_{\ell}(k'r)
\vs&\quad
- \frac{\dd}{\dd r}\left(r^2\,\frac{\dd g_{\ell}(k'r)}{\dd r}\right)g_{\ell}(kr)\bigg\}\,.
\label{eq:integrate_glgl_step1}
\ea
Partial integration with the terms on the right hand side (r.h.s.)
yields
\ba
&\int\dd r\,\frac{\dd}{\dd r}\!\(kr^2 g'_\ell(kr)\)g_\ell(k'r)
\vs
&=
\left.kr^2g'_\ell(kr)g_\ell(k'r)\right|_{r_\min}^{r_\max}
- kk'\int\dd r\,r^2\,g'_\ell(kr)g'_\ell(k'r)\,.
\ea
Then, \cref{eq:integrate_glgl_step1} becomes
\ba
&\big[k'^2 - k^2\big]\int_{r_\min}^{r_\max}\dd r\,r^2
\,g_\ell(kr)\,g_{\ell}(k'r)
\vs
&=
\left.kr^2g'_\ell(kr)g_\ell(k'r)\right|_{r_\min}^{r_\max}
-\left.k'r^2g'_\ell(k'r)g_\ell(kr)\right|_{r_\min}^{r_\max}\,.
\label{eq:glgl_kk_integral}
\ea
The r.h.s.\  will vanish for any $k$ whenever
\ba
0
&=
Akr_\max^2g'_\ell(kr_\max)
- Br_\max^2g_\ell(kr_\max)
\vs&\quad
- akr_\min^2g'_\ell(kr_\min)
+ br_\min^2g_\ell(kr_\min)\,.
\label{eq:glgl_orthogonality_condition}
\ea
for any constants $a$, $b$, $A$, and $B$.

To choose $a$, $b$, $A$, and $B$, we note that the representable field
$\delta(\vr)$ is written as a sum of the solutions to \cref{eq:helmholtz}
inside the SFB volume, and we have some freedom to choose the desired behavior
outside of it. Since the field inside the SFB volume satisfies the Poisson
equation, it is natural to have it satisfy Laplace's equation\footnote{Laplace's equation is Poisson's equation without a source term.} outside it, and
demand that the solution is continuous and smooth at the boundaries. That is,
\begin{widetext}
\ba
\delta(\vr)
&=
\begin{cases}
  \sum_{\ell m}\left[a_{\ell m}\(\frac{r}{r_\min}\)^\ell
  + b_{\ell m}\(\frac{r_\min}{r}\)^{\ell+1} \right] Y_{\ell m}(\rhat)\,,
  & \text{for } r < r_\min\,,
  \\
  \sum_{n\ell m} \Big[c_{n\ell}\, j_\ell(k_{n\ell}\,r) + d_{n\ell}\,y_\ell(k_{n\ell}\,r)\Big]Y_{\ell m}(\rhat)\,\delta_{n\ell m}\,,
  & \text{for } r_\min \leq r \leq r_\max\,,
  \\
  \sum_{\ell m}\left[A_{\ell m}\(\frac{r}{r_\max}\)^\ell
  + B_{\ell m}\(\frac{r_\max}{r}\)^{\ell+1} \right] Y_{\ell m}(\rhat)\,,
  & \text{for } r > r_\max\,,
\end{cases}
\ea
\end{widetext}
where we defined the constants $a_{\ell m}$, $b_{\ell m}$, $c_{n\ell}$,
$d_{n\ell}$, $A_{\ell m}$, and $B_{\ell m}$, and we explicitly wrote
\ba
g_\ell(kr)=c_{n\ell}\,j_\ell(kr) + d_{n\ell}\, y_\ell(kr)
\,,
\ea
and we anticipate that the function $g_\ell$ will also depend on $n$.
Continuity at the boundaries requires
\ba
a_{\ell m} + b_{\ell m} &= \sum_n g_\ell(k_{n\ell}\,r_\min)\,\delta_{n\ell m}\,,\\
A_{\ell m} + B_{\ell m} &= \sum_n g_\ell(k_{n\ell}\,r_\max)\,\delta_{n\ell m}\,.
\ea
Smoothness further requires
\ba
\ell\,\frac{a_{\ell m}}{r_\min}
-(\ell+1)\,\frac{b_{\ell m}}{r_\min}
&=
\sum_n k_{n\ell}\,g'_\ell(k_{n\ell}\,r_\min)\,\delta_{n\ell m}
\,,\\
\ell\,\frac{A_{\ell m}}{r_\max}
-(\ell+1)\,\frac{B_{\ell m}}{r_\max}
&=
\sum_n k_{n\ell}\,g'_\ell(k_{n\ell}\,r_\max)\,\delta_{n\ell m}\,,
\ea
Now requiring $\delta(\vr)$ to be finite at $r=0$ and $r=\infty$ sets $b_{\ell
m}=A_{\ell m}=0$, and requiring continuity and smoothness for any
$\delta_{n\ell m}$, we get
\ba
\label{eq:knl_rmin_condition_v1}
\ell\,g_\ell(k_{n\ell}\,r_\min)
&=
k_{n\ell}\,r_\min\,g'_\ell(k_{n\ell}\,r_\min)
\,,\\
\label{eq:knl_rmax_condition_v1}
- (\ell+1)\,g_\ell(k_{n\ell}\,r_\max)
&=
k_{n\ell}\,r_\max\,g'_\ell(k_{n\ell}\,r_\max)
\,.
\ea
These choices lead to $a=1$, $b=\ell/r_\min$, $A=1$, and $B=-(\ell+1)/r_\max$
in \cref{eq:glgl_orthogonality_condition}, which shows that the conditions
\cref{eq:knl_rmin_condition_v1,eq:knl_rmax_condition_v1} on $k_{n\ell}$ lead to
an orthogonality relation for the $g_\ell$.

Both $j_\ell$ and $y_\ell$ satisfy the two recurrence relations
\ba
j'_\ell(kr) &= -j_{\ell+1}(kr) + \frac{\ell}{kr}\,j_\ell(kr)\,,
\\
j'_\ell(kr) &= j_{\ell-1}(kr) - \frac{\ell+1}{kr}\,j_\ell(kr)\,.
\ea
Then, \cref{eq:knl_rmin_condition_v1,eq:knl_rmax_condition_v1} simplify to
\ba
\label{eq:knl_rmin_condition_v2}
c_{n\ell}\,j_{\ell+1}(k_{n\ell}\,r_\min)
+ d_{n\ell}\,y_{\ell+1}(k_{n\ell}\,r_\min)
&= 0
\,,\\
\label{eq:knl_rmax_condition_v2}
c_{n\ell}\,j_{\ell-1}(k_{n\ell}\,r_\max)
+ d_{n\ell}\,y_{\ell-1}(k_{n\ell}\,r_\max)
&= 0
\,.
\ea
The normalization of $g_\ell$ is obtained by dividing
\cref{eq:glgl_kk_integral} by $k'^2-k^2$, and taking the limit $k'\to
k=k_{n\ell}$,
\ba
1
&=
\int_{r_\min}^{r_\max}\dd r\,r^2
\,g^2_\ell(kr)
\vs
&=
\lim_{k'\to k}
\frac{\left.kr^2g'_\ell(kr)g_\ell(k'r)\right|_{r_\min}^{r_\max}
    -\left.k'r^2g'_\ell(k'r)g_\ell(kr)\right|_{r_\min}^{r_\max}}
    {k'^2 - k^2}
    \,.
    \label{eq:gl_normlization_definition}
\ea
When $kr=k_{n\ell}r_\min$ as in \cref{eq:knl_rmin_condition_v1},
\ba
&kr\,g'_\ell(kr)g_\ell(k'r) - k'r\,g'_\ell(k'r)g_\ell(kr)
\vs
&=
k'r\,g_\ell(kr)\big[c_{n\ell}\,j_{\ell+1}(k'r) + d_{n\ell}\,y_{\ell+1}(k'r)\big]
\,,
\ea
and when $kr=k_{n\ell}r_\max$ as in \cref{eq:knl_rmax_condition_v1},
\ba
&kr\,g'_\ell(kr)g_\ell(k'r) - k'r\,g'_\ell(k'r)g_\ell(kr)
\vs
&=
-k'r\,g_\ell(kr)\big[c_{n\ell}\,j_{\ell-1}(k'r) + d_{n\ell}\,y_{\ell-1}(k'r)\big]
\,.
\ea
The terms in brackets vanish in the limit $k'\to k=k_{n\ell}$ as per
\cref{eq:knl_rmin_condition_v2,eq:knl_rmax_condition_v2}. That is, we need
limits
\ba
\lim_{q'\to q} \frac{c_{n\ell}\,j_{\ell+1}(q') + d_{n\ell}\,y_{\ell+1}(q')}{q'^2-q^2}
&=
\frac{g_\ell(q)}{2q}
\,,\\
\lim_{q'\to q} \frac{c_{n\ell}\,j_{\ell-1}(q') + d_{n\ell}\,y_{\ell-1}(q')}{q'^2-q^2}
&=
-\frac{g_\ell(q)}{2q}
\,,
\ea
for $q=kr_\min$ and $q=kr_\max$, respectively.
Then, \cref{eq:gl_normlization_definition} becomes
\ba
\label{eq:gl_normalization}
1
&=
\frac{r^3_\max}{2}\,\,g^2_\ell(k_{n\ell}\,r_\max)
- \frac{r^3_\min}{2}\,g^2_\ell(k_{n\ell}\,r_\min)
\,.
\ea

Choosing $k_{n\ell}$, $c_{n\ell}$, and $d_{n\ell}$ that satisfy
\cref{eq:knl_rmin_condition_v2,eq:knl_rmax_condition_v2,eq:gl_normalization}
guarantees the orthonormality of the $g_\ell$,
\ba
\int_{r_\min}^{r_\max}\dd r\,r^2\,g_\ell(k_{n\ell}\,r)\,g_{\ell}(k_{n'\ell}\,r)
&=
\delta^K_{nn'}
\,.
\ea
Note that the condition $\ell=\ell'$ is \emph{not} enforced by the $g_\ell$.
Instead, $\ell=\ell'$ comes from the spherical harmonics, i.e., \cref{eq:YlmYlmDelta}.

\subsection{Phase factor}
We are free to introduce a phase factor for the $g_{n\ell}(r)$, and we choose
it so that the sign of $g_{n\ell}(r_\min)$ alternates with $n$, but stays
constant with $\ell$,
\ba
g_{n\ell}(r)
&=
(-1)^{\left[1-\mathrm{floor}\(\frac{1}{\ell+1}\)\right] \left[1-\mathrm{floor}\(\frac{1}{n}\)\right]}
\, \tilde g_{n\ell}(r)
\,,
\ea
where the tilde indicates that we have not included the phase factor. This
flips the sign unless either $\ell\neq0$ or $n\neq 1$. Thus, the basis
functions in \cref{fig:sfb_gnl_basis_potential} are obtained.

\subsection{Numerical concerns}
To calculate $k_{n\ell}$, solve each of
\cref{eq:knl_rmin_condition_v2,eq:knl_rmax_condition_v2} for the ratio
$d_{n\ell}/c_{n\ell}$, and set them equal to get
\ba
0
&=
j_{\ell-1}(k_{n\ell}\,r_\max)\,y_{\ell+1}(k_{n\ell}\,r_\min)
\vs&\quad
-
y_{\ell-1}(k_{n\ell}\,r_\max)
\,j_{\ell+1}(k_{n\ell}\,r_\min)
\,.
\ea
Examples for the resulting zeros and the first few basis functions are shown in
\cref{fig:sfb_gnl_basis_potential}. The ratio $d_{n\ell}/c_{n\ell}$ then
follows from \cref{eq:knl_rmin_condition_v2} or
\cref{eq:knl_rmax_condition_v2}.
Finally, the overall normalization is fixed by \cref{eq:gl_normalization} up to
a sign.

When $r_\max$ is large and $r_\min$ is small, the $k_{n\ell}$ may need to be
computed using arbitrary precision floats, and the $g_{n\ell}$ may need to be
calculated with arbitrary precision as well. Caching the result in double
precision should then provide for sufficient speed for the actual transform.

\begin{widetext}
\section{Covariance matrix simplification}
\label{sec:sfb_covariance_simplification}
In this appendix we simplify the covariance matrix
\cref{eq:Clnnobs_covariance_raw}. For simplicity we ignore the local average
effect. We explicitly treat the shot noise \cref{eq:shotnoise} because it is
inhomogeneous and anisotropic. We get
\ba
\<\delta^{W,A}_{NLM}\,\delta^{W,A,*}_{N'L'M'}\>
&=
\sum_{n_1\ell_1 m_1}
W_{NLM}^{n_1\ell_1 m_1}
\sum_{n_2\ell_2m_2}
W_{N'L'M'}^{n_2\ell_2m_2,*}
\bigg[
\delta^K_{\ell_1\ell_2}\delta^K_{m_1m_2} \,C_{\ell_1 n_1n_2}
+ \frac{1}{\nbar}\(W^{-1}\)_{n_1 \ell_1 m_1}^{n_2 \ell_2 m_2}
\bigg]
\\
&=
\sum_{\ell_1 n_1 n_2}
C_{\ell_1 n_1 n_2}
\sum_{m_1}
W_{NLM}^{n_1 \ell_1 m_1}
W^{N'L'M'}_{n_2 \ell_1 m_1}
+
\frac{1}{\nbar}
W^{N'L'M'}_{NLM}\,,
\ea
where we used \cref{eq:sfb_window_symmetry}.
Similarly, when neither density contrast has the complex conjugate attached,
\ba
\<\delta^{W,A}_{NLM}\,\delta^{W,A}_{N'L'M'}\>
&=
(-1)^{M'}
\sum_{\ell_1 n_1 n_2}
C_{\ell_1 n_1n_2}
\sum_{m_1}
W_{NLM}^{n_1\ell_1 m_1}
W^{N'L',-M'}_{n_2\ell_1m_1}
+
\frac{(-1)^{M'}}{\nbar}
W^{N'L',-M'}_{NLM}\,,
\ea
where we used \cref{eq:Ylm_conjugate}.
Therefore, both terms in \cref{eq:Clnnobs_covariance_raw} are of the form
\ba
&\frac{1}{(2\ell+1)(2L+1)}\sum_{mM}
\<\delta^{W,A}_{n\ell m}\delta^{W,A,*}_{NLM}\>
\<\delta^{{W,A}}_{n'\ell m}\delta^{{W,A},*}_{N'LM}\>
\vs
&=
\frac{1}{(2\ell+1)(2L+1)}\sum_{mM}
\left[\sum_{\ell_1 n_1 n_2}
C_{\ell_1 n_1 n_2}
\sum_{m_1}
W_{n \ell m}^{n_1 \ell_1 m_1}
W^{NLM}_{n_2 \ell_1 m_1}
+
\frac{1}{\nbar}
W^{NLM}_{n \ell m}
\right]
\vs&\quad\times
\left[\sum_{\ell_3 n_3 n_4}
C_{\ell_3 n_3 n_4}
\sum_{m_3}
W_{n' \ell m}^{n_3 \ell_3 m_3,*}
W^{N'LM,*}_{n_4 \ell_3 m_3}
+
\frac{1}{\nbar}
W^{N'LM,*}_{n' \ell m}
\right]
\displaybreak[0]\\
&=
\frac{1}{(2\ell+1)(2L+1)}\sum_{mM}
\Bigg[
\sum_{\ell_1 n_1 n_2}
C_{\ell_1 n_1 n_2}
\sum_{m_1}
W_{n \ell m}^{n_1 \ell_1 m_1}
W^{NLM}_{n_2 \ell_1 m_1}
\sum_{\ell_3 n_3 n_4}
C_{\ell_3 n_3 n_4}
\sum_{m_3}
W_{n' \ell m}^{n_3 \ell_3 m_3,*}
W^{N'LM,*}_{n_4 \ell_3 m_3}
\vs&\quad
+
\frac{1}{\nbar}
W^{NLM}_{n \ell m}
\frac{1}{\nbar}
W^{N'LM,*}_{n' \ell m}
+
\frac{1}{\nbar}
W^{N'LM,*}_{n' \ell m}
\sum_{\ell_1 n_1 n_2}
C_{\ell_1 n_1 n_2}
\sum_{m_1}
W_{n \ell m}^{n_1 \ell_1 m_1}
W^{NLM}_{n_2 \ell_1 m_1}
+\<n\leftrightarrow n',N\leftrightarrow N'\>^*
\Bigg]
\\
&=
A_1{}^{LNN'}_{\ell nn'}
+ A_2{}^{LNN'}_{\ell nn'} + A_2{}^{LN'N}_{\ell n'n}
+ A_3{}^{LNN'}_{\ell nn'}
\,.
\ea
Using \cref{eq:sfb_window_symmetry}, we get
\ba
\label{eq:sfb_A1}
A_1{}^{LNN'}_{\ell nn'}
&=
\frac{1}{(2\ell+1)(2L+1)}
\sum_{\ell_1 n_1 n_2}
C_{\ell_1 n_1 n_2}
\sum_{\ell_3 n_3 n_4}
C_{\ell_3 n_3 n_4}
\,W_4\!\!\begin{pmatrix}
  \ell_1 & L  & \ell_3 & \ell \\
  n_1    & N  & n_4    & n' \\
  n_2    & N' & n_3    & n
\end{pmatrix},
\ea
and
\ba
\label{eq:sfb_A2}
A_2{}^{LNN'}_{\ell nn'}
&=
\frac{1}{\nbar}
\,\frac{1}{(2\ell+1)(2L+1)}
\sum_{\ell_1 n_1 n_2}
C_{\ell_1 n_1 n_2}
\,W_3\!\!\begin{pmatrix}
  \ell_1 & L  & \ell \\
  n_1    & N  & n' \\
  n_2    & N' & n
\end{pmatrix},
\ea
and
\ba
\label{eq:sfb_A3}
A_3{}^{LNN'}_{\ell nn'}
&=
\frac{1}{\nbar^2}
\,\frac{1}{(2\ell+1)(2L+1)}
\,W_2\!\!\begin{pmatrix}
  L & \ell \\
  N & n' \\
  N' & n
\end{pmatrix}
=
\frac{1}{\nbar^2}
\,\frac{1}{2\ell+1}
\,\mathcal{M}^{LNN'}_{\ell nn'}\,.
\ea
The $W_k$ symbols are defined in \cref{eq:sfb_Wk} and discussed in
\cref{sec:w_chains}.
Then \cref{eq:Clnnobs_covariance_raw} becomes
\ba
V^W{}_{\ell nn'}^{LNN'}
&=
A_1{}^{LNN'}_{\ell nn'}
+ A_2{}^{LNN'}_{\ell nn'} + A_2{}^{LN'N}_{\ell n'n}
+ A_3{}^{LNN'}_{\ell nn'}
+ \<N \leftrightarrow N'\>.
\ea
The $A_1$ term dominates if the power spectrum is much larger than the shot
noise, and $A_3$ dominates if shot noise is larger.

\section{Chains of window functions}
\label{sec:w_chains}
Throughout the paper, we find that traces of density-contrast window coupling
matrices $W$ appear with summations over the azimuthal modes $m$. That is, we
find tensors $W_k$ of the form
\ba
\label{eq:sfb_Wk}
W_k\!\!\begin{pmatrix}
  \ell_1 & \ell_2 & \ldots & \ell_k \\
  n_1    & n_2    & \ldots & n_k \\
  n'_1   & n'_2   & \ldots & n'_k
\end{pmatrix}
&=
\sum_{m_1 m_2 \ldots m_k}
W_{n'_k \ell_k m_k}^{n_1 \ell_1 m_1}
\,W_{n'_1 \ell_1 m_1}^{n_2 \ell_2 m_2}
\cdots
W_{n'_{k-1} \ell_{k-1} m_{k-1}}^{n_k \ell_k m_k}\,.
\ea
This starts with a single window function \cref{eq:Nshot_lnn} ($k=1$) for shot
noise, two window functions \cref{eq:sfb_cmix,eq:sfb_A3} ($k=2$) for the pseudo
power spectrum mixing matrix and shot noise covariance, three window functions
\cref{eq:sfb_WWW,eq:sfb_A2} for part of the local average effect and covariance
calculations, four window functions \cref{eq:sfb_A1}. If we were to include
the local average effect in the covariance, then $k=5$ and $k=6$ would also
occur.

The $W_k$ are real, which is trivially shown by substituting each window with
its definition \cref{eq:delta_mixing_matrix_discrete} and using
\cref{eq:legendre_spherical_harmonics}. Also, cyclical permutations of the
argument columns leave $W_k$ invariant, and anticyclical permutations leave it
invariant if all $n_i$ and $n'_i$ are switched as well. That is,
\ba
\label{eq:sfb_Wk_cyclical}
W_k\!\!\begin{pmatrix}
  \ell_1 & \ell_2 & \ldots & \ell_k \\
  n_1    & n_2    & \ldots & n_k \\
  n'_1   & n'_2   & \ldots & n'_k
\end{pmatrix}
=
W_k\!\!\begin{pmatrix}
  \ell_k & \ell_1 & \ldots & \ell_{k-1} \\
  n_k    & n_1    & \ldots & n_{k-1} \\
  n'_k   & n'_1   & \ldots & n'_{k-1}
\end{pmatrix}
=
W_k\!\!\begin{pmatrix}
  \ell_k & \ell_{k-1} & \ldots & \ell_1 \\
  n'_k   & n'_{k-1}   & \ldots & n'_1   \\
  n_k    & n_{k-1}    & \ldots & n_1
\end{pmatrix}.
\ea
The first equality trivially follows from the definition \cref{eq:sfb_Wk}, and
the second from applying \cref{eq:sfb_window_symmetry} to all the windows.

\subsection{Evaluation for separable window function}
For a separable window $W(\vr)=\phi(r)\,W(\rhat)$,
\cref{eq:delta_mixing_matrix_discrete_sht} becomes
\ba
W_{n'\ell'm'}^{n \ell m}
&=
I_{n'\ell'}^{n\ell}
\,W_{\ell'm'}^{\ell m}\,,
\ea
where
\ba
I_{n'\ell'}^{n\ell}
&=
\int_{r_\min}^{r_\max}\dd r\,r^2
\,g_{n' \ell'}(r)
\,g_{n \ell}(r)
\,\phi(r)\,,
\\
W_{\ell'm'}^{\ell m}
&=
(-1)^{m'}
\sum_{LM}
\mathcal{G}^{\ell'\ell L}_{-m',m,M}
\,W_{L M}\,,
\ea
where $W_{LM}$ was defined in \cref{eq:sfb_Wlm}.
Thus, \cref{eq:sfb_Wk} is
\ba
\label{eq:sfb_Wk_separable}
W_k\!\!\begin{pmatrix}
  \ell_1 & \ell_2 & \ldots & \ell_k \\
  n_1    & n_2    & \ldots & n_k \\
  n'_1   & n'_2   & \ldots & n'_k
\end{pmatrix}
&=
I_{n'_k \ell_k}^{n_1 \ell_1}
\,I_{n'_1 \ell_1}^{n_2 \ell_2}
\cdots
I_{n'_{k-1} \ell_{k-1}}^{n_k \ell_k}
\sum_{m_1 m_2 \ldots m_k}
\,W_{\ell_k m_k}^{\ell_1 m_1}
\,W_{\ell_1 m_1}^{\ell_2 m_2}
\cdots
W_{\ell_{k-1} m_{k-1}}^{\ell_k m_k}\,.
\ea
\cref{eq:sfb_Wk_separable} contains $\orderof\!\(\ell^k\)$ terms.

\end{widetext}

\end{document}